\documentclass{aa}  

\usepackage{graphicx}
\usepackage{txfonts}
\newcommand{\um}{$\mu$m}

\newcommand{\kms}{km~s$^{-1}$}
\newcommand{\Msun}{M$_{\odot}$}

\newcommand{\water}{H$_2$O}
\newcommand{\meth}{CH$_3$OH}
\newcommand{\degreesym}{$^{\circ}$}

\begin{document} 

\title{Star Formation Sites toward the Galactic Center Region}
\subtitle{The Correlation of \meth\, Masers, \water\, Masers, and
  Near-IR Green Sources}

\author{E. T. Chambers
  \inst{1},
  F. Yusef-Zadeh
  \inst{2},
  \and
  J. Ott
  \inst{3}
}

\institute{I. Physikalisches Institut, Universit\"at zu K\"oln,
  Z\"ulpicher Str. 77, 50937 K\"oln, Germany\\
  \email{chambers@ph1.uni-koeln.de}
  \and
  Department of Physics and Astronomy and Center for Interdisciplinary 
  Research in Astronomy, Northwestern University, Evanston, IL 60208
  \and
  National Radio Astronomy Observatory, 1003 Lopezville Road, 
  Socorro, NM 87801
}

   \date{Received xxxx; accepted xxxx}

 
\abstract {} 
{We present a study of star formation in the Central
  Molecular Zone (CMZ) of our Galaxy through the association of three
  star formation indicators: 6.7~GHz \meth\, masers, 22~GHz \water\,
  masers, and enhanced 4.5~\um\, emission (`green') sources.  We
  explore how star formation in the Galactic center
  ($|\ell|~<$~1.3\degreesym, $|b|~<$~10\arcmin, where $\ell$ and $b$
  are Galactic longitude and Galactic latitude) compares with that of
  the Galactic disk (6~\degreesym$<~\ell~<~345$\degreesym,
  $|b|~<~2$\degreesym).}  
{Using an automated algorithm, we search for
  enhanced 4.5~\um\, emission sources toward 6.7~GHz \meth\, masers
  detected in the Parkes Methanol Multibeam Survey.  We combine these
  results with our 22~GHz \water\, maser survey of the CMZ carried out
  with the Mopra telescope.}  
{We find that the correlation of \meth\,
  masers with green sources is a function of Galactic latitude, with a
  minimum close to $b$=0 and increasing with $|b|$ (toward the central
  part of the Galaxy, 6~\degreesym$<~\ell~<~345$\degreesym,
  $|b|~<~2$\degreesym). We find no significant difference between the
  correlation rate of \meth\, masers with green sources in the CMZ and
  the disk.  This suggests that although the physical conditions of
  the gas are different in the Galactic center from that
  of the Galactic disk, once gravitational instability sets in at
  sufficiently high densities, signatures of star formation appear to
  be similar in both regions.  Moreover, the detection of green
  sources, even at the distance of the Galactic center, shows that our
  technique can easily identify the early stages of star formation,
  especially in low extinction regions of the Galaxy.  Through the
  association of \water\, and \meth\, masers, we identify 15
  star-forming sites in the CMZ.  We find no coincident \water\, and
  \meth\, masers outside the CMZ (with limited \water\, maser survey
  coverage outside the CMZ), possibly indicating a difference in the
  maser evolutionary sequence for star-forming cores in the Galactic
  center region and the disk.}  
{}

   \keywords{ISM: clouds--ISM: molecules--stars: formation--Galaxy: center}

   \maketitle
%

\section{Introduction}

\begin{figure*}
\includegraphics[angle=-90,width=\textwidth,clip=true]{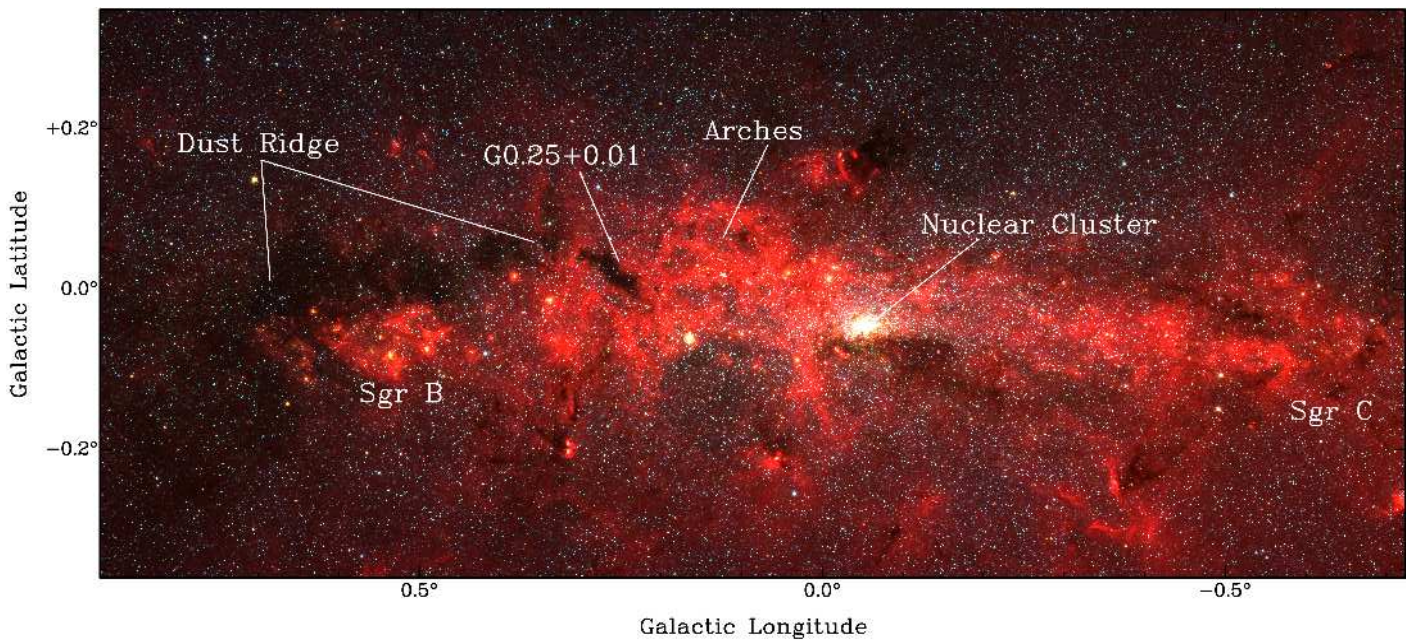}\\
\caption{{\it Spitzer}/IRAC 3-color image (with 8.0~\um\, in red, 4.5~\um\,
  in green, and 3.6~\um\, in blue) of the Central Molecular Zone.
  Some of the more prominent Galactic center features are
  labelled. Sgr~A$^*$\, is centered on the nuclear cluster. \label{galcen_color}}
\end{figure*}

As the nearest galactic nucleus, our Galactic center (GC) provides us
with the opportunity to study in detail the star formation process in
the extreme conditions at the centers of galaxies.  The inner few
hundred pc of the Galactic center, known as the Central Molecular Zone
(CMZ) contains few times 10$^7$~\Msun\, of molecular gas
\citep{pier00,ferr07,long12} and hosts several prominent sites of star
formation regions (e.g., Sgr A, Sgr B2 and Sgr C; see Fig.~\ref{galcen_color}).  By studying sites
of star formation in this complex region, we can learn global
characteristics of star formation process and compare with nuclei of
other galaxies.

The CMZ is characterized to be different from the Galactic disk in
several ways, such as its chemistry
\citep{mart00,oka05,riqu10,jone11,jone12,jone13,yuse13a}, a
two-temperature distribution of molecular gas \citep[e.g.,][]{hutt98,
  mill13}, its dust temperature being lower than its gas temperature
\citep{oden84,cox89,pier00,ferr07,moli11,imme12}, stronger turbulence
\citep[cf.][]{morr96}, and evidence that it is a region dominated by
cosmic rays \citep{yuse13a, yuse13b, ao13}.  Using thermal dust
emission data from the {\it Herschel}\, Hi-GAL Key-Project,
\citet{moli11}\, have shown that much of this mass
($\sim$~3~$\times$~10$^{7}$~\Msun) resides in a filamentary, twisted
ring-like structure that encircles the center of the Galaxy.  This
ring has a projected extent of $\sim$~180~pc, is comprised of a dense,
cold (T~$\leq$~20~K) dust cloud with high column density that encloses
warmer dust (T~$\geq$~25~K) with lower column density in the interior
of the ring. The recent results of \citet{jone12}, who used the Mopra
telescope to map the CMZ in $\sim$~20 molecular transitions at
$\sim$~3~mm, show that the clouds in the region display bright
emission with large linewidths and profiles consisting of several
velocity components, all indicating the complex kinematic structure
and composition of the gas in the Galactic center zone.

While it is clear that physical characteristics of the gas are unique
in the CMZ, it is not clear how star formation in this region compares
to that of the Galactic disk.  Assuming that the gas is distributed
uniformly throughout the inner 400~pc of the Galaxy rather than in a
ring geometry, the Kennicut law \citep{kenn98}\, still holds in a
region where tidal shear could be important \citep{yuse09}.  To make
the comparison between star formation in the CMZ and in the Galactic
disk, we look at the correlation of two star formation
indicators--maser emission (from both \meth\, and \water) and
extended, enhanced 4.5~\um\, emission.  Because of their different
emission mechanisms, the correlation between the two can help
constrain the evolutionary states of protostars.

The Class~II \meth\, maser transition at 6.7~GHz is one of the
brightest known maser lines and is known to be an excellent signpost
of star formation \citep[e.g.,][]{ment91, mini03,casw10}. The
population inversion that gives rise to the transition is thought to
be radiatively pumped, presumably by a central high-mass
($\geq$~8~\Msun) protostellar object \citep{crag92}.  It has been
shown that this maser traces only high-mass star formation
\citep{wals01, mini03}, and is typically found close to protostars
\citep{casw97, elli05}.  As such, it is used as a marker of high-mass
star formation throughout the Galaxy.  Another well-known signpost of
star formation is the \water\, maser transition at 22.23~GHz
\citep{genz78,gwin94,clau98,furu01}.  The population inversion is
created in shocks and outflows \citep{norm79, elit89, holl13}, and the
maser is associated with both low- and high-mass star formation (as
well as AGB stars and super-massive black holes), and has been the
target of many recent observations and surveys \citep[e.g.,
][]{casw11, wals11}.  Many young star forming sites in the Galaxy
harbor both \meth\, and \water\, masers.

A more recently discovered tracer of star formation activity is
extended, enhanced 4.5~\um\, emission, commonly known as `green
fuzzies' \citep{cham09}\, or extended green objects
\citep[EGOs;][]{cyga08}.  These sources appear green in {\it
  Spitzer}/IRAC \citep{fazi04} 3-color images (8.0~\um\, in red,
4.5~\um\, in green, and 3.6~\um\, in blue).  The 4.5~\um\, enhancement
is likely due to a shock-excited H$_2$ feature in the 4.5~\um\, band
\citep{nori04, debu10, fost12}, but it could also be from a
shocked CO feature \citep{mars04}.  While its exact nature is still
uncertain, this 4.5~\um\, emission, frequently referred to as `green'
emission throughout this paper, is a reliable tracer of the early
stages of star formation \citep{cham09,cyga08,cyga09}.

There have been several studies that show the correlation of green
sources with maser emission.  \citet{yuse07}\, found an association of
green sources with 6.7~GHz \meth\, masers toward the Galactic center,
and other studies have since found similar correlations in the
Galactic disk, with both 6.7~GHz \meth\, and 22~GHz \water\, masers
\citep{cham09, chen09, cyga09}.  Using a sample of 31 green sources
(15 in the CMZ, and 16 foreground to the Galactic center region),
\citet{cham11}\, searched for a difference in the correlation rate of
6.7~GHz \meth\, masers with green sources between the CMZ and the
Galactic disk.  They found no significant difference in the
correlation rate, but that study suffered from small-number
statistics.  

In this paper, we search for differences in the correlation of masers
with green sources between the Galactic center and the Galactic disk
using a different approach.  While the studies mentioned above search
for maser emission toward samples of green sources, we instead search
for extended, enhanced 4.5~\um\, emission toward 6.7~GHz \meth\, maser
emission.  We then correlate the results with a list of \water\,
masers identified in a 22.23~GHz Mopra \water\, maser survey of the
CMZ.  This method has several advantages, such as a large sample of
\meth\, masers from the Parkes Methanol Multibeam Survey
\citep{casw10}.  Moreover, it uses an automated, uniform approach to
identifying enhanced 4.5~\um\, emission. Finally, it makes use of
\water\, masers as a third tracer of star formation indicator.  Using
this method, we explore how star formation in the Galactic center
compares with that of the Galactic disk and identify sites of star
formation using three different tracers.

The structure of this paper is as follows.  In Section~\ref{data}, we
present our Mopra \water\, maser survey of the Galactic center region,
along with brief descriptions of the \meth\, maser survey carried out
by \citet{casw10}\, and the {\it Spitzer}/IRAC survey of the GC
region.  We describe our source detection and correlation
algorighthms in Section~\ref{methods}.  In Sections~\ref{results} and
\ref{results_water}, we present the results of applying the methods in
Section~\ref{methods}\, to the data in Section~\ref{data}.  We discuss
these results in Section~\ref{discussion}, and summarize our findings
in Section~\ref{conclusions}.

\section{Data \label{data}}

\subsection{6.7 GHz \meth\, Masers}
The 6.7~GHz \meth\, masers we use in our analysis were identified by
\citet[][C10 hereafter]{casw10} as part of the Parkes Methanol
Multibeam Survey. Initial detections of the 6.7~GHz \meth\, masers
were made with the Parkes Observatory \footnote{\label{foot_atnf} The
  Parkes and Mopra telescopes are part of the Australia Telescope,
  which is funded by the Commonwealth of Australia for operation as a
  National Facility managed by CSIRO.}, and subsequent follow-up
observations with the Australia Telescope Compact Array to pinpoint
the maser locations to 0.4\arcsec\, accuracy.  The full survey covers
the entire Galactic plane with a latitude range of
$|b|~<~2$\degreesym.  C10 published a subset of this survey, covering
6~\degreesym$<~\ell~<~345$\degreesym\,, $|b|~<~2$\degreesym\, with a
1$\sigma$ detection limit of 0.07~Jy.

\subsection{\water\, Maser Data \label{water_data}}

We observed the Galactic center region with the CSIRO/CASS
Mopra\footnote{See Footnote~\ref{foot_atnf}.} telescope in the
period 2006 Sep 13 to 2006 Oct 15 and 2007 July 24 to September
17. Data were taken in the on-the-fly mode, dumping data every 2\,s
with the then newly installed 12mm receiver, dual polarization. We
observed the Galactic longitude range of -1.5$^{\circ}~<~\ell~<~2^{\circ}$
and Galactic latitude range of $|b|~<~0.5^{\circ}$. The entire area was
split into smaller, typically 18$\arcmin$ on-the-fly subfields that
were repeatedly and alternately observed in Galactic latitude and
longitude directions, at least four times.

As a reference position we observed at ($\ell, b$) = (4.301$^{\circ}$,
1.667$^{\circ}$) after every row. The data were calibrated with an
internal noise diode. At the time of the observations in 2006,
however, the noise diode was not fully calibrated itself and we used
the 2007 observations to obtain pointed observations in each
on-the-fly field from which we re-calibrated the 2006 data. The
absolute calibration is estimated to be accurate to $\sim 20$\%. We
used the MOPS backend in the mops\_2208\_8192\_4f wideband mode to
cover a large range of molecular lines, including the 22.235120\,GHz
rest frequency of the $J=6_{16}-5_{23}$ water (H$_{2}$O)
transition. The full frequency coverage is 8.8\,GHz (4 subbands) with
8096 channels within a 2.2\,GHz subband. The data were processed with
the ATNF packages Livedata and Gridzilla. Livedata calibrates the
spectra using ON and OFF scans and gridzilla grids all spectra into
datacubes that we specified to be 0.5$\arcmin$ per pixel, with a FWHM
of 2.4$\arcmin$ of the Mopra beam.  The final channel width after
regridding on a common velocity axis was 3.6~\kms. The rms noise is
typically 0.01\,K per channel. For Mopra, the Jy/K conversion in the
K-band is 12.3 \citep{urqu10}.  The velocity coverage of this survey
is $-$250 to 300~\kms.

\subsection{{\it Spitzer} Data \label{irac_data}}
To identify enhanced 4.5~\um\, sources, we use data obtained with the
{\it Spitzer Space Telescope} using the Infrared Array Camera
\citep[IRAC;][]{fazi04}. In particular, we use a combination of data
from the Galactic Legacy Infrared Midplane Survey Extraordinaire
\citep[GLIMPSE;][]{benj03} and another IRAC survey toward the Galactic
center \citep{stol06, aren08, rami08}. Both data sets include data at
all 4 IRAC wavelengths: 3.6, 4.5, 5.8, and 8.0~\um. These data have an
angular resolution of $\lesssim$~2\arcsec, and a pixel size of
1.2\arcsec.  In total, the coverage of the survey in Galactic
coordinates is $|\ell|~<~65$\degreesym, $|b|~<~1$\degreesym, with
extended $b$ coverage (up to $|b|~\sim$~2\degreesym) within
5\degreesym\, of the Galactic center.

\section{Methods \label{methods}}

\subsection{Association of \meth\, masers with Green Sources}
To determine which 6.7~GHz \meth\, masers are associated with green
sources, we use an automated source detection algorithm.  This
algorithm uses an input list of \meth\, maser positions to search for
sources with enhanced 4.5~\um\, emission (`green sources') in
{\it Spitzer}/IRAC data.  The algorithm first generates a list of candidate
green sources; if the candidate green sources meet certain size and
`green' thresholds, we identify them as green sources.  This new
technique is based on combining two methods in which green sources are
detected.  One method uses ratio maps \citep{yuse09}\, and the other
uses an automated algorithm \citep{cham09}.  The details of this new
algorithm are described below.

We search for green sources in a `green ratio' image.  We generate the
green ratio image using: $I(4.5)/[I(3.6)^{1.2} \times I(5.8)]^{0.5}$,
an empirical relation identifying sources of enhanced 4.5~\um\,
emission \citep{yuse09}.  

\begin{figure}
\centering
G352.111$+$0.176\\
\includegraphics[angle=-90,width=0.6\hsize,clip=true]{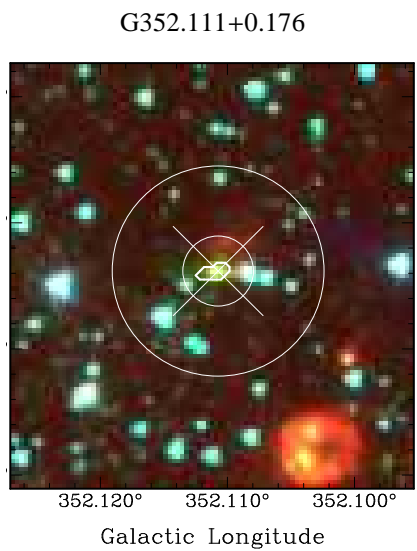}\\
\caption{{\it Spitzer}/IRAC 3-color image (with 8.0~\um\, in red, 4.5~\um\,
  in green, and 3.6~\um\, in blue) toward the position of a \meth\,
  maser (G352.111$+$0.176, $v_{LSR}~=~-54.8$~\kms), which is marked by
  the white cross sign ($\times$).  Also shown on the image are: (1)
  the search radius (smaller circle) used to identify the peak `green'
  value ($R_{peak}$), (2) the search radius (larger circle) used to
  identify find the background value ($R_{bg}$) (3) the contour that
  defines the candidate green source in the image, and (4) the
  position of the \meth\, maser ($\times$).  Note that the cross
  ($\times$) indicates only the position of the maser, and the size of
  the cross has no meaning.
\label{example}}
\end{figure}

To identify green sources associated with \meth\, masers, we find
contiguous pixels of enhanced 4.5~\um\, emission that are
above the local background and close to the maser position.  6.7~GHz
\meth\, masers are distributed in close proximity to the protostellar
objects that generate them \citep{casw97, elli05}.  As a result, we
use a search radius that is comparable to the typical green source
size of $\sim$~10--20\arcsec\, \citep[e.g.,][]{cyga08, cham09}.  Thus,
for each 6.7~GHz maser, we search within a 10\arcsec\, radius for a
local peak in the green ratio image ($R_{peak}$).  This same search
radius of 10\arcsec\, was also used by \citet{cham11} to associate
these two star formation indicators. An example in which a green
source is detected based on this algorithm is shown in Figure~\ref{example}.

To determine a local background ratio value near the position of the
6.7~GHz maser, we draw an additional aperture with a 30\arcsec\,
radius centered on the maser position (again, see Fig.~\ref{example}).
We define the local ratio background, $R_{bg}$, as the median value of
the pixels within the annulus defined by the 10\arcsec\, and
30\arcsec\, apertures.

With $R_{peak}$\, and $R_{bg}$, we calculate the mimimum ratio value,
$R_{cut}$, required to be included as part of the candidate green
source, by taking their average: $R_{cut} = [R_{peak}-R_{bg}]/2$.  To
find candidate green sources, we find all pixels ($N_{pix}$) with
$R~\geq~R_{cut}$\, that are contiguous with the peak green value
($R_{peak}$).  In this way, we find a {\it candidate} green source for
each 6.7~GHz maser. The contour in Figure~\ref{example} shows the
pixels identified as the candidate green source.

\begin{figure}
\centering
\includegraphics[angle=-90,width=\hsize,clip=true]{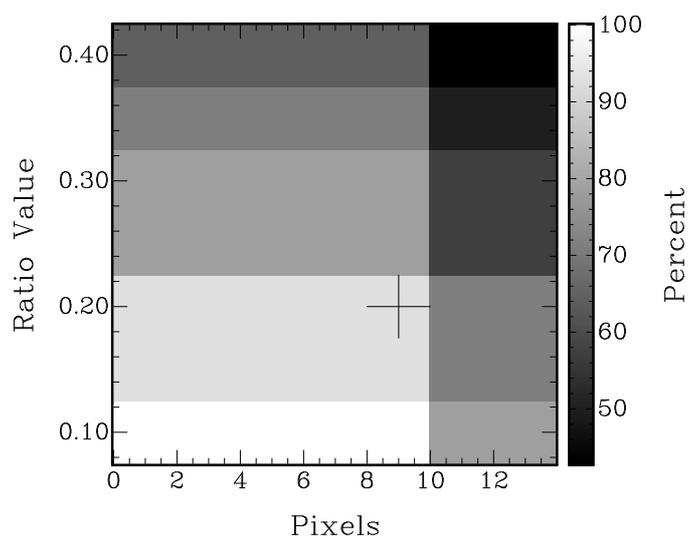}\\
\caption{Image showing the average green value vs. the number of
  pixels contained in candidate green sources.  To determine the
  values used as selection criteria in our detection
  algorithm, we apply different combinations of these values to a
  known sample of green source/maser pairs to see how many the
  algorithm recovers.  The grayscale shows the percent of known pairs
  recovered for each set of values.  In our algorithm, we use a
  minimum number of 9 pixels and a mean ratio value of 0.2 as the
  cutoffs for green sources (marked with a black $+$ in the
  image). With these values, we recover 13 of 14 (93\%) green
  source/maser pairs.\label{galcen_test}}
\end{figure}

To determine which candidate green sources are bonafide green sources,
we require that they meet two more criteria: (1) a minimum number of
pixels, and (2) a minimum green value threshold, calculated by taking
the average green value of the background-subtracted pixels
($R_{thresh}$).  To arrive at a value for this threshold, we use a
prior known sample of green source/\meth\, maser pairs from
\citet{cham11}.  Using green sources that were identified by eye,
\citet{cham11} found that 14 of the masers from C10 were associated
with their visually selected sample of green sources.  To optimize our
algorithm, we use a combination of number of pixels and R$_{thresh}$
that identifies most or all of the associations found by
\citet{cham11}, while minimizing the number of spurious associations.
We find that a minimum of 9 pixels and $R_{thresh}~\geq$~0.2 properly
identifies 13 of the 14 (93\%) green source/\meth\, maser pairs from
\citet{cham11} and finds few spurious detections.  This result is
shown graphically in Figure~\ref{galcen_test}.  Thus, we use these
values to determine our final green source list from the candidate
green sources.

To examine the bias in this method, we search for green sources
toward random positions.  These results are found in Section~\ref{random}.

\subsection{Identification of \water\, Masers \label{find_water}}
To identify \water\, masers in the Mopra CMZ map, we use the Clumpfind
algorithm \citep{will94}.  Clumpfind identifies clumps by finding
local peaks within a contoured data cube.  Contour intervals (used to
separate blended features) and lower detection limits are provided by
the user, and are typically a multiple of the signal-to-noise ratio
(SNR) of the data.

\begin{figure}
\centering
\includegraphics[angle=-90,width=\hsize,clip=true]{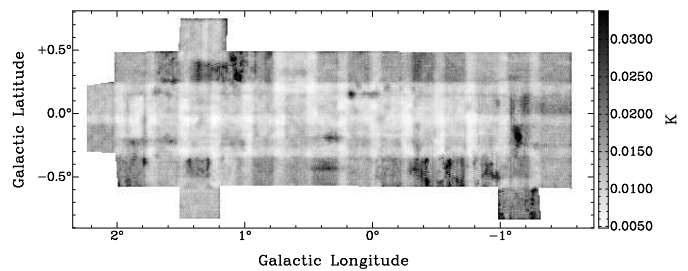}\\
\caption{Image displaying the noise in the spectrum at each position
  in the Mopra data cube.  Because of the uneven noise throughout the
  data, we divide the original data cube by this 2-D image, resulting
  in a signal-to-noise data cube. The grayscale bar on the right shows
  the noise temperature in units of K (the conversion to Jy is
  12.3~Jy/K). \label{noise}}
\end{figure}

In order to determine an appropriate contour interval and lower limit
for our Mopra data, we first analyze the SNR of the input data cube.
After masking the line-free channels at each position, we calculate
the noise in each spectrum, resulting in the 2-d 'noise' image shown
in Figure~\ref{noise}.  This 2-d image clearly shows that the noise in
the Mopra \water\, data is not uniform throughout the entire cube,
which is likely a result of having in different integration times in
some regions, as well as varying weather during the observations.
Because Clumpfind works best on data with uniform noise, we divide the
original data cube by the 2-d SNR image, resulting in a data cube
containing SNR values.  This cube is then used as the input for
Clumpfind.  We find that, for the SNR data cube, a minimum threshold
(i.e., detection limit) of 4 and a contour interval (used to separate
blended features) of 4 work best for identifying the \water\, masers
in the data.

In addition to the threshold and contour levels, we also select a
minimum number of volume elements (voxels) required for a source to be
identified.  Based on the Mopra beam size at the frequency of the
\water\, maser transition (FWHM~$\sim$~144\arcsec\, at 22.2~GHz) and the
angular size of one pixel (30\arcsec by 30\arcsec), we require a
source to have at least 18~voxels to be included in our final list of
masers.

\section{Results \label{results}}

\subsection{\meth\, Masers and Green Sources \label{dist-results}}
\begin{figure}
\includegraphics[angle=0,width=\hsize,clip=true]{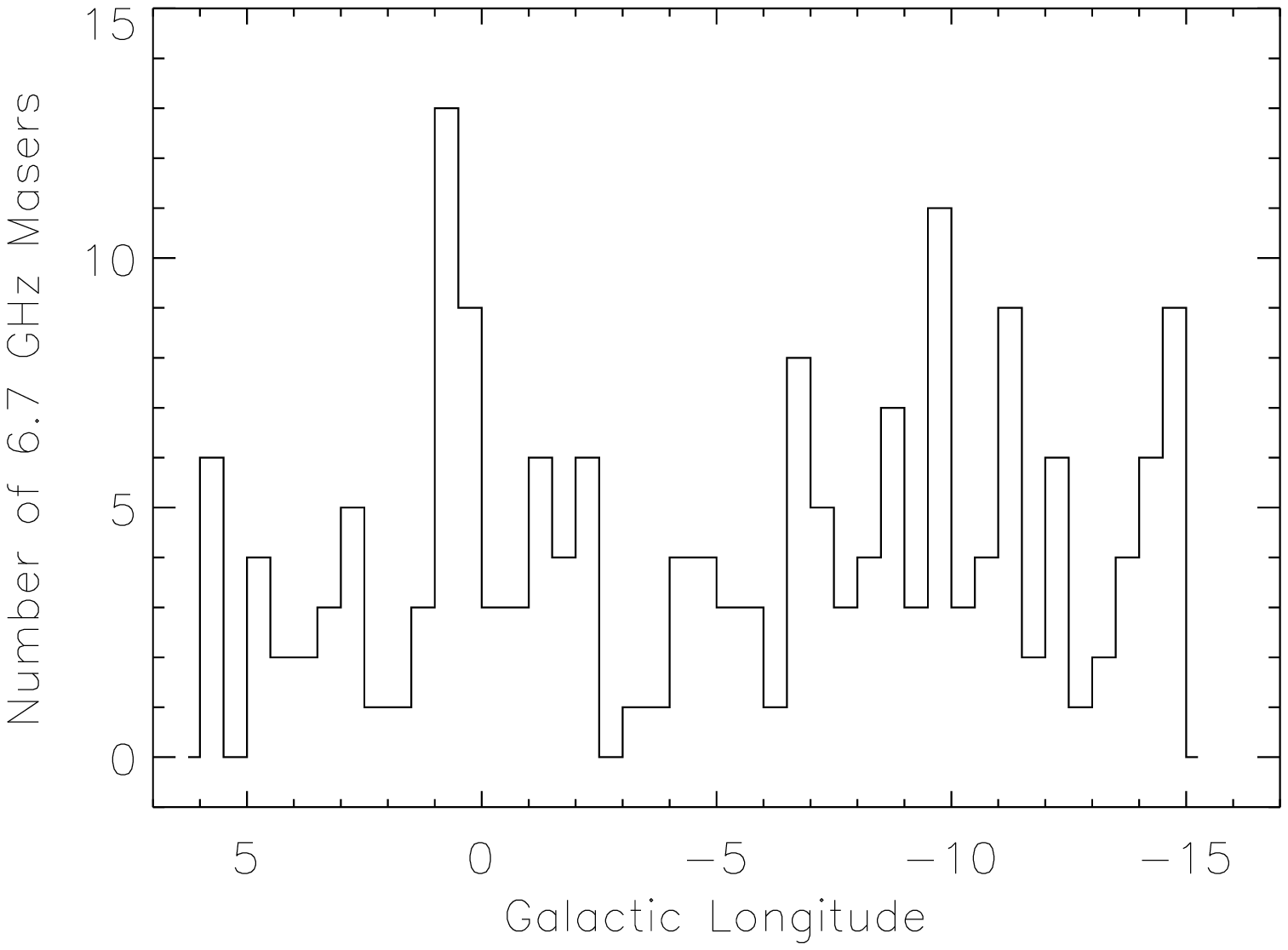}\\
\includegraphics[angle=0,width=\hsize,clip=true]{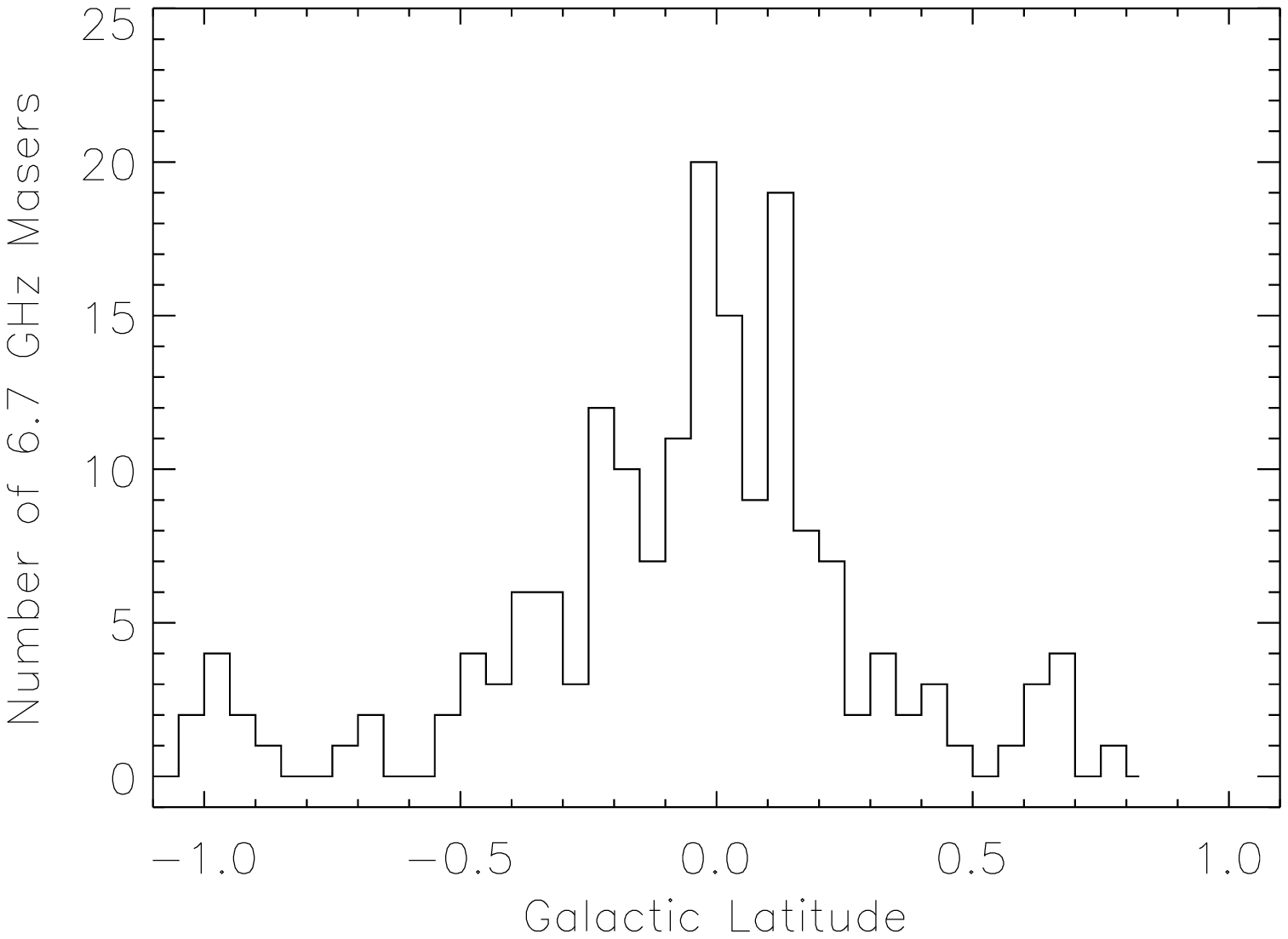}\\
\caption{{\it Top}: Histogram of the distribution of 6.7~GHz masers
  in Galactic longitude (integrated over all Galactic latitudes).  The bin size is 1.5\degreesym.  The peak of
  the distribution is largely due to the cluster of 11 masers in
  Sgr~B2.  {\it Bottom}: Histogram of the distribution of 6.7~GHz
  masers in Galactic latitude (integrated over all Galactic longitudes).  The bin size is 0.15\degreesym. The
  masers are distributed around the Galactic
  equator. \label{ell_bee_histos}}
\end{figure}

\begin{figure*}
\includegraphics[angle=-90,width=\textwidth,clip=true]{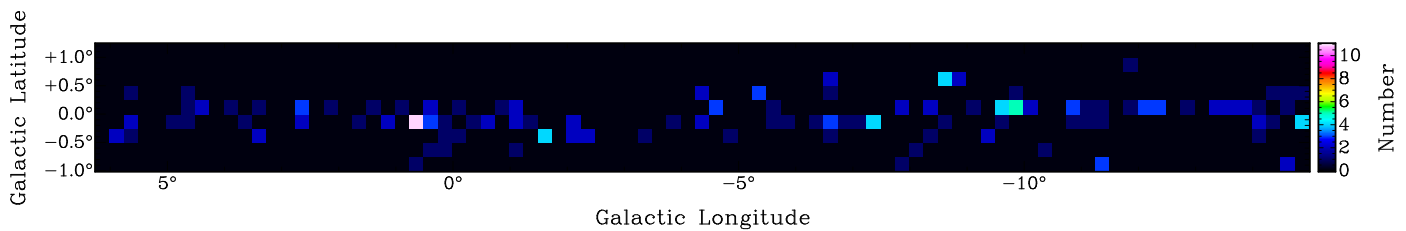}\\
\caption{Distribution of the 6.7~GHz masers from C10.  The pixel
  size is 0.25\degreesym.  The bright peak at ($\ell,b$)~=~(0.66,$-$0.04) is from a cluster of
  masers in Sgr~B2.  \label{caswell_dist}}
\end{figure*}

C10 detect 183 \meth\, masers in their survey region
(6~\degreesym$<~\ell~<~ 345$\degreesym\,, $|b|~<~2$\degreesym), with a
1$\sigma$ detection limit of 0.07~Jy.  Histograms of the maser
distribution as a function of $\ell$ and $b$ can be seen in
Figure~\ref{ell_bee_histos}.  The distribution in Galactic latitude
shows that the masers have a roughly Gaussian distribution, centered
on the Galactic plane.  The Galactic longitude distribution displays a
peak toward the Galactic center, mostly due to a large number of
masers in the Sgr~B star forming region.  Figure~\ref{caswell_dist}
contains a 2-D image of the maser distribution, and clearly shows the
peak of the maser distribution in Sgr~B.

\begin{figure}
\centering
\includegraphics[angle=0,width=\hsize,clip=true]{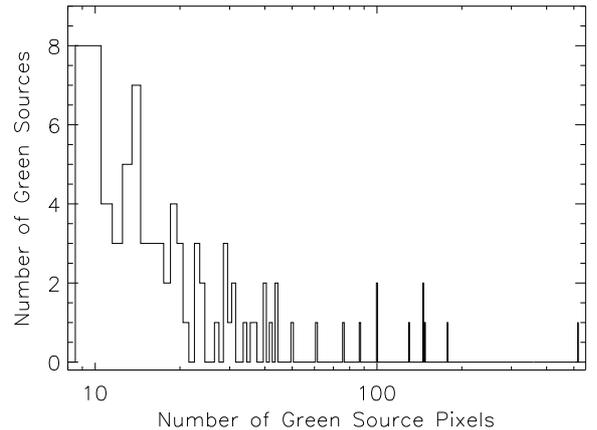}\\
\caption{Histogram of the number of pixels in the green sources associated
  with 6.7~GHz \meth\, masers.  The sizes (in IRAC pixels) range from 9 (the
  minimum based on our detection method) to 517, with a median of
  17.\label{pix_histo}}
\end{figure}

We run the green source detection algorithm (described above) on 175
of the 183 masers in the C10 catalog (8 were exluded because we do not
have corresponding IRAC data).  We find that 86/175 (49.1~$\pm$~5.3\%)
of the 6.7~GHz masers are associated with a green source. These green
sources range in size (in the 1.2\arcsec\, by 1.2\arcsec\, IRAC
pixels) from 9 (the minimum based on our detection method) to 517,
with a median of 17 IRAC pixels (see Fig.~\ref{pix_histo}). 

A summary of these results can be found in Table~\ref{green-summary},
which lists, for each \meth\, maser, the maser name (Column~1; from
C10), the velocity of peak maser emission (Column~2; from C10), the
Galactic longitude, latitude and value for the peak green pixel in
each green source candidate, $R_{peak}$ (Columns~3, 4, and 5), the
background green value near each green source candidate, $R_{bg}$
(Column~6), the green value used to define the boundary of the green
source candidate, $R_{cut}$ (Column~7), the average green value for
the pixels within the green source candidate, $R_{thresh}$ (Column~8),
and the number of pixels, $N_{pix}$, in the green source candidate
(Column~9).  The final column in Table~\ref{green-summary} contains a
flag ('Y' or 'N'; Column~10) to indicate if a green source is
associated with a \meth\, maser based on the values in Columns~3
through 9.

\begin{longtab}
\begin{longtable}{cccccccccc}
\caption{\label{green-summary}Association of Green Sources with \meth\, Masers}\\
\hline\hline
Maser  & $V_{peak}$\tablefootmark{a}  & \multicolumn{3}{c}{$R_{peak}$}  & 
$R_{bg}$   & $R_{cut}$      & $R_{thresh}$  & $N_{pix}$    & Green \\
\cline{3-5}
Name\tablefootmark{a}     & (\kms) & $\ell$\tablefootmark{b} & 
$b$\tablefootmark{b}   & Value                             &   &
    & &    & Source\\
\hline
\endfirsthead
\caption{continued.}\\
\hline\hline
Maser  & $V_{peak}$\tablefootmark{a}  & \multicolumn{3}{c}{$R_{peak}$}  & 
$R_{bg}$   & $R_{cut}$      & $R_{thresh}$  & $N_{pix}$    & Green \\
\cline{3-5}
Name\tablefootmark{a}     & (\kms) & $\ell$\tablefootmark{b} & 
$b$\tablefootmark{b}   & Value                             &   &
    & &    & Source\\
\hline
\endhead
\hline
\endfoot
G345.003$-$0.223 & $-$23.1 &  345.0038 &  $-$0.2258 &   1.178 &   0.495 &   0.342 &   0.463 &          100 & Y \\
G345.003$-$0.224 & $-$26.2 &  345.0038 &  $-$0.2258 &   1.178 &   0.495 &   0.342 &   0.463 &          100 & Y \\
G345.131$-$0.174 & $-$28.9 &  345.1308 &  $-$0.1742 &   0.968 &   0.270 &   0.349 &   0.476 &           11 & Y \\
G345.198$-$0.030 & $-$0.5 &  345.1978 &  $-$0.0302 &   0.521 &   0.272 &   0.124 &   0.170 &           13 & N \\
G345.205$+$0.317 & $-$63.5 &  345.2048 &   0.3175 &   0.606 &   0.260 &   0.173 &   0.266 &            5 & N \\
G345.407$-$0.952 & $-$14.3 &  345.4065 &  $-$0.9518 &   0.429 &   0.170 &   0.130 &   0.167 &            4 & N \\
G345.424$-$0.951 & $-$13.2 &  345.4245 &  $-$0.9512 &   0.326 &   0.168 &   0.079 &   0.116 &            4 & N \\
G345.441$+$0.205 & 0.9 &  345.4412 &   0.2048 &   0.695 &   0.276 &   0.209 &   0.312 &            7 & N \\
G345.487$+$0.314 & $-$22.6 &  345.4875 &   0.3152 &   0.513 &   0.240 &   0.137 &   0.184 &           52 & N \\
G345.505$+$0.348 & $-$17.8 &  345.5042 &   0.3465 &   1.029 &   0.351 &   0.339 &   0.437 &           87 & Y \\
G345.576$-$0.225 & $-$126.8 &  345.5762 &  $-$0.2255 &   0.683 &   0.292 &   0.196 &   0.291 &           13 & Y \\
G345.807$-$0.044 & $-$2 &  345.8068 &  $-$0.0442 &   0.828 &   0.287 &   0.271 &   0.416 &            7 & N \\
G345.824$+$0.044 & $-$10.3 &  345.8235 &   0.0432 &   0.912 &   0.268 &   0.322 &   0.435 &           10 & Y \\
G345.949$-$0.268 & $-$21.9 &  345.9495 &  $-$0.2678 &   0.671 &   0.294 &   0.189 &   0.295 &           12 & Y \\
G345.985$-$0.020 & $-$83.2 &  345.9852 &  $-$0.0202 &   1.191 &   0.249 &   0.471 &   0.605 &           13 & Y \\
G346.036$+$0.048 & $-$6.4 &  346.0362 &   0.0485 &   1.070 &   0.252 &   0.409 &   0.581 &            6 & N \\
G346.231$+$0.119 & $-$95 &  346.2325 &   0.1195 &   0.722 &   0.271 &   0.225 &   0.319 &           21 & Y \\
G346.480$+$0.221 & $-$18.9 &  346.4805 &   0.2212 &   1.162 &   0.307 &   0.427 &   0.620 &           13 & Y \\
G346.481$+$0.132 & $-$5.5 &  346.4815 &   0.1322 &   0.897 &   0.280 &   0.308 &   0.441 &           14 & Y \\
G346.517$+$0.117 & $-$1.7 &  346.5172 &   0.1175 &   0.402 &   0.223 &   0.090 &   0.129 &            3 & N \\
G346.522$+$0.085 & 5.7 &  346.5225 &   0.0842 &   0.443 &   0.220 &   0.111 &   0.153 &           22 & N \\
G347.230$+$0.016 & $-$68.9 &  347.2302 &   0.0155 &   0.769 &   0.212 &   0.278 &   0.418 &            3 & N \\
G347.583$+$0.213 & $-$102.3 &  347.5835 &   0.2125 &   0.358 &   0.244 &   0.057 &   0.114 &            1 & N \\
G347.628$+$0.149 & $-$96.5 &  347.6278 &   0.1485 &   0.707 &   0.227 &   0.240 &   0.340 &           16 & Y \\
G347.631$+$0.211 & $-$91.9 &  347.6342 &   0.2082 &   0.343 &   0.232 &   0.056 &   0.101 &            2 & N \\
G347.817$+$0.018 & $-$24.1 &  347.8172 &   0.0178 &   0.926 &   0.264 &   0.331 &   0.449 &            8 & N \\
G347.863$+$0.019 & $-$34.7 &  347.8632 &   0.0185 &   0.632 &   0.234 &   0.199 &   0.291 &            9 & Y \\
G347.902$+$0.052 & $-$27.4 &  347.9042 &   0.0515 &   0.428 &   0.205 &   0.112 &   0.152 &          123 & N \\
G348.027$+$0.106 & $-$121.2 &  348.0268 &   0.1058 &   0.954 &   0.267 &   0.343 &   0.473 &           17 & Y \\
G348.195$+$0.768 & $-$0.8 &  348.1918 &   0.7665 &   1.076 &   0.332 &   0.372 &   0.744 &            1 & N \\
G348.550$-$0.979 & $-$10.6 &  348.5502 &  $-$0.9785 &   0.914 &   0.271 &   0.322 &   0.426 &          146 & Y \\
G348.550$-$0.979n & $-$20 &  348.5502 &  $-$0.9785 &   0.914 &   0.271 &   0.322 &   0.426 &          146 & Y \\
G348.579$-$0.920 & $-$15.1 &  348.5785 &  $-$0.9198 &   1.062 &   0.440 &   0.311 &   0.411 &          148 & Y \\
G348.654$+$0.244 & 16.9 &  348.6542 &   0.2445 &   0.747 &   0.275 &   0.236 &   0.355 &           11 & Y \\
G348.723$-$0.078 & 11.5 &  348.7232 &  $-$0.0775 &   0.602 &   0.305 &   0.149 &   0.209 &            7 & N \\
G348.703$-$1.043 & $-$3.5 &  348.7032 &  $-$1.0422 &   0.309 &   0.177 &   0.066 &   0.085 &           36 & N \\
G348.727$-$1.037 & $-$7.4 &  348.7272 &  $-$1.0372 &   0.860 &   0.158 &   0.351 &   0.532 &            8 & N \\
G348.884$+$0.096 & $-$74.5 &  348.8845 &   0.0965 &   0.863 &   0.274 &   0.295 &   0.367 &           17 & Y \\
G348.892$-$0.180 & 1.5 &  348.8918 &  $-$0.1795 &   1.140 &   0.282 &   0.429 &   0.630 &            5 & N \\
G349.067$-$0.017 & 11.6 &  349.0672 &  $-$0.0168 &   0.593 &   0.262 &   0.166 &   0.241 &           12 & Y \\
G349.092$+$0.105 & $-$76.6 &  349.0915 &   0.1052 &   1.357 &   0.236 &   0.560 &   0.827 &           18 & Y \\
G349.092$+$0.106 & $-$81.5 &  349.0915 &   0.1052 &   1.357 &   0.236 &   0.560 &   0.827 &           18 & Y \\
G349.151$+$0.021 & 14.6 &  349.1525 &   0.0222 &   0.697 &   0.268 &   0.215 &   0.279 &            7 & N \\
G349.579$-$0.679 & $-$25 &  349.5792 &  $-$0.6792 &   0.568 &   0.285 &   0.141 &   0.210 &            3 & N \\
G349.799$+$0.108 & $-$64.7 &  349.7985 &   0.1085 &   0.806 &   0.268 &   0.269 &   0.421 &            9 & Y \\
G349.884$+$0.231 & 16.2 &  349.8832 &   0.2302 &   0.767 &   0.263 &   0.252 &   0.378 &           17 & Y \\
G350.015$+$0.433 & $-$30.3 &  350.0155 &   0.4332 &   0.615 &   0.241 &   0.187 &   0.261 &           16 & Y \\
G350.104$+$0.084 & $-$68.1 &  350.1048 &   0.0838 &   0.492 &   0.223 &   0.134 &   0.181 &           31 & N \\
G350.105$+$0.083 & $-$74.1 &  350.1048 &   0.0838 &   0.492 &   0.223 &   0.134 &   0.181 &           31 & N \\
G350.116$+$0.084 & $-$68 &  350.1168 &   0.0842 &   0.716 &   0.237 &   0.240 &   0.320 &            8 & N \\
G350.116$+$0.220 & 4.2 &  350.1155 &   0.2205 &   0.997 &   0.279 &   0.359 &   0.541 &            4 & N \\
G350.189$+$0.003 & $-$62.4 &  350.1868 &   0.0022 &   0.711 &   0.268 &   0.221 &   0.343 &            4 & N \\
G350.299$+$0.122 & $-$62.1 &  350.2982 &   0.1198 &   0.700 &   0.231 &   0.234 &   0.334 &           31 & Y \\
G350.340$+$0.141 & $-$58.4 &  350.3402 &   0.1412 &   0.499 &   0.246 &   0.126 &   0.163 &           10 & N \\
G350.344$+$0.116 & $-$65.4 &  350.3438 &   0.1165 &   0.611 &   0.274 &   0.168 &   0.226 &            8 & N \\
G350.356$-$0.068 & $-$67.6 &  350.3562 &  $-$0.0678 &   0.874 &   0.263 &   0.306 &   0.431 &           19 & Y \\
G350.470$+$0.029 & $-$6.3 &  350.4702 &   0.0288 &   0.491 &   0.270 &   0.111 &   0.186 &            4 & N \\
G350.520$-$0.350 & $-$24.6 &  350.5195 &  $-$0.3492 &   1.017 &   0.269 &   0.374 &   0.547 &           14 & Y \\
G350.686$-$0.491 & $-$13.7 &  350.6855 &  $-$0.4905 &   1.349 &   0.312 &   0.519 &   0.749 &           10 & Y \\
G350.776$+$0.138 & 38.7 &  350.7755 &   0.1382 &   0.826 &   0.255 &   0.286 &   0.380 &            4 & N \\
G351.161$+$0.697 & $-$5.2 &  351.1602 &   0.6968 &   0.853 &   0.392 &   0.231 &   0.321 &          517 & Y \\
G351.242$+$0.670 & 2.5 &  351.2425 &   0.6708 &   0.377 &   0.194 &   0.092 &   0.136 &           54 & N \\
G351.251$+$0.652 & $-$7.1 &  351.2522 &   0.6515 &   0.620 &   0.166 &   0.227 &   0.290 &           61 & Y \\
G351.382$-$0.181 & $-$59.7 &  351.3822 &  $-$0.1808 &   0.807 &   0.277 &   0.265 &   0.407 &           10 & Y \\
G351.417$+$0.645 & $-$10.4 &  351.4168 &   0.6462 &   0.587 &   0.187 &   0.200 &   0.273 &           23 & Y \\
G351.417$+$0.646 & $-$11.1 &  351.4168 &   0.6462 &   0.587 &   0.187 &   0.200 &   0.273 &           23 & Y \\
G351.445$+$0.660 & $-$7.1 &  351.4442 &   0.6602 &   0.548 &   0.243 &   0.152 &   0.204 &           27 & Y \\
G351.581$-$0.353 & $-$94.2 &  351.5818 &  $-$0.3532 &   1.574 &   0.337 &   0.619 &   0.885 &           10 & Y \\
G351.611$+$0.172 & $-$43.7 &  351.6115 &   0.1698 &   0.284 &   0.195 &   0.044 &   0.075 &          309 & N \\
G351.688$+$0.171 & $-$36.1 &  351.6878 &   0.1708 &   0.726 &   0.258 &   0.234 &   0.343 &           13 & Y \\
G351.775$-$0.536 & 1.3 &  351.7752 &  $-$0.5345 &   1.790 &   0.501 &   0.644 &   0.797 &           20 & Y \\
G352.083$+$0.167 & $-$66 &  352.0835 &   0.1672 &   0.615 &   0.259 &   0.178 &   0.259 &           13 & Y \\
G352.111$+$0.176 & $-$54.8 &  352.1108 &   0.1762 &   0.852 &   0.259 &   0.296 &   0.404 &           16 & Y \\
G352.133$-$0.944 & $-$7.7 &  352.1325 &  $-$0.9425 &   0.977 &   0.359 &   0.309 &   0.378 &           50 & Y \\
G352.517$-$0.155 & $-$51.2 &  352.5175 &  $-$0.1552 &   0.398 &   0.239 &   0.079 &   0.127 &            6 & N \\
G352.525$-$0.158 & $-$53 &  352.5248 &  $-$0.1582 &   0.854 &   0.238 &   0.308 &   0.483 &            7 & N \\
G352.584$-$0.185 & $-$85.7 &  352.5835 &  $-$0.1848 &   0.694 &   0.252 &   0.221 &   0.334 &            6 & N \\
G352.604$-$0.225 & $-$81.7 &  352.6042 &  $-$0.2252 &   0.829 &   0.284 &   0.272 &   0.368 &           14 & Y \\
G352.855$-$0.201 & $-$51.3 &  352.8548 &  $-$0.2005 &   0.616 &   0.231 &   0.193 &   0.295 &            4 & N \\
G353.216$-$0.249 & $-$22.9 &  353.2132 &  $-$0.2478 &   0.394 &   0.215 &   0.089 &   0.122 &            8 & N \\
G353.273$+$0.641 & $-$4.4 &  353.2725 &   0.6418 &   0.920 &   0.181 &   0.370 &   0.463 &           19 & Y \\
G353.363$-$0.166 & $-$79 &  353.3635 &  $-$0.1678 &   0.352 &   0.203 &   0.075 &   0.102 &            7 & N \\
G353.370$-$0.091 & $-$44.7 &  353.3705 &  $-$0.0905 &   0.462 &   0.251 &   0.105 &   0.154 &           16 & N \\
G353.378$+$0.438 & $-$15.7 &  353.3782 &   0.4375 &   0.419 &   0.252 &   0.084 &   0.115 &            4 & N \\
G353.410$-$0.360 & $-$20.3 &  353.4105 &  $-$0.3602 &   0.566 &   0.258 &   0.154 &   0.235 &           11 & Y \\
G353.429$-$0.090 & $-$61.8 &  353.4268 &  $-$0.0892 &   0.837 &   0.307 &   0.265 &   0.408 &            7 & N \\
G353.464$+$0.562 & $-$50.3 &  353.4635 &   0.5628 &   1.161 &   0.294 &   0.434 &   0.605 &           29 & Y \\
G353.537$-$0.091 & $-$56.6 &  353.5372 &  $-$0.0912 &   0.971 &   0.269 &   0.351 &   0.518 &           14 & Y \\
G354.206$-$0.038 & $-$37.1 &  354.2052 &  $-$0.0382 &   0.613 &   0.227 &   0.193 &   0.386 &            1 & N \\
G354.308$-$0.110 & 18.8 &  354.3075 &  $-$0.1098 &   0.575 &   0.279 &   0.148 &   0.231 &            5 & N \\
G354.496$+$0.083 & 27 &  354.4955 &   0.0825 &   0.390 &   0.238 &   0.076 &   0.098 &           14 & N \\
G354.615$+$0.472 & $-$24.4 &  354.6132 &   0.4718 &   0.619 &   0.284 &   0.168 &   0.228 &           37 & Y \\
G354.701$+$0.299 & 102.8 &  354.6998 &   0.3002 &   0.975 &   0.323 &   0.326 &   0.537 &            4 & N \\
G354.724$+$0.300 & 93.9 &  354.7232 &   0.3012 &   1.118 &   0.341 &   0.388 &   0.574 &            4 & N \\
G355.184$-$0.419 & $-$1.4\tablefootmark{c} &  355.1825 &  $-$0.4188 &   1.557 &   0.403 &   0.577 &   0.720 &           15 & Y \\
G355.343$+$0.148 & 5.8 &  355.3442 &   0.1475 &   0.485 &   0.270 &   0.107 &   0.157 &           14 & N \\
G355.344$+$0.147 & 19.9 &  355.3442 &   0.1475 &   0.485 &   0.270 &   0.107 &   0.157 &           14 & N \\
G355.346$+$0.149 & 10.5 &  355.3455 &   0.1492 &   0.510 &   0.273 &   0.118 &   0.184 &            4 & N \\
G355.538$-$0.105 & 3.8 &  355.5382 &  $-$0.1045 &   0.603 &   0.272 &   0.165 &   0.245 &           29 & Y \\
G355.545$-$0.103 & $-$28.2 &  355.5452 &  $-$0.1035 &   0.566 &   0.267 &   0.150 &   0.210 &           20 & Y \\
G355.642$+$0.398 & $-$7.9 &  355.6418 &   0.3978 &   1.064 &   0.286 &   0.389 &   0.778 &            1 & N \\
G355.666$+$0.374 & $-$3.3 &  355.6665 &   0.3735 &   0.428 &   0.264 &   0.082 &   0.123 &           11 & N \\
G356.054$-$0.095 & 16.7 &  356.0545 &  $-$0.0972 &   0.442 &   0.281 &   0.080 &   0.139 &            2 & N \\
G356.662$-$0.263 & $-$53.8 &  356.6612 &  $-$0.2628 &   1.188 &   0.356 &   0.416 &   0.570 &            9 & Y \\
G357.558$-$0.321 & $-$3.9 &  357.5575 &  $-$0.3215 &   0.776 &   0.328 &   0.224 &   0.307 &            8 & N \\
G357.559$-$0.321 & 16.2 &  357.5575 &  $-$0.3215 &   0.776 &   0.328 &   0.224 &   0.307 &            8 & N \\
G357.922$-$0.337 & $-$4.6 &  357.9252 &  $-$0.3368 &   0.800 &   0.279 &   0.260 &   0.385 &           24 & Y \\
G357.924$-$0.337 & $-$2.1 &  357.9255 &  $-$0.3372 &   0.828 &   0.279 &   0.274 &   0.389 &           23 & Y \\
G357.965$-$0.164 & $-$8.6 &  357.9642 &  $-$0.1638 &   0.911 &   0.274 &   0.318 &   0.476 &           10 & Y \\
G357.967$-$0.163 & $-$3.1 &  357.9645 &  $-$0.1642 &   0.848 &   0.274 &   0.287 &   0.425 &           14 & Y \\
G358.371$-$0.468 & 1.3 &  358.3712 &  $-$0.4685 &   0.799 &   0.231 &   0.284 &   0.425 &           20 & Y \\
G358.386$-$0.483 & $-$6 &  358.3868 &  $-$0.4812 &   0.585 &   0.281 &   0.152 &   0.226 &          178 & Y \\
G358.460$-$0.391 & 1.3 &  358.4598 &  $-$0.3935 &   0.981 &   0.329 &   0.326 &   0.458 &           34 & Y \\
G358.460$-$0.393 & $-$7.3 &  358.4602 &  $-$0.3938 &   0.975 &   0.324 &   0.326 &   0.456 &           36 & Y \\
G358.721$-$0.126 & 10.6 &  358.7215 &  $-$0.1262 &   0.917 &   0.316 &   0.301 &   0.410 &            6 & N \\
G358.809$-$0.085 & $-$56.2 &  358.8085 &  $-$0.0852 &   1.466 &   0.492 &   0.487 &   0.673 &           12 & Y \\
G358.841$-$0.737 & $-$20.7 &  358.8405 &  $-$0.7365 &   0.707 &   0.309 &   0.199 &   0.282 &           31 & Y \\
G358.906$+$0.106 & $-$18.1 &  358.9058 &   0.1045 &   0.530 &   0.363 &   0.083 &   0.107 &            4 & N \\
G358.931$-$0.030 & $-$15.9 &  358.9318 &  $-$0.0305 &   0.519 &   0.299 &   0.110 &   0.158 &            7 & N \\
G358.980$+$0.084 & 6.2 &  358.9798 &   0.0842 &   1.330 &   0.515 &   0.407 &   0.537 &           29 & Y \\
G359.138$+$0.031 & $-$3.9 &  359.1378 &   0.0325 &   0.503 &   0.306 &   0.099 &   0.142 &           23 & N \\
G359.436$-$0.104 & $-$47.8 &  359.4368 &  $-$0.1018 &   0.942 &   0.292 &   0.325 &   0.434 &           40 & Y \\
G359.436$-$0.102 & $-$53.3 &  359.4368 &  $-$0.1018 &   0.942 &   0.292 &   0.325 &   0.434 &           40 & Y \\
G359.615$-$0.243 & 19.3 &  359.6125 &  $-$0.2415 &   1.154 &   0.328 &   0.413 &   0.559 &            9 & Y \\
G359.938$+$0.170 & $-$0.5 &  359.9382 &   0.1705 &   1.401 &   0.511 &   0.445 &   0.590 &           30 & Y \\
G359.970$-$0.457 & 23 &  359.9692 &  $-$0.4588 &   0.735 &   0.263 &   0.236 &   0.313 &           42 & Y \\
G0.092$+$0.663 & 23.8 &    0.0902 &  $-$0.6632 &   1.918 &   0.459 &   0.729 &   0.995 &           19 & Y \\
G0.167$-$0.446 & 13.8 &    0.1672 &  $-$0.4438 &   1.330 &   0.362 &   0.484 &   0.690 &           14 & Y \\
G0.212$-$0.001 & 49.5 &    0.2118 &  $-$0.0008 &   0.314 &   0.229 &   0.042 &   0.066 &            6 & N \\
G0.315$-$0.201 & 19.4 &    0.3155 &  $-$0.2002 &   0.536 &   0.190 &   0.173 &   0.239 &           44 & Y \\
G0.316$-$0.201 & 21.1 &    0.3155 &  $-$0.2002 &   0.536 &   0.190 &   0.173 &   0.239 &           44 & Y \\
G0.376$+$0.040 & 37.1 &    0.3762 &   0.0402 &   1.114 &   0.342 &   0.386 &   0.550 &           14 & Y \\
G0.409$-$0.504 & 25.4 &    0.4085 &  $-$0.5042 &   0.699 &   0.211 &   0.244 &   0.357 &           15 & Y \\
G0.475$-$0.010 & 28.8 &    0.4748 &  $-$0.0098 &   0.374 &   0.317 &   0.029 &   0.048 &            4 & N \\
G0.496$+$0.188 & 0.9 &    0.4965 &   0.1885 &   0.522 &   0.244 &   0.139 &   0.225 &            8 & N \\
G0.546$-$0.852 & 11.8 &    0.5462 &  $-$0.8505 &   0.679 &   0.190 &   0.245 &   0.345 &            9 & Y \\
G0.645$-$0.042 & 49.5 &    0.6438 &  $-$0.0405 &   0.847 &   0.293 &   0.277 &   0.366 &            6 & N \\
G0.647$-$0.055 & 51\tablefootmark{c} &    0.6462 &  $-$0.0552 &   0.510 &   0.282 &   0.114 &   0.183 &            2 & N \\
G0.651$-$0.049 & 48.3 &    0.6528 &  $-$0.0498 &   0.430 &   0.285 &   0.073 &   0.108 &            3 & N \\
G0.657$-$0.041 & 49.9\tablefootmark{c} &    0.6578 &  $-$0.0415 &   0.856 &   0.291 &   0.282 &   0.385 &           19 & Y \\
G0.665$-$0.036 & 60.4 &    0.6668 &  $-$0.0352 &   1.845 &   0.301 &   0.772 &   1.024 &            9 & Y \\
G0.666$-$0.029 & 70.5 &    0.6662 &  $-$0.0288 &   0.483 &   0.283 &   0.100 &   0.145 &            4 & N \\
G0.667$-$0.034 & 55\tablefootmark{c} &    0.6668 &  $-$0.0352 &   1.845 &   0.301 &   0.772 &   1.024 &            9 & Y \\
G0.672$-$0.031 & 58.2 &    0.6735 &  $-$0.0308 &   0.429 &   0.306 &   0.061 &   0.123 &            1 & N \\
G0.673$-$0.029 & 66\tablefootmark{c} &    0.6722 &  $-$0.0285 &   0.454 &   0.287 &   0.084 &   0.111 &            6 & N \\
G0.677$-$0.025 & 73.3 &    0.6782 &  $-$0.0242 &   0.477 &   0.288 &   0.094 &   0.152 &            2 & N \\
G0.695$-$0.038 & 68.6 &    0.6948 &  $-$0.0378 &   0.562 &   0.307 &   0.127 &   0.197 &            5 & N \\
G0.836$+$0.184 & 3.5 &    0.8352 &   0.1848 &   0.314 &   0.245 &   0.035 &   0.070 &            1 & N \\
G1.008$-$0.237 & 1.6 &    1.0085 &  $-$0.2378 &   0.450 &   0.266 &   0.092 &   0.135 &           19 & N \\
G1.147$-$0.124 & $-$15.3 &    1.1485 &  $-$0.1268 &   0.704 &   0.226 &   0.239 &   0.328 &            4 & N \\
G1.329$+$0.150 & $-$12.2 &    1.3285 &   0.1502 &   0.287 &   0.191 &   0.048 &   0.096 &            1 & N \\
G1.719$-$0.088 & $-$8.1 &    1.7188 &  $-$0.0878 &   1.203 &   0.374 &   0.414 &   0.543 &            6 & N \\
G2.143$+$0.009 & 62.6 &    2.1448 &   0.0098 &   0.676 &   0.319 &   0.178 &   0.267 &           24 & Y \\
G2.521$-$0.220 & $-$6.1 &    2.5205 &  $-$0.2198 &   0.857 &   0.297 &   0.280 &   0.425 &           10 & Y \\
G2.536$+$0.198 & 3.1 &    2.5358 &   0.1998 &   1.259 &   0.489 &   0.385 &   0.493 &          130 & Y \\
G2.591$-$0.029 & $-$8.3 &    2.5892 &  $-$0.0282 &   0.472 &   0.263 &   0.104 &   0.209 &            1 & N \\
G2.615$+$0.134 & 94.1 &    2.6165 &   0.1345 &   0.380 &   0.264 &   0.058 &   0.084 &            6 & N \\
G2.703$+$0.040 & 93.5 &    2.7062 &   0.0395 &   0.800 &   0.302 &   0.249 &   0.498 &            1 & N \\
G3.253$+$0.018 & 2.2 &    3.2532 &   0.0185 &   0.727 &   0.306 &   0.211 &   0.346 &            5 & N \\
G3.312$-$0.399 & 0.4 &    3.3118 &  $-$0.3965 &   1.649 &   0.413 &   0.618 &   0.939 &           10 & Y \\
G3.442$-$0.348 & $-$35.1 &    3.4418 &  $-$0.3482 &   2.928 &   0.388 &   1.270 &   1.946 &            4 & N \\
G3.502$-$0.200 & 43.9 &    3.5008 &  $-$0.2008 &   1.081 &   0.333 &   0.374 &   0.517 &            6 & N \\
G3.910$+$0.001 & 17.8 &    3.9102 &   0.0008 &   0.734 &   0.258 &   0.238 &   0.347 &            9 & Y \\
G4.393$+$0.079 & 1.9 &    4.3935 &   0.0788 &   0.475 &   0.240 &   0.117 &   0.151 &            7 & N \\
G4.434$+$0.129 & $-$1 &    4.4342 &   0.1288 &   0.595 &   0.261 &   0.167 &   0.258 &            5 & N \\
G4.569$-$0.079 & 9.5 &    4.5685 &  $-$0.0792 &   0.729 &   0.268 &   0.230 &   0.302 &            7 & N \\
G4.586$+$0.028 & 26.1 &    4.5848 &   0.0252 &   0.609 &   0.280 &   0.164 &   0.329 &            1 & N \\
G4.676$+$0.276 & 4.5 &    4.6752 &   0.2775 &   0.842 &   0.299 &   0.272 &   0.434 &            4 & N \\
G4.866$-$0.171 & 5.4 &    4.8658 &  $-$0.1708 &   0.488 &   0.306 &   0.091 &   0.132 &            9 & N \\
G5.618$-$0.082 & $-$27.1 &    5.6188 &  $-$0.0812 &   1.113 &   0.294 &   0.410 &   0.523 &           15 & Y \\
G5.630$-$0.294 & 10.5 &    5.6302 &  $-$0.2938 &   0.586 &   0.286 &   0.150 &   0.231 &           10 & Y \\
G5.657$+$0.416 & 20 &    5.6555 &   0.4158 &   0.775 &   0.288 &   0.244 &   0.366 &           11 & Y \\
G5.677$-$0.027 & $-$11.7 &    5.6772 &  $-$0.0272 &   0.531 &   0.274 &   0.128 &   0.208 &            8 & N \\
G5.885$-$0.393 & 6.7 &    5.8832 &  $-$0.3935 &   0.641 &   0.208 &   0.217 &   0.263 &           76 & Y \\
G5.900$-$0.430 & 10.4 &    5.9002 &  $-$0.4318 &   0.224 &   0.153 &   0.035 &   0.049 &            5 & N \\
\end{longtable}
\tablefoot{
\tablefoottext{a}{From C10.}
\tablefoottext{b}{Derived from IRAC data, which have a pixel size of 1.2\arcsec.}
\tablefoottext{c}{For \meth\, masers not detected in the C10 follow-up data, we use their survey cube data
or the value they list from \citet{houg95}.}
}
\end{longtab}

\begin{figure}
\includegraphics[angle=0,width=\hsize,clip=true]{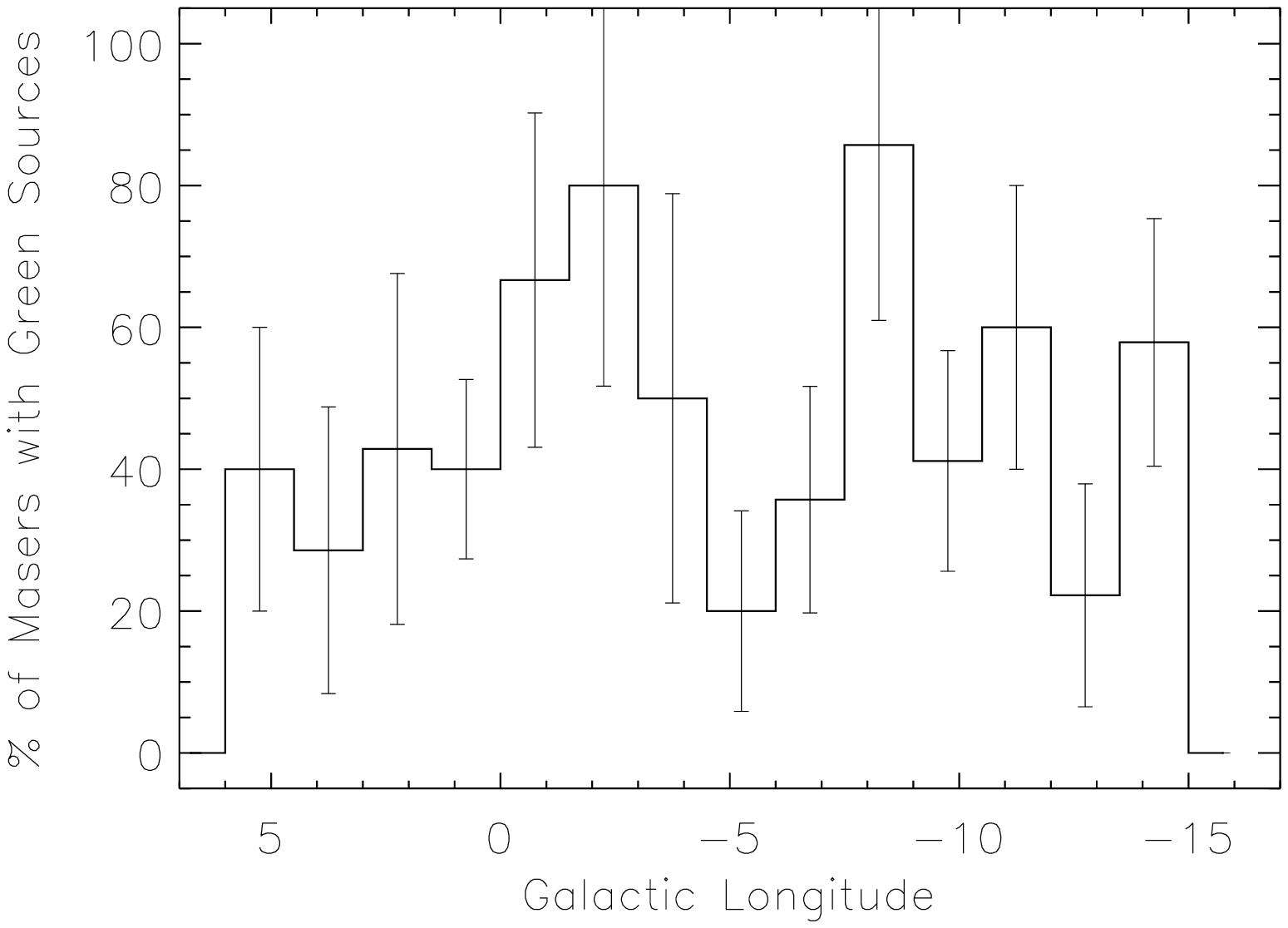}
\includegraphics[angle=0,width=\hsize,clip=true]{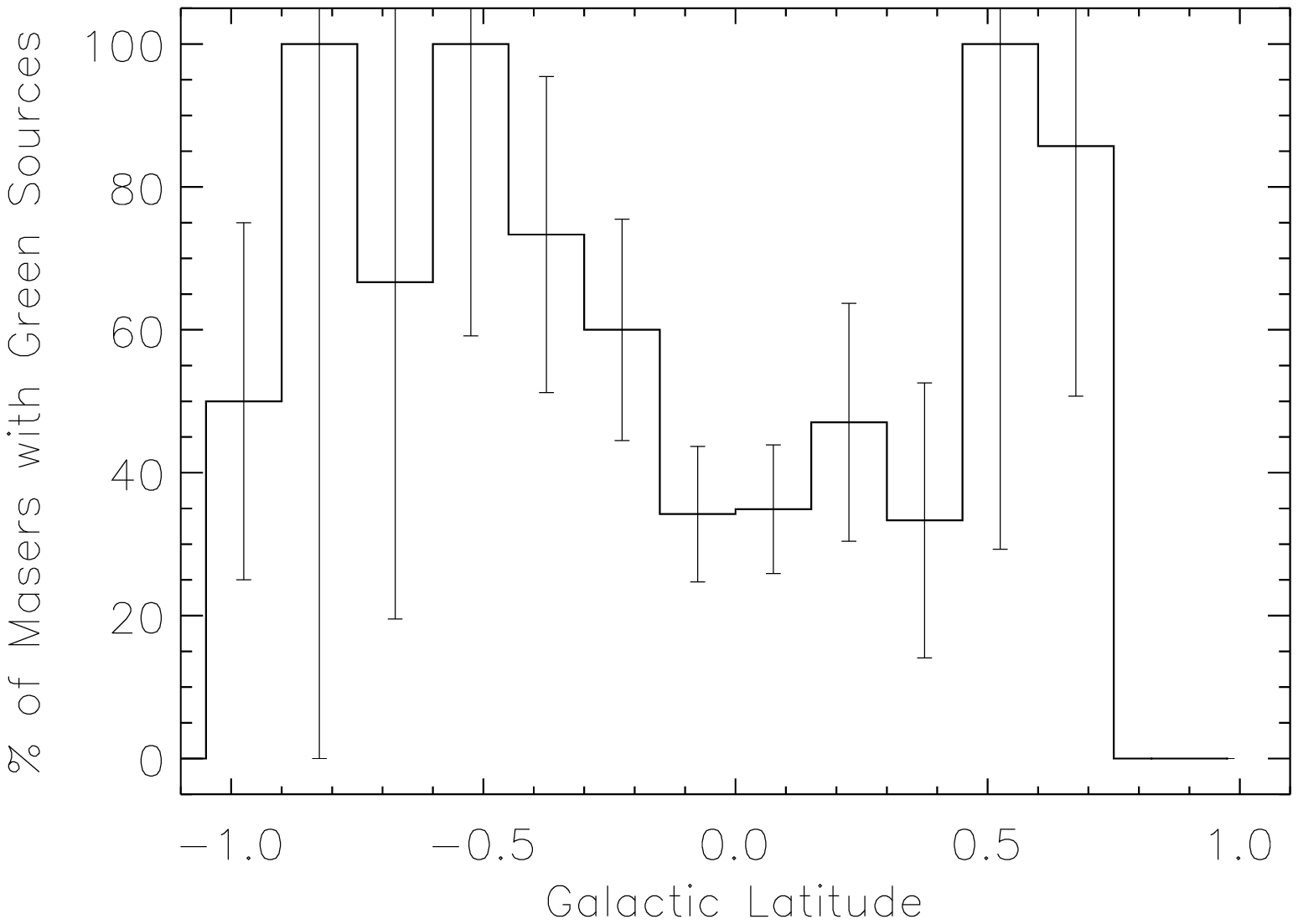}\\
\caption{{\it Top}: Histogram of the percentage of 6.7 GHz masers
  associated with green sources as a function of Galactic longitude.
  {\it Bottom}: Histogram of the percentage of 6.7 GHz masers
  associated with green sources as a function of Galactic
  latiitude. The bin size is 0.15\degreesym. Errors for the detection
  rates in both histograms are calculated using $\sqrt{N}$ counting
  statistics. \label{hit_pct}}
\end{figure}

Using these detected sources, we calculate the detection rates for green
sources toward 6.7~GHz masers as a function of Galactic longitude and
latitude.  In Figure~\ref{hit_pct}, we show that the longitude
distribution of detection rates varies between 20\% and 86\%, with
typical errors of $\sim$~20\%.  The distribution shows possible
enhancements at $\ell~\sim~-2$\degreesym and $\ell~\sim~-$8\degreesym,
and a depressed rate at $\ell~\sim~-$5\degreesym.  The distribution of
green source detection rates as a function of Galactic latitude, shown
in Figure~\ref{hit_pct}, shows a clear deficit of green sources
detected near the midplane ($b~\sim~0$\degreesym) of the Galaxy.  The
central 0.3\degreesym\, has a detection rate of $\sim$~35~$\pm$~10\%,
while the regions with $|b|~>$~0.5\degreesym\, have detection rates of
$>$~50\%, with some bins at 100\%.

\begin{figure*}
\includegraphics[angle=-90,width=\textwidth,clip=true]{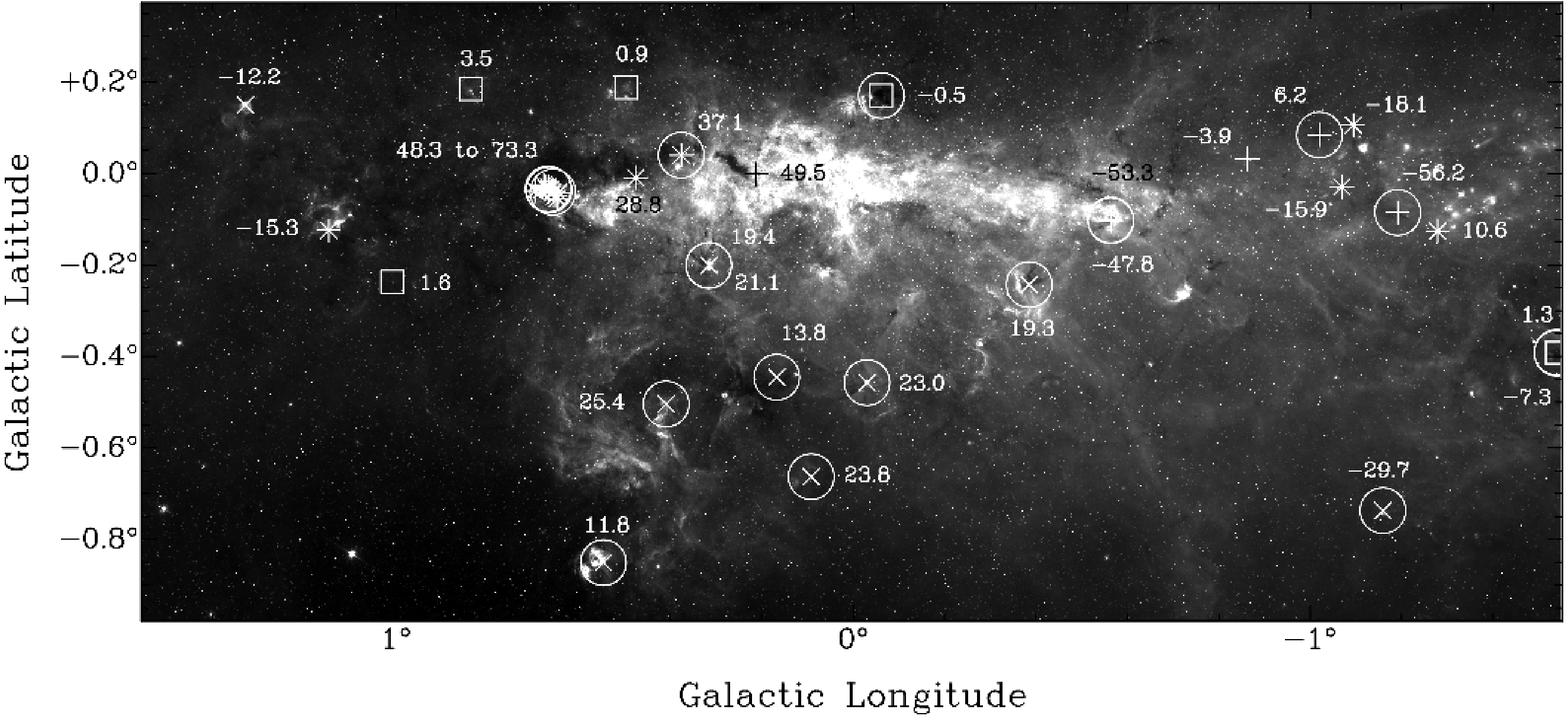}\\
\includegraphics[angle=-90,width=\textwidth,clip=true]{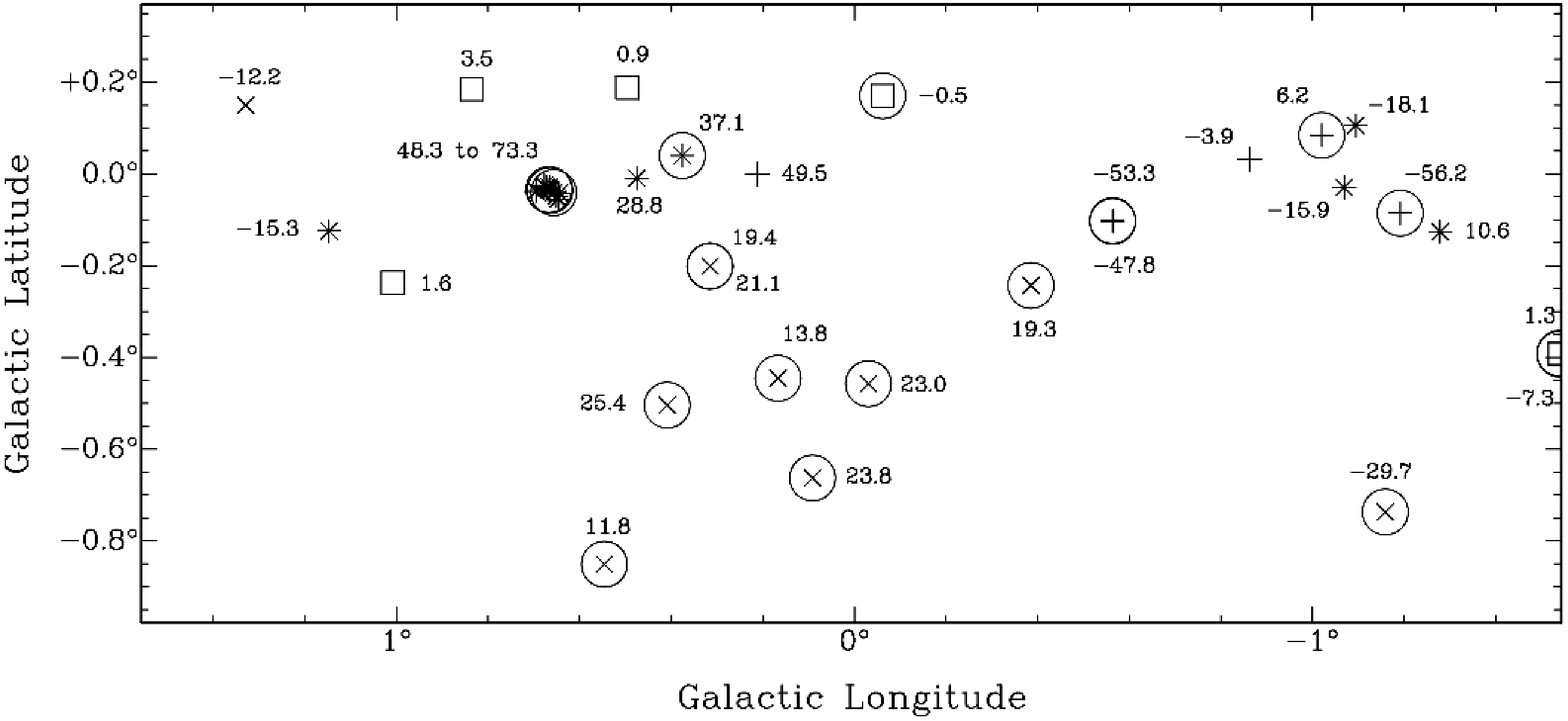}\\
\caption{{\it Top}: {\it Spitzer}/IRAC 8~\um\, image of the Galactic
  center region showing 6.7~GHz \meth\, masers.  Crosses (+) indicate
  the positions of 6.7~GHz \meth\, masers placed in the Galactic
  center region based on position.  The $\times$ symbols indicate the
  positions of 6.7~GHz \meth\, masers placed in the Galactic center
  region based on velocity.  The asterisks (*) indicate the positions
  of 6.7~GHz \meth\, masers placed in the Galactic center region based
  on both distance criteria.  The squares show the locations of the
  masers thought to reside in the disk.  Sources that are circled are
  associated with an enhanced 4.5~\um\, emission source. Each maser is
  labelled with the velocity (in \kms) of the peak maser
  emission. {\it Bottom}: Same as the {\it top} figure, but with a
  blank backround instead of the 8~\um\, image.
  \label{gc_disk_dist}}
\end{figure*}

Because we want to compare Galactic center candidate sources to
Galactic disk sources, we need a method (or methods) to estimate the
distances to the sources.  To determine if the green sources are
foreground or Galactic center sources, we use their Galactic
coordinates with respect to the Galactic plane and their corresponding
maser velocities. 
First, we use the Galactic coordinates of the sources to separate our
sources into GC sources and non-GC sources.  To cover the CMZ, and to
match the scale height of young stellar objects in the Galactic center
region \citep{yuse09}, we define the GC region to be
$|\ell|~<$~1.3\degreesym, $|b|~<$~10\arcmin.  Using these criteria for
distance, we find that 8/23 GC \meth\, masers (34.8~$\pm$~12.3\%) are
associated with green sources, and 78/152 (51.3~$\pm$~5.8\%) non-GC
\meth\, masers are associated with green sources.  These association
rates are listed in Table~\ref{hit-results}, which lists the method of
distance estimation in Column~1, the location of the masers in
Column~2, the number of \meth\, masers at that distance in Column~3,
the number of those masers that are associated with green sources in
Column~4, and the association rate calculated on those numbers in
Column~5. Figure~\ref{gc_disk_dist}\, shows the positions of the
\meth\, masers, their assigned locations (based on their Galactic
coordinates), and the velocities of the peak maser emission.

\begin{table}
\caption{\label{hit-results}Association Rates of \meth\, Masers with Green Sources}
\begin{tabular}{lcccc}
\hline\hline
Distance  & Galactic & \meth & Green Source & Association \\
Method    & Location & Masers & Associations & Rate \\ 
\hline
Position\tablefootmark{a}  & GC   & 23  & 8  & 34.8~$\pm$~12.3\% \\
               & Disk & 152 & 78 & 51.3~$\pm$~5.8\%  \\
Velocity\tablefootmark{b} & GC   & 27  & 13 & 48.1~$\pm$~13.3\% \\
               & Disk & 148 & 73 & 49.3~$\pm$~5.8\%  \\
Both           & GC   & 17  & 4  & 23.5~$\pm$~11.8\% \\
               & Disk & 158 & 82 & 51.9~$\pm$~5.7\%  \\
\cline{3-5}
Total          &      & 175 & 86 & 49.1~$\pm$~5.3\%  \\
\hline
\end{tabular}
\tablefoot{ 
\tablefoottext{a}{Based on Galactic coordinates, with
    $|\ell|~<$~1.3\degreesym\, and $|b|~<$~10\arcmin\, defining the GC
    region (see Section~\ref{dist-results}).}  
\tablefoottext{b}{Based on the velocities of the masers and a model
  of circular Galactic rotation (see Section~\ref{dist-results}).}
}
\end{table}

Next, we use the distance estimates made by C10, who based their
estimates on the velocities of the \meth\, masers and a model of
circular Galactic rotation (and allowing for small deviations; see C10
for more details).  In short, some maser velocities are consistent
with Galactic rotation models, and are placed outside the GC region.
Other masers have velocities consistent with a location within the
3~kpc arms, and are assigned that location.  Those masers that fit
neither of these categories, and have $|\ell|~<$1.3\degreesym\, (with
no restriction in Galactic latitude) are placed at the distance of the
GC. Using these criteria for distance, we find that 13/27 GC \meth\,
masers (48.1~$\pm$~13.3\%) are associated with green sources, and
73/148 (49.3~$\pm$~5.8\%) non-GC \meth\, masers are associated with
green sources.

Finally, we also look at the subset of sources that meet the position
(by Galactic coordinates) {\it and} velocity criteria described above.
These masers all have positions with $|\ell|~<$~1.3\degreesym,
$|b|~<$~10\arcmin, and velocities that are inconsistent with both
standard Galactic rotation and placement in the 3~kpc arms.  Using
these criteria for distance, we find that 4/17 GC \meth\, masers
(23.5~$\pm$~11.8\%) are associated with green sources, and 82/158
(51.9~$\pm$~5.7\%) non-GC \meth\, masers are associated with green
sources.

\subsubsection{Random Test \label{random}}
To determine if the association rates are true associations of green
sources with 6.7~GHz masers or simply chance alignments of the two
star formation indicators, we placed test masers in the survey
longitude and latitude ranges and searched for green sources using the
same method described above.  Because the C10 \meth\, masers are
roughly evenly distributed in longitude, our test masers have a
uniform distribution in longitude.  To match the latitude distribution
of masers seen in Figure~\ref{ell_bee_histos}, we fit a Gaussian to
the distribution and used its parameters to generate the Galactic
latitude distribution of test masers.  We placed 175 test masers in
the survey region, to match the number of masers in the C10 catalog.
To increase the statistics, we ran the test a total of ten times.  We
find a test maser-green source association rate of 1.3~$\pm$~0.3\%\,
for our total sample of 1,750 test masers.  Thus, we consider the
chances of our detection rates being contaminated by random chance
associations to be small.

\subsection{\water\, Masers \label{results_water}}
\begin{figure*}
\includegraphics[width=0.32\textwidth,clip=true]{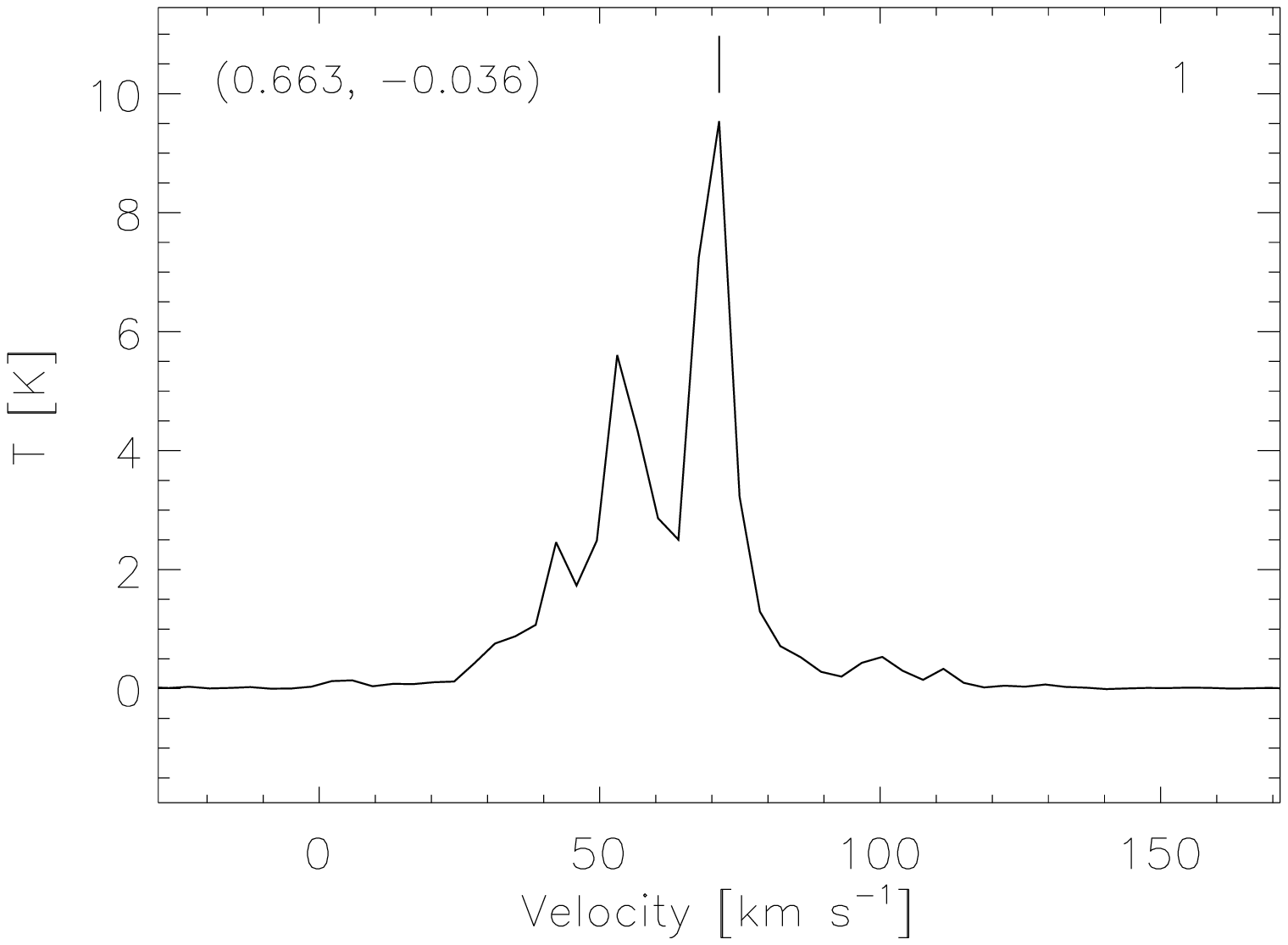}
\includegraphics[width=0.32\textwidth,clip=true]{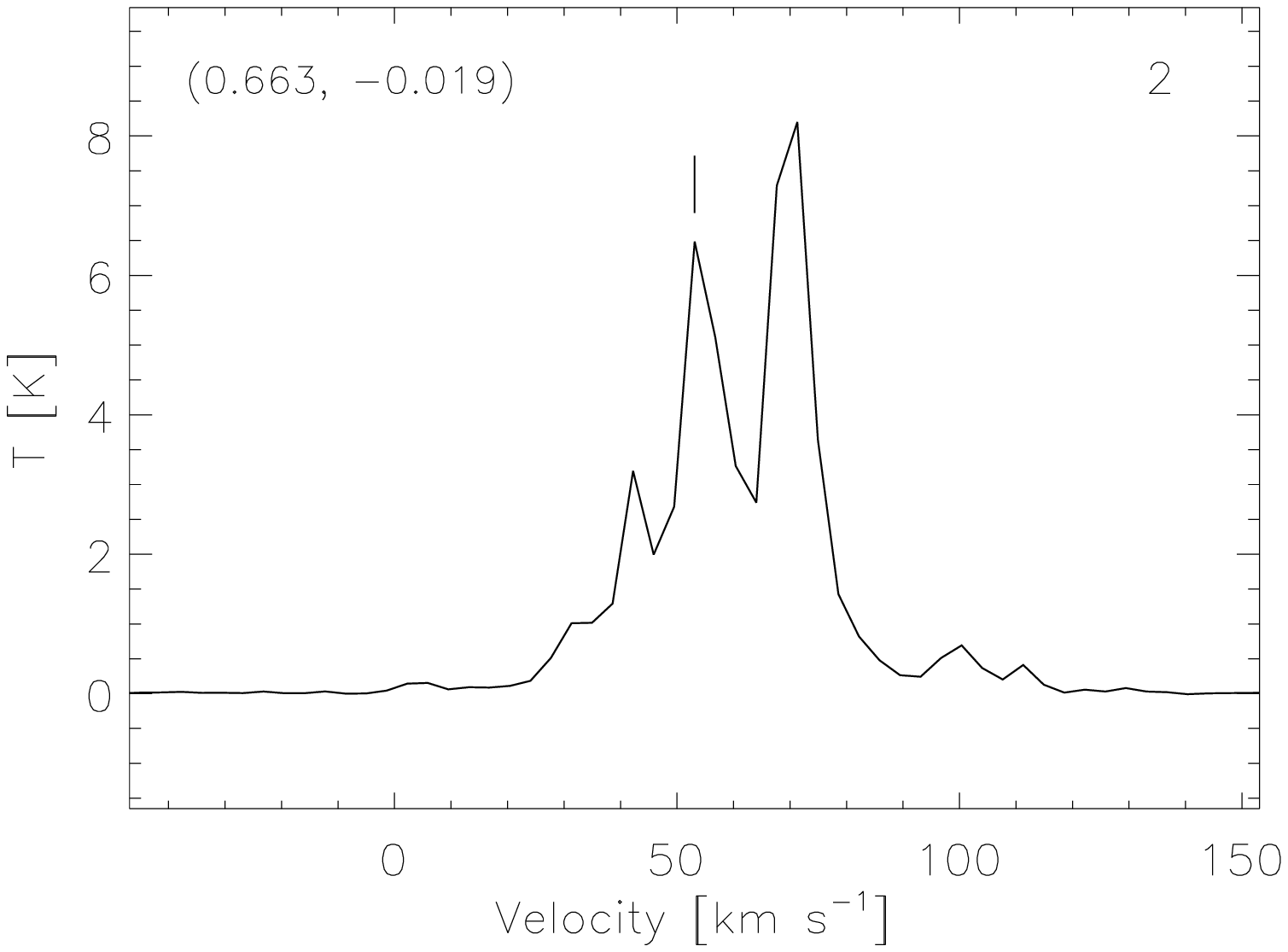}
\includegraphics[width=0.32\textwidth,clip=true]{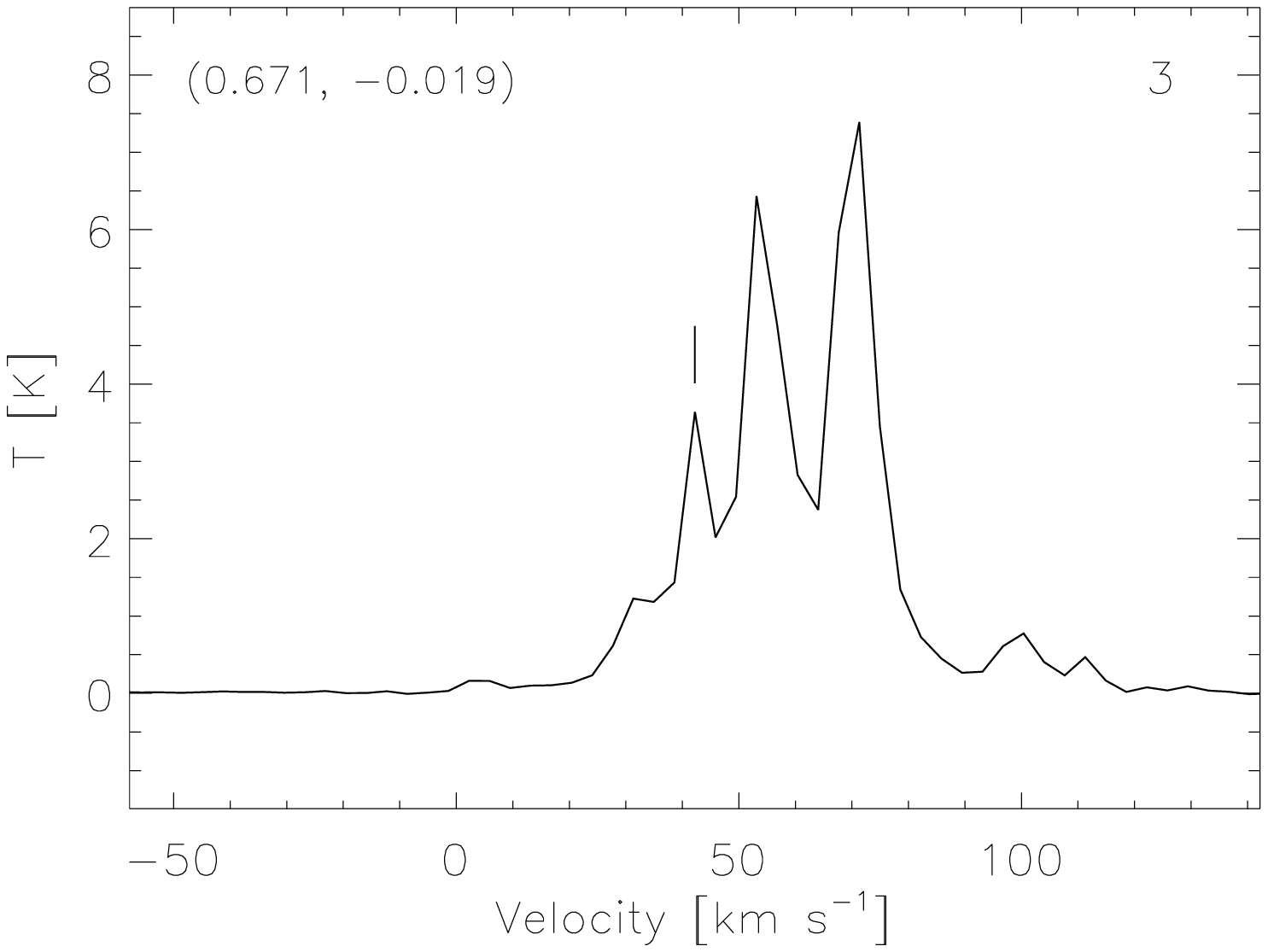}\\ 
\includegraphics[width=0.32\textwidth,clip=true]{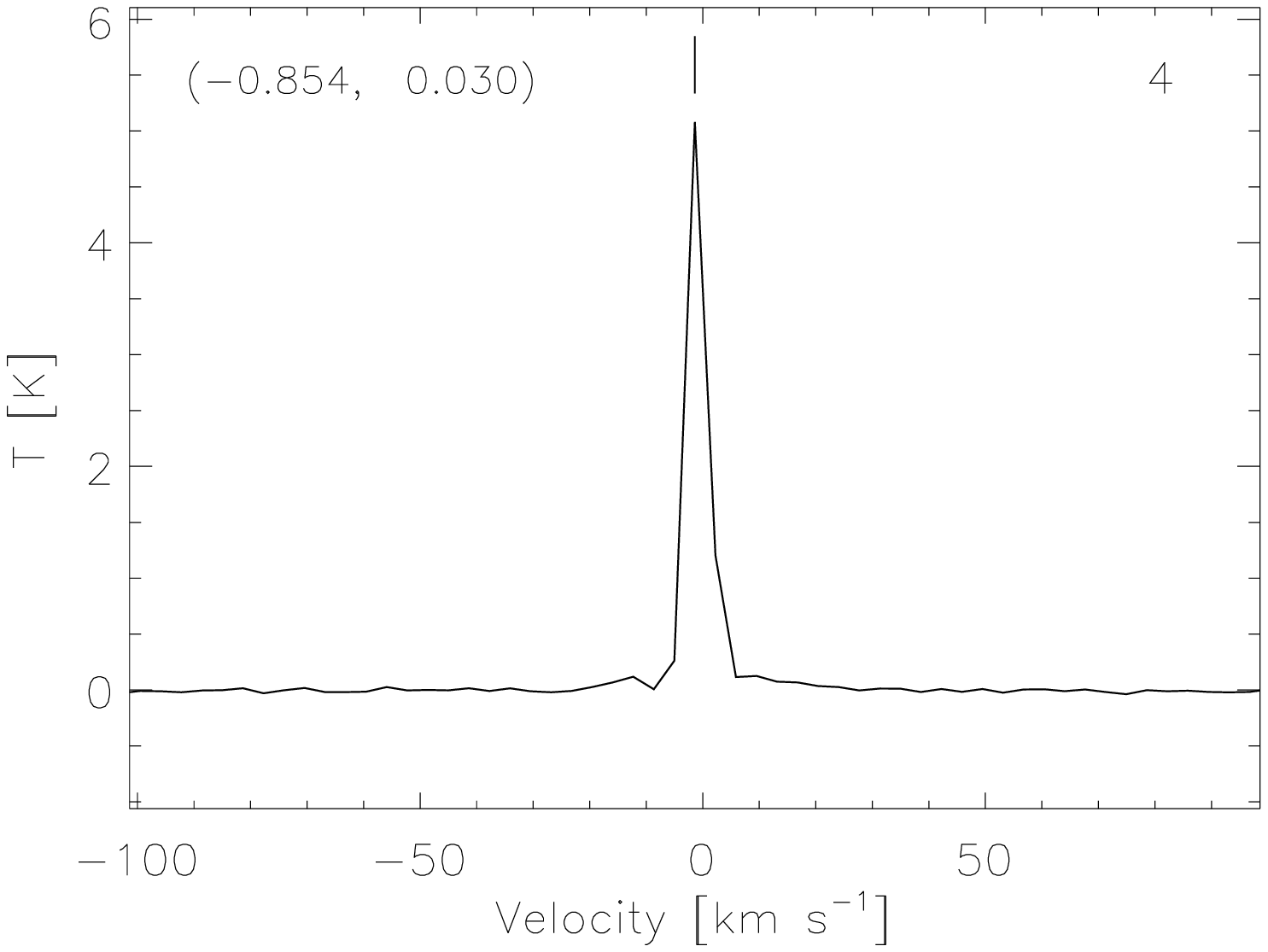}
\includegraphics[width=0.32\textwidth,clip=true]{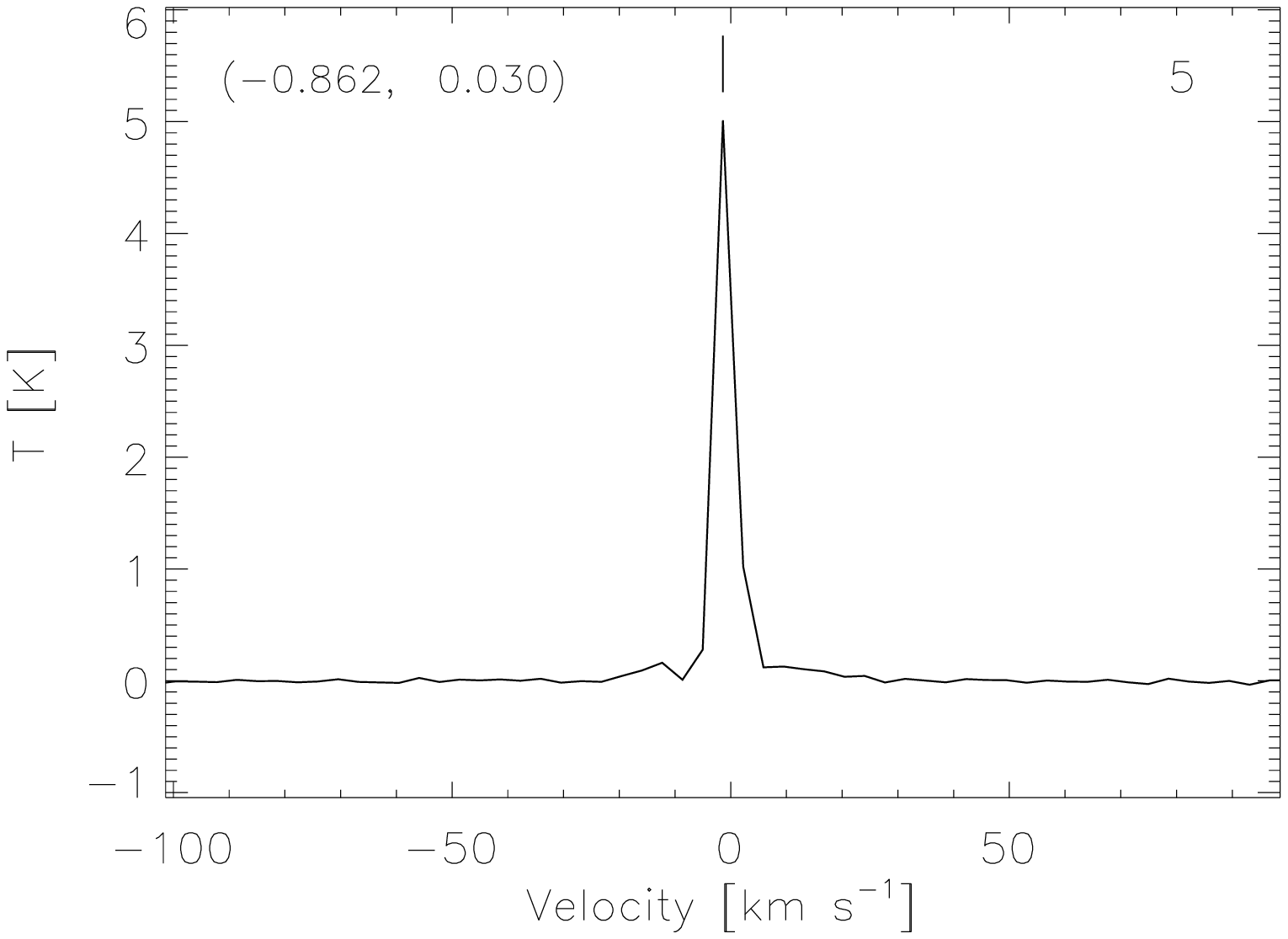}
\includegraphics[width=0.32\textwidth,clip=true]{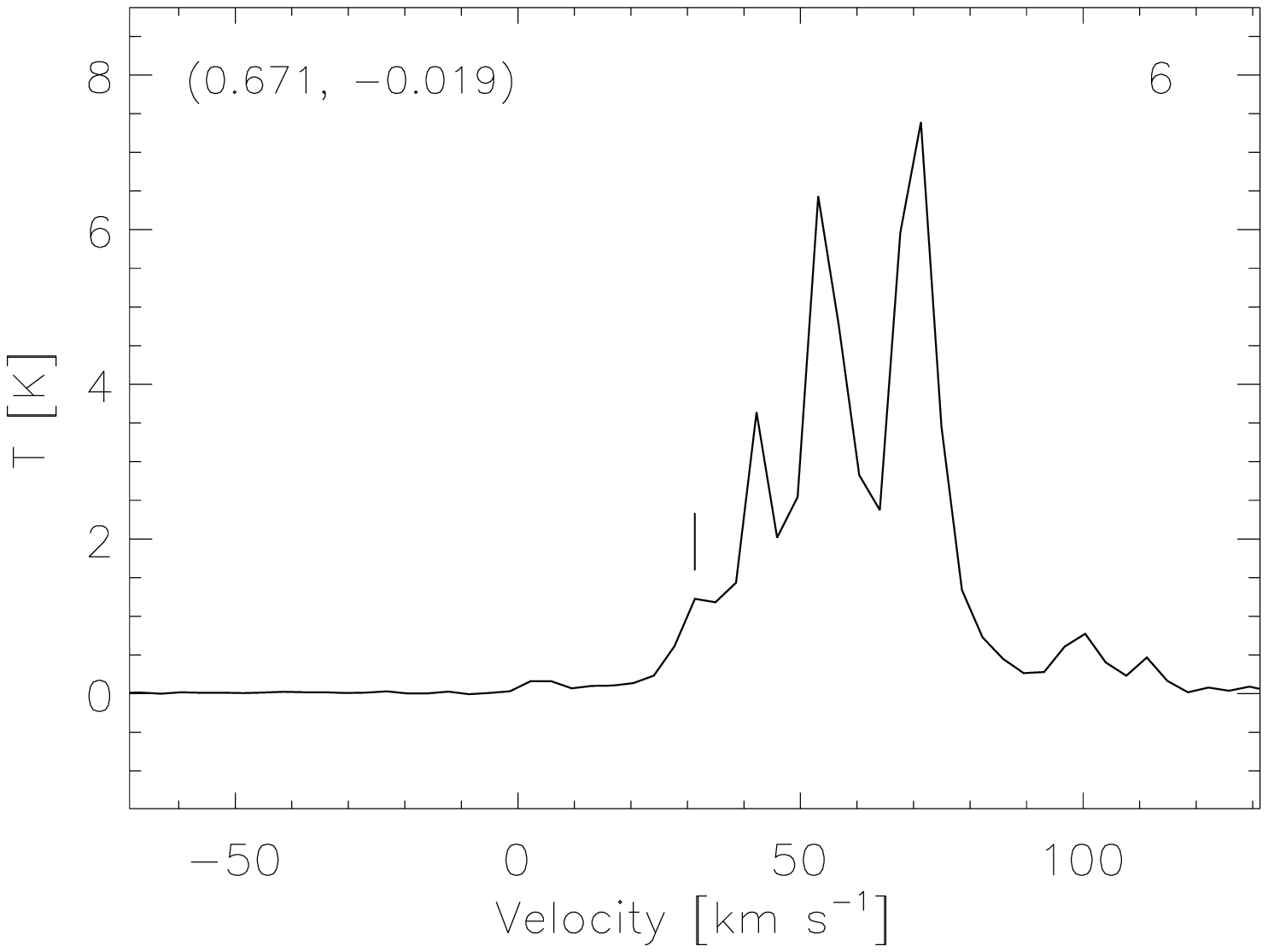}\\ 
\includegraphics[width=0.32\textwidth,clip=true]{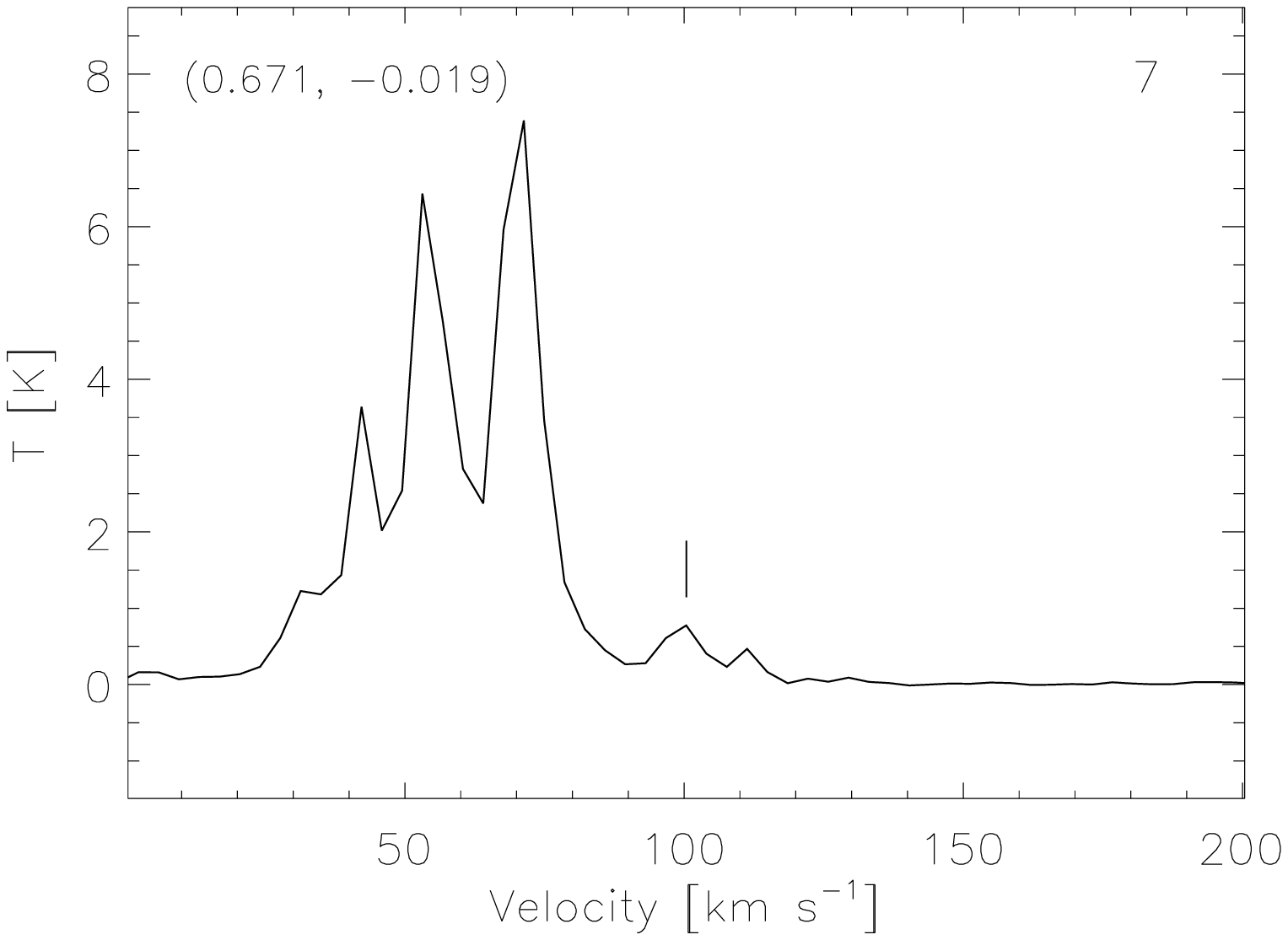}
\includegraphics[width=0.32\textwidth,clip=true]{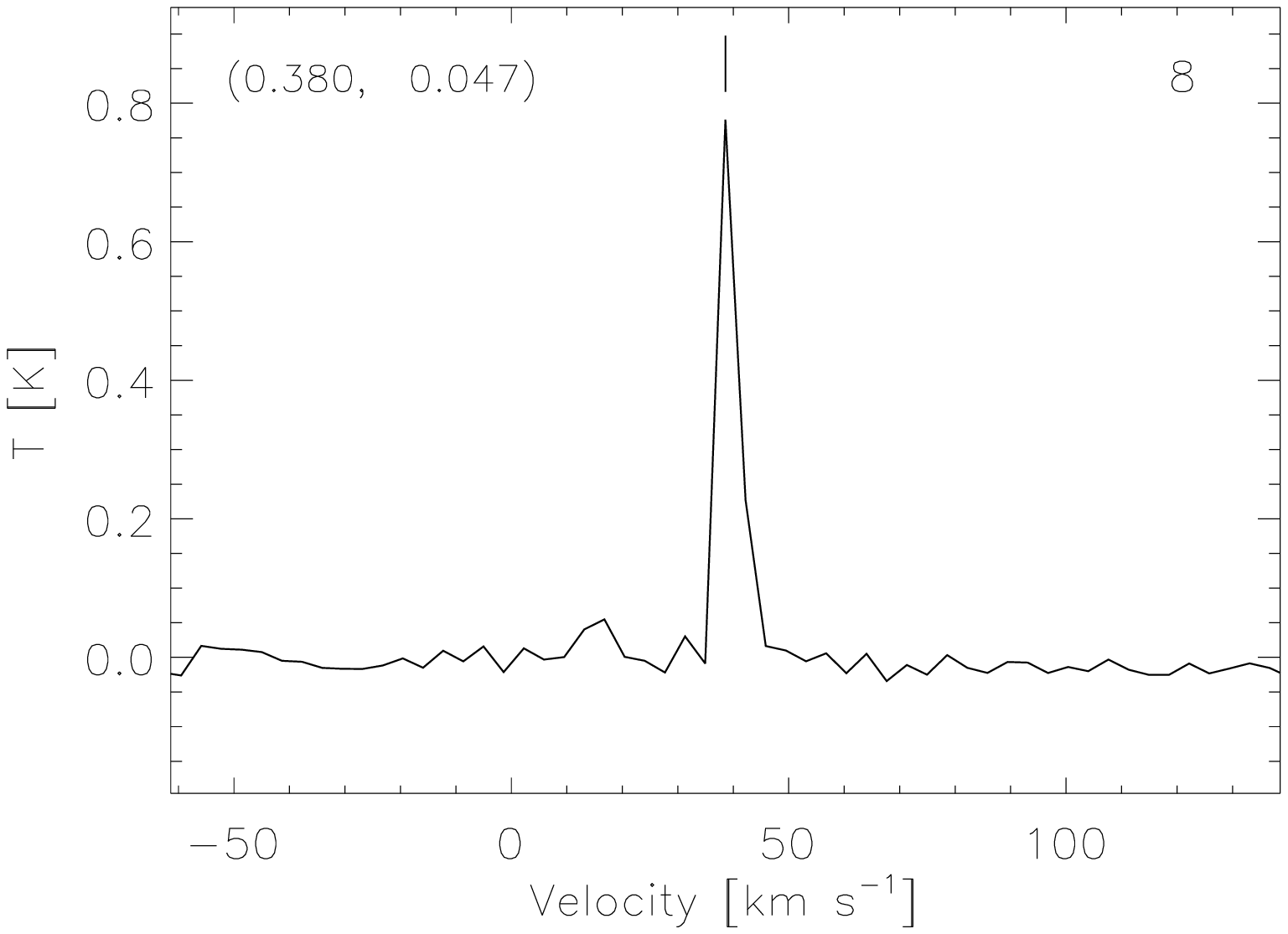}
\includegraphics[width=0.32\textwidth,clip=true]{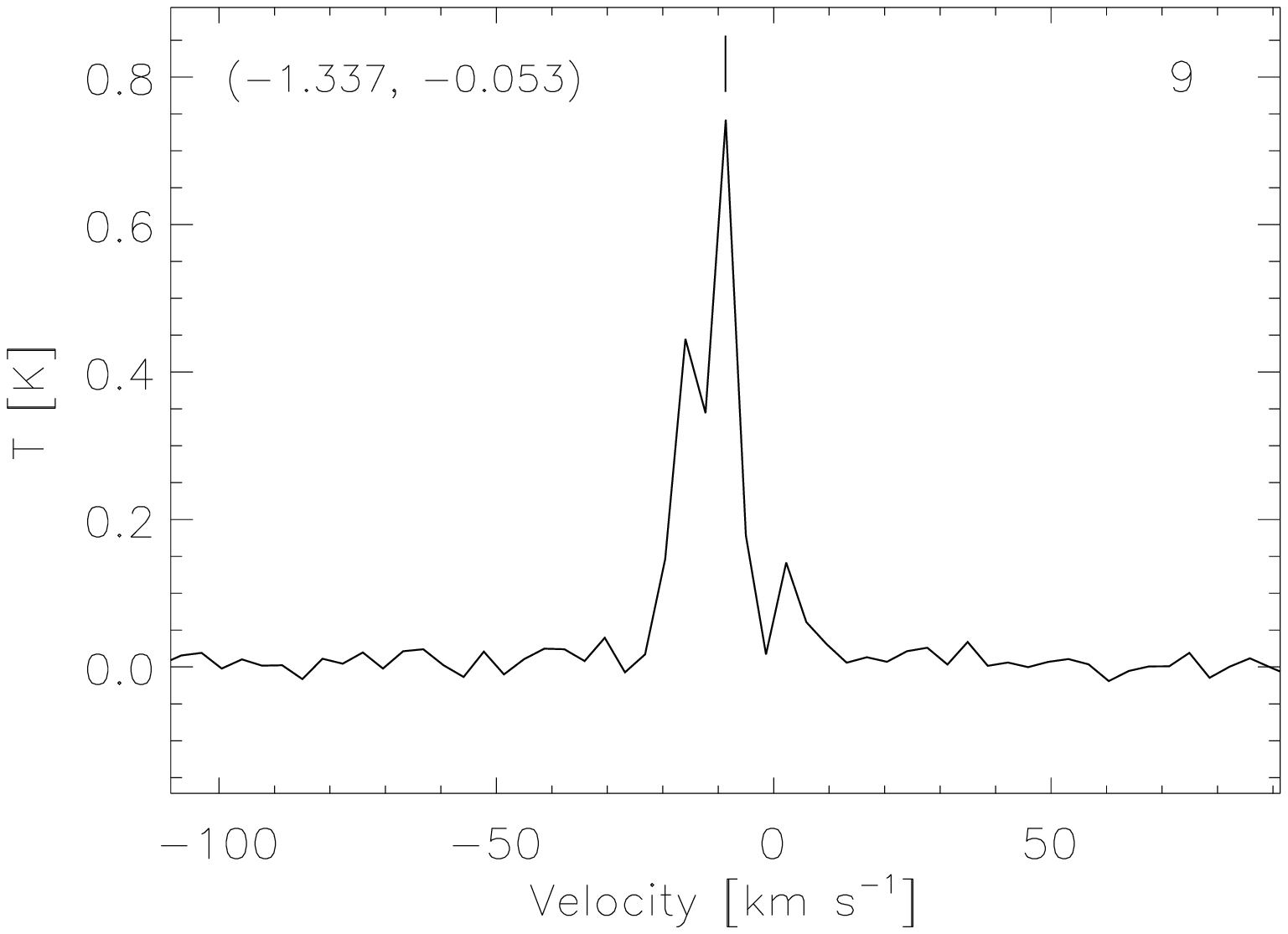}\\ 
\includegraphics[width=0.32\textwidth,clip=true]{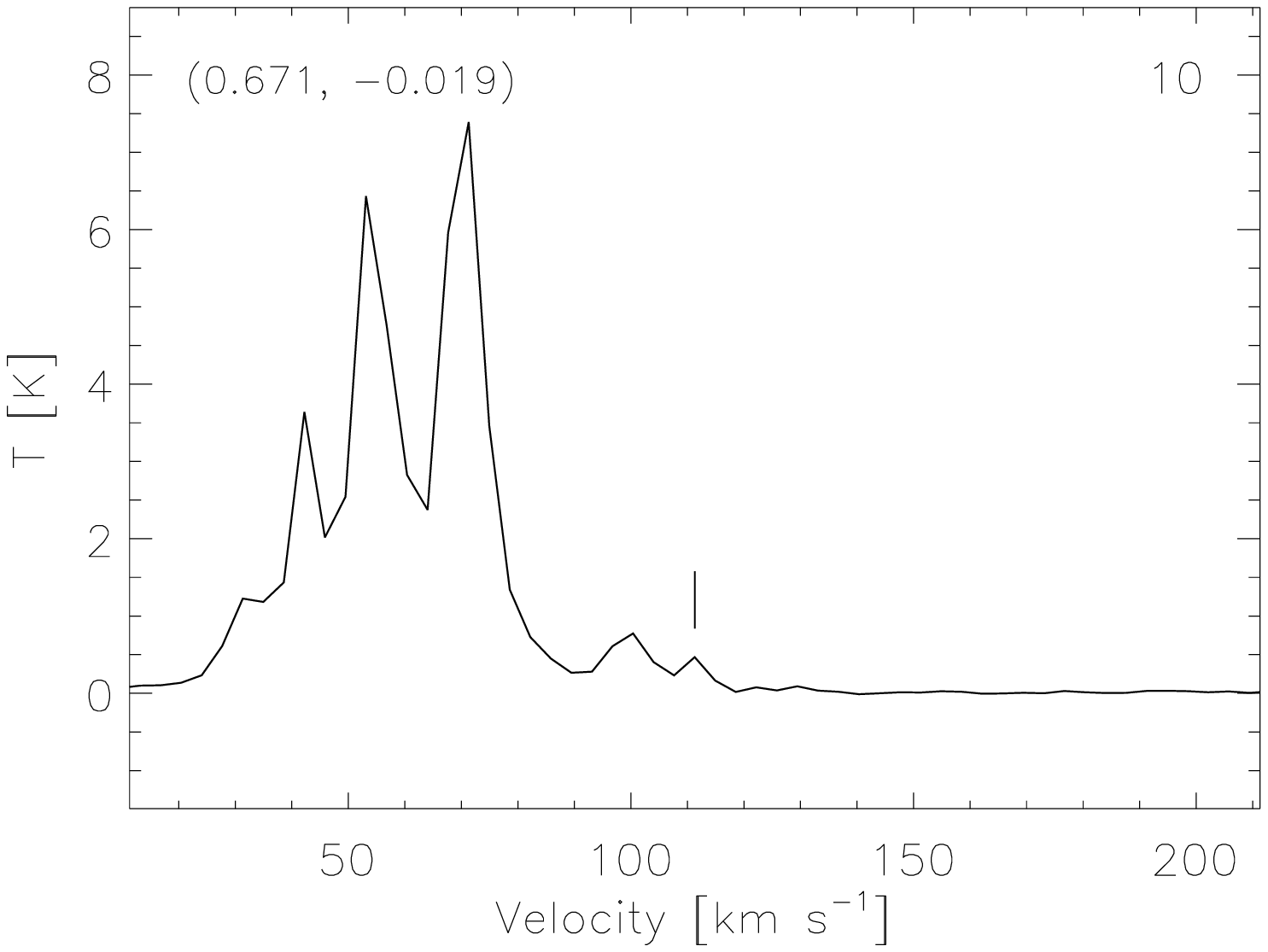}
\includegraphics[width=0.32\textwidth,clip=true]{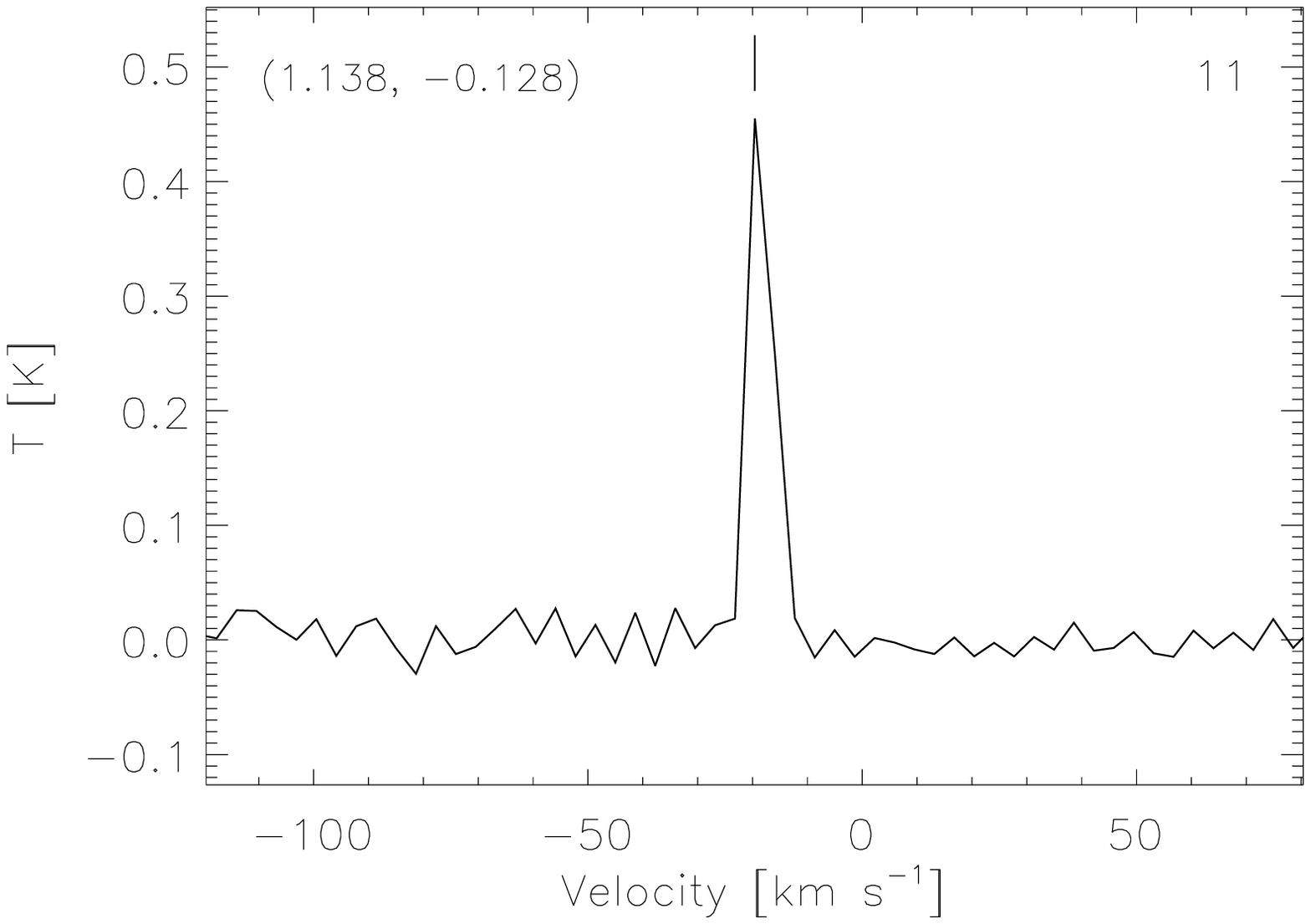}
\includegraphics[width=0.32\textwidth,clip=true]{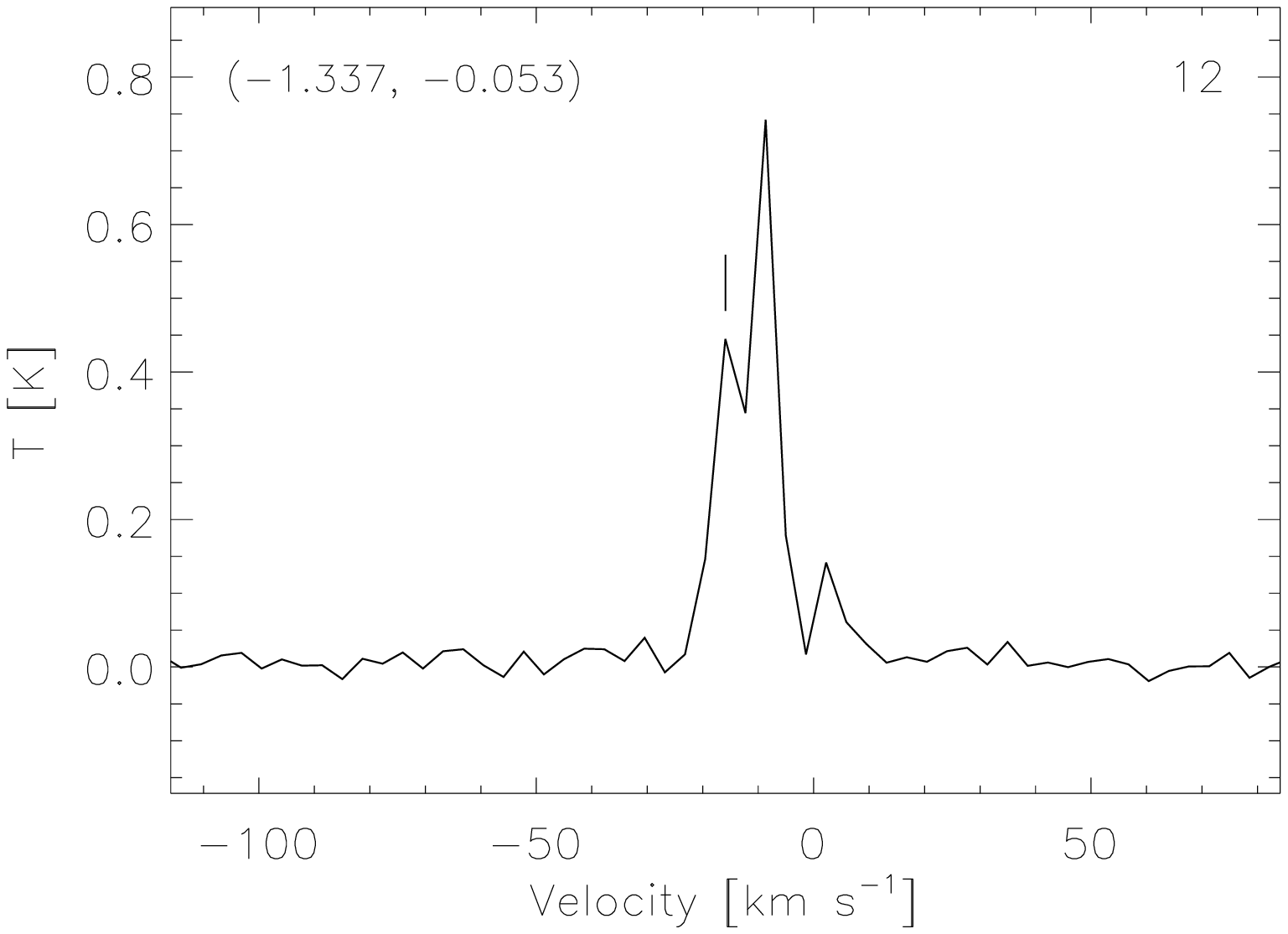}\\ 
\includegraphics[width=0.32\textwidth,clip=true]{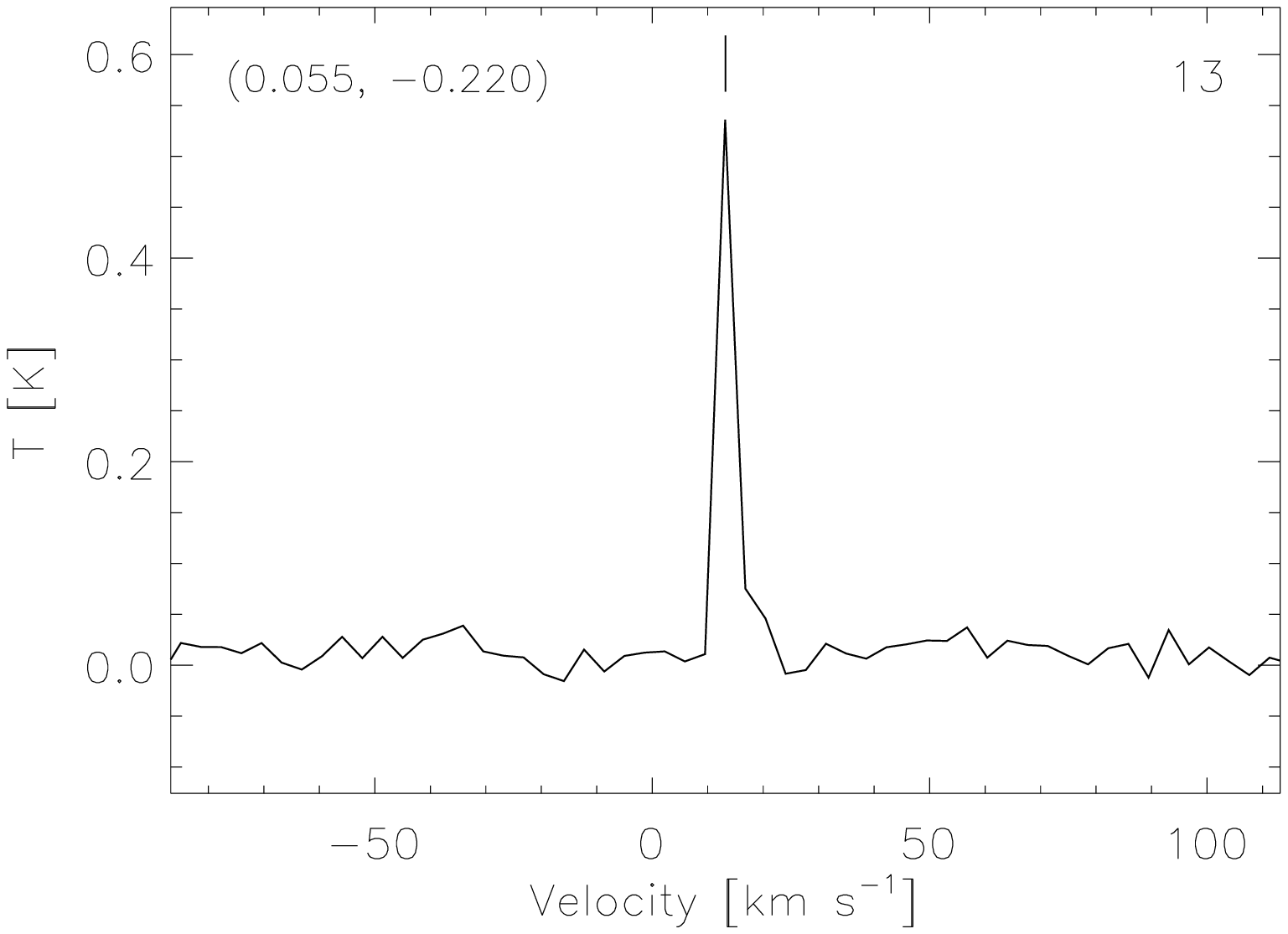}
\includegraphics[width=0.32\textwidth,clip=true]{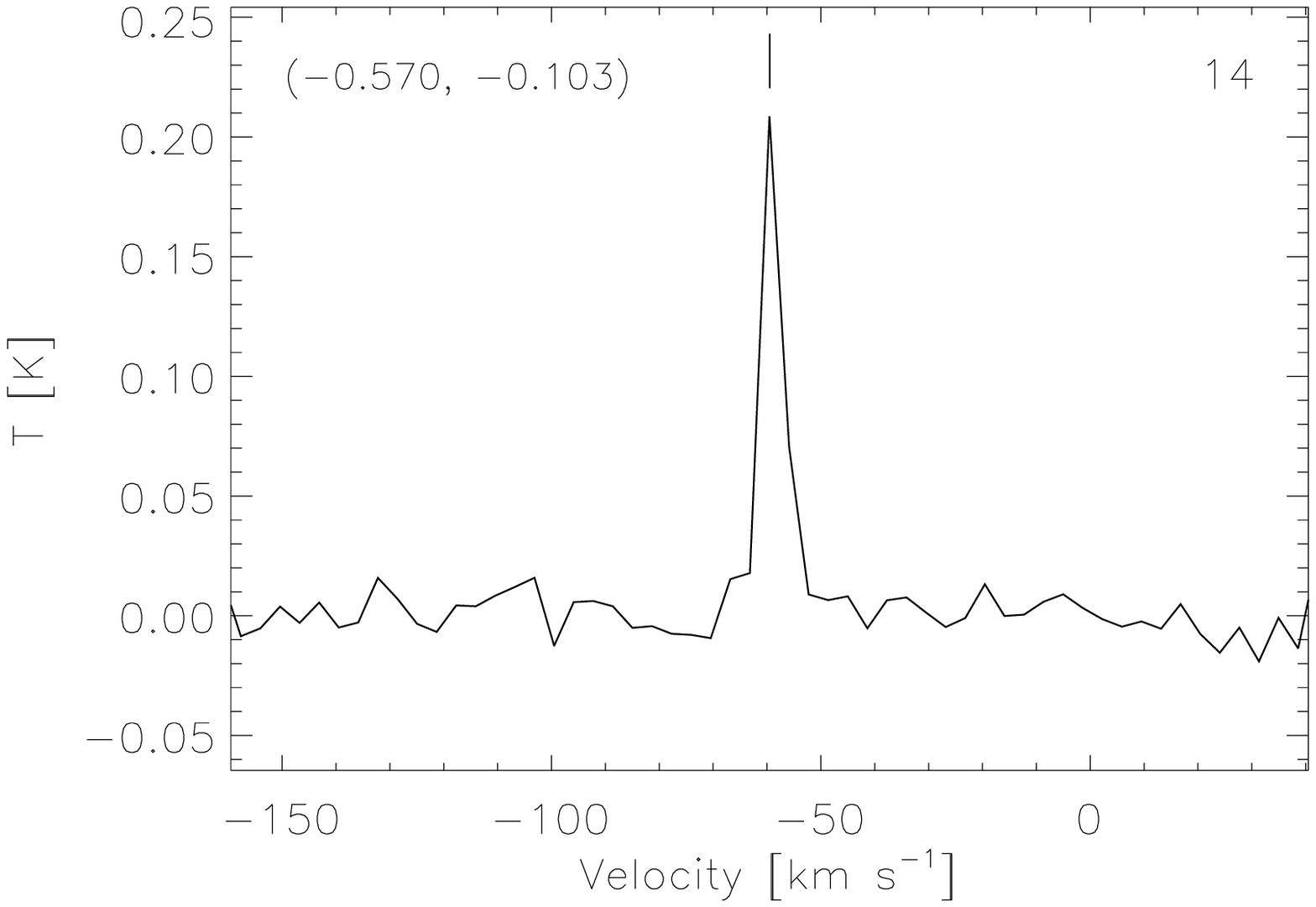}
\includegraphics[width=0.32\textwidth,clip=true]{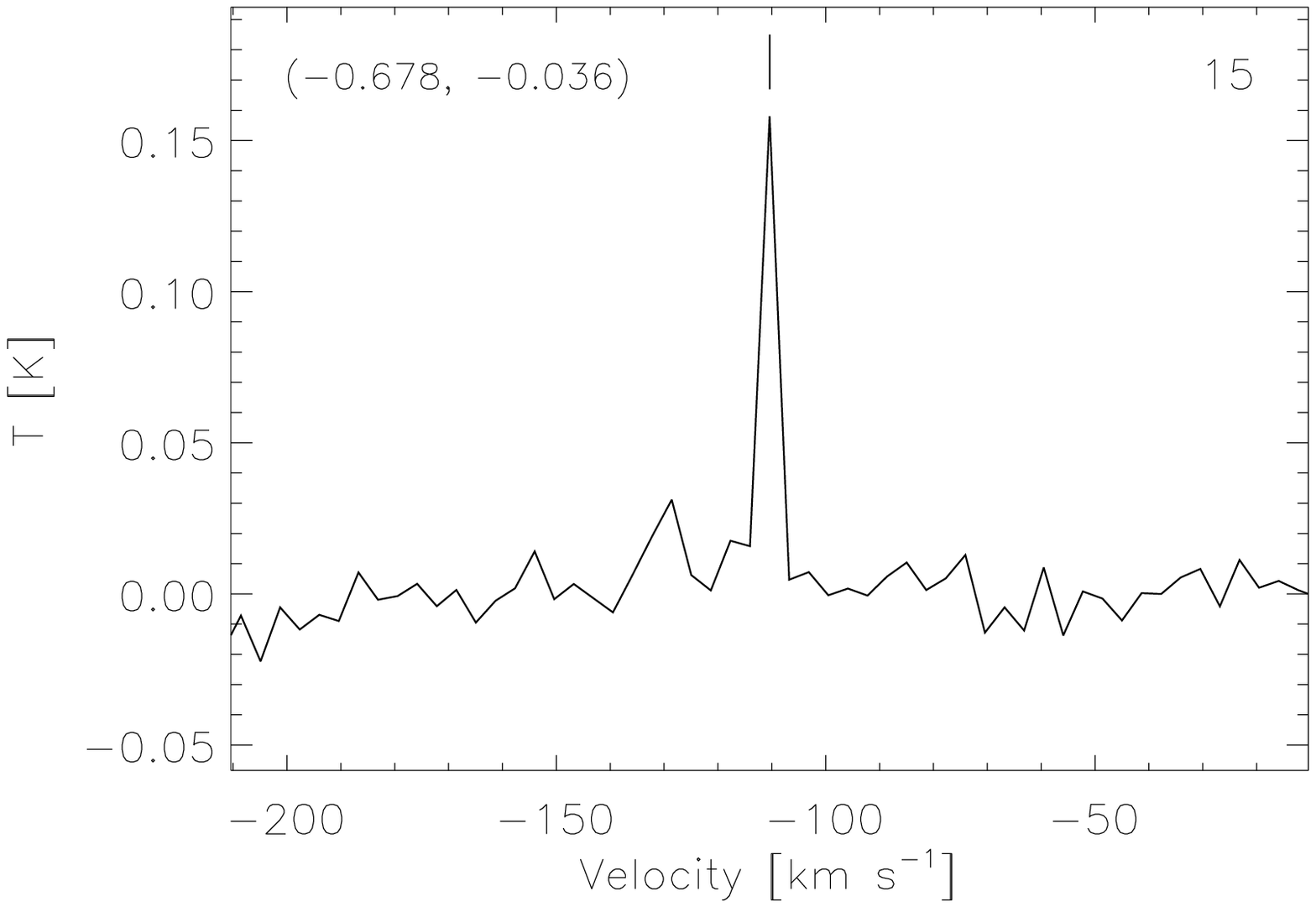}\\
\caption{Spectra of \water\, masers 1--15 detected in the survey region.
  The positions of the peak emission are printed on the spectra, and
  the numbers in the top-right of each plot correspond to the numbers
  in Table~\ref{water-tot}.  The vertical line in each spectrum
  indicates the velocity of that maser.
  \label{first_spec}}
\end{figure*}

\begin{figure}
\includegraphics[width=\hsize,clip=true]{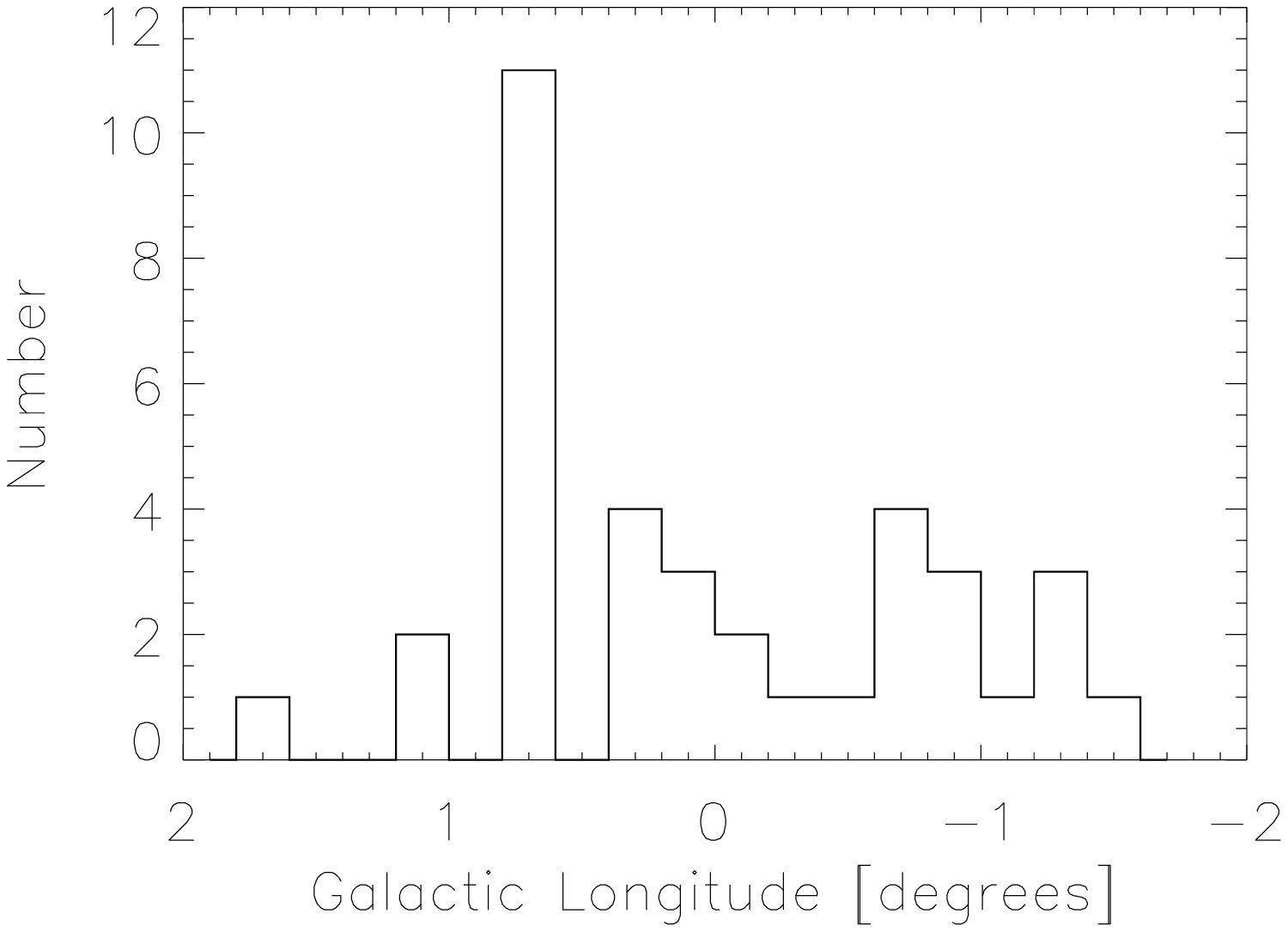}\\
\includegraphics[width=\hsize,clip=true]{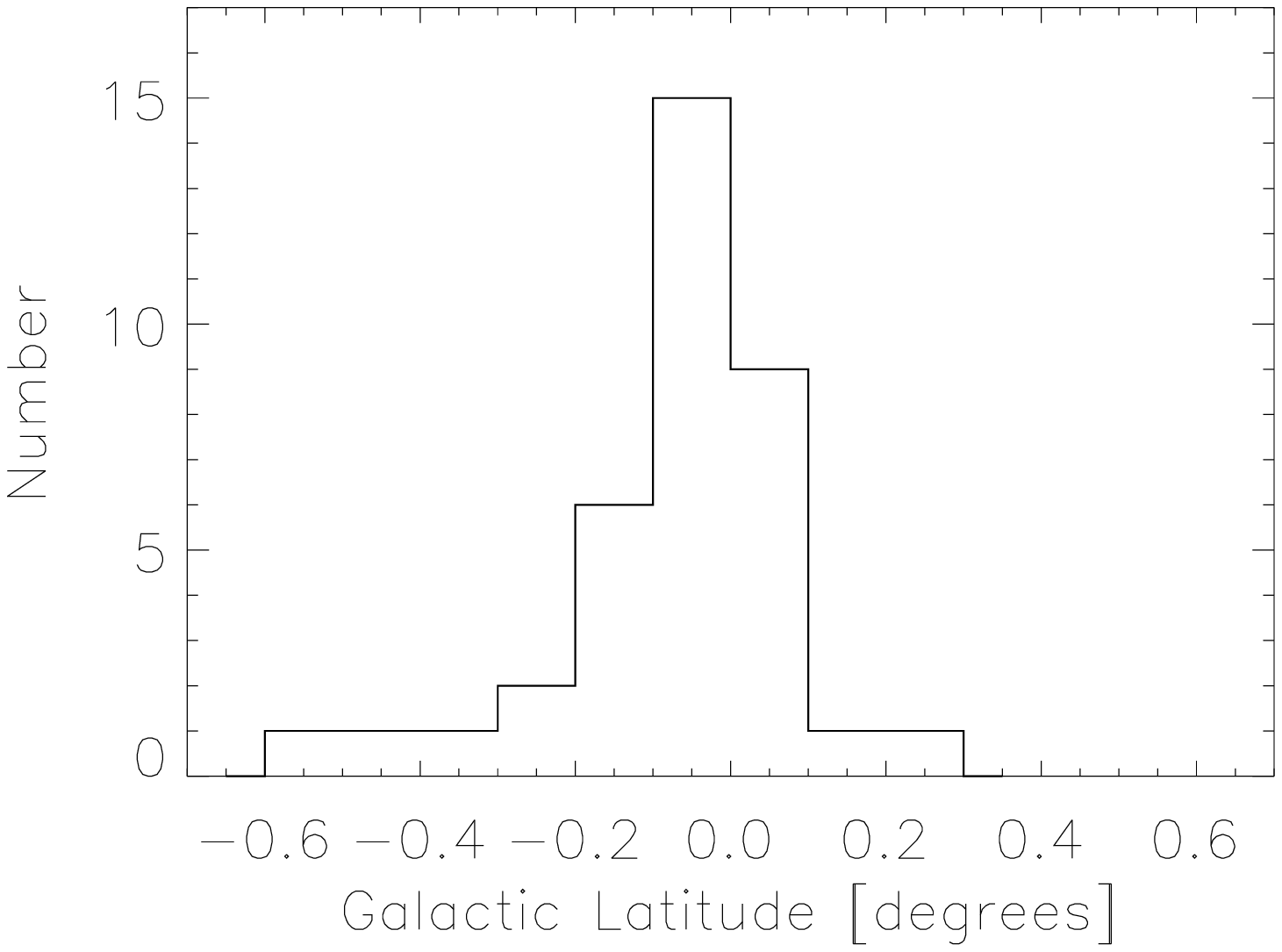}\\
\caption{Distribution of the Mopra \water\, masers as a function of Galactic
longitude ({\it top}) and latitude ({\it bottom}).  The peak of the longitude
distribution is due to the 9 masers associated with Sgr~B.
  \label{water_dist}}
\end{figure}

\begin{figure*}
\includegraphics[angle=-90,width=\textwidth,clip=true]{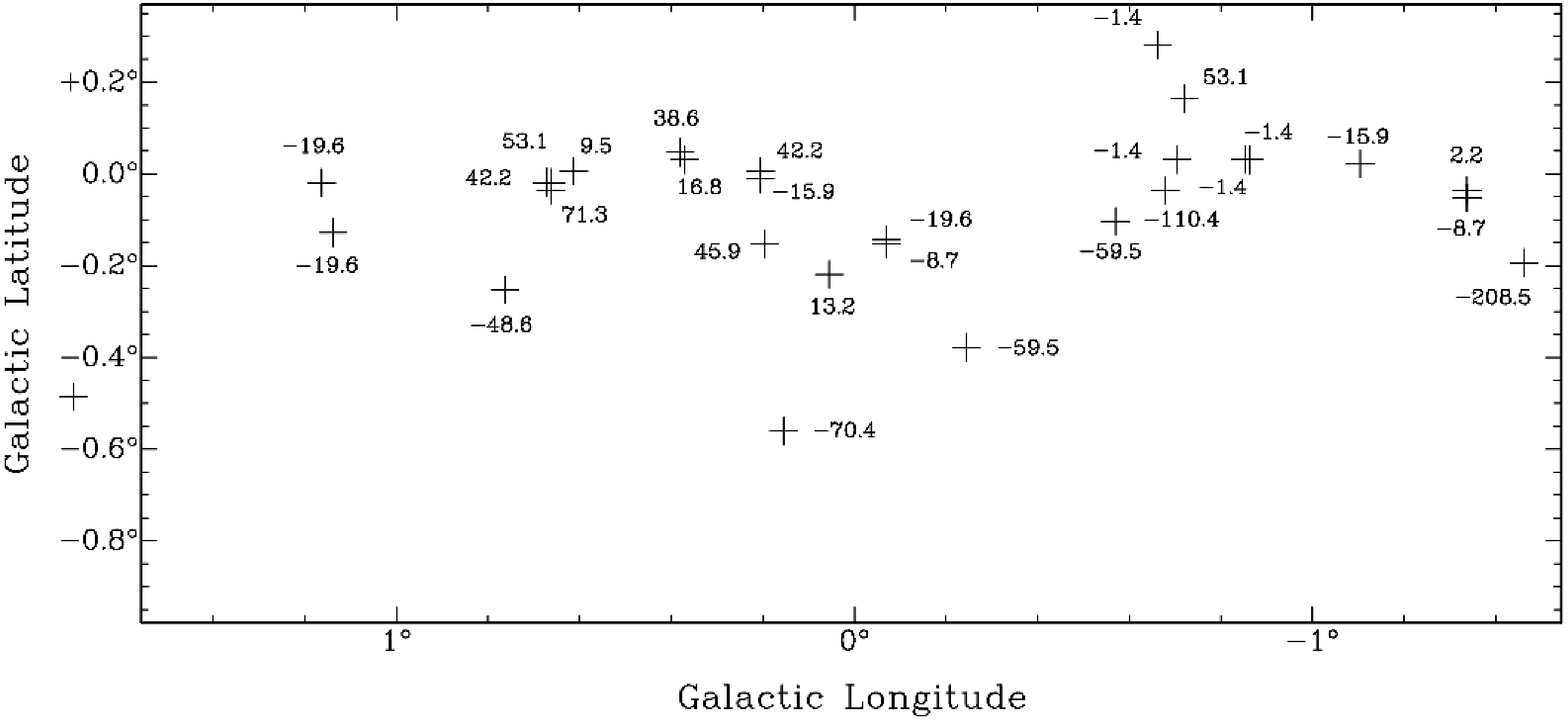}\\
\includegraphics[angle=-90,width=\textwidth,clip=true]{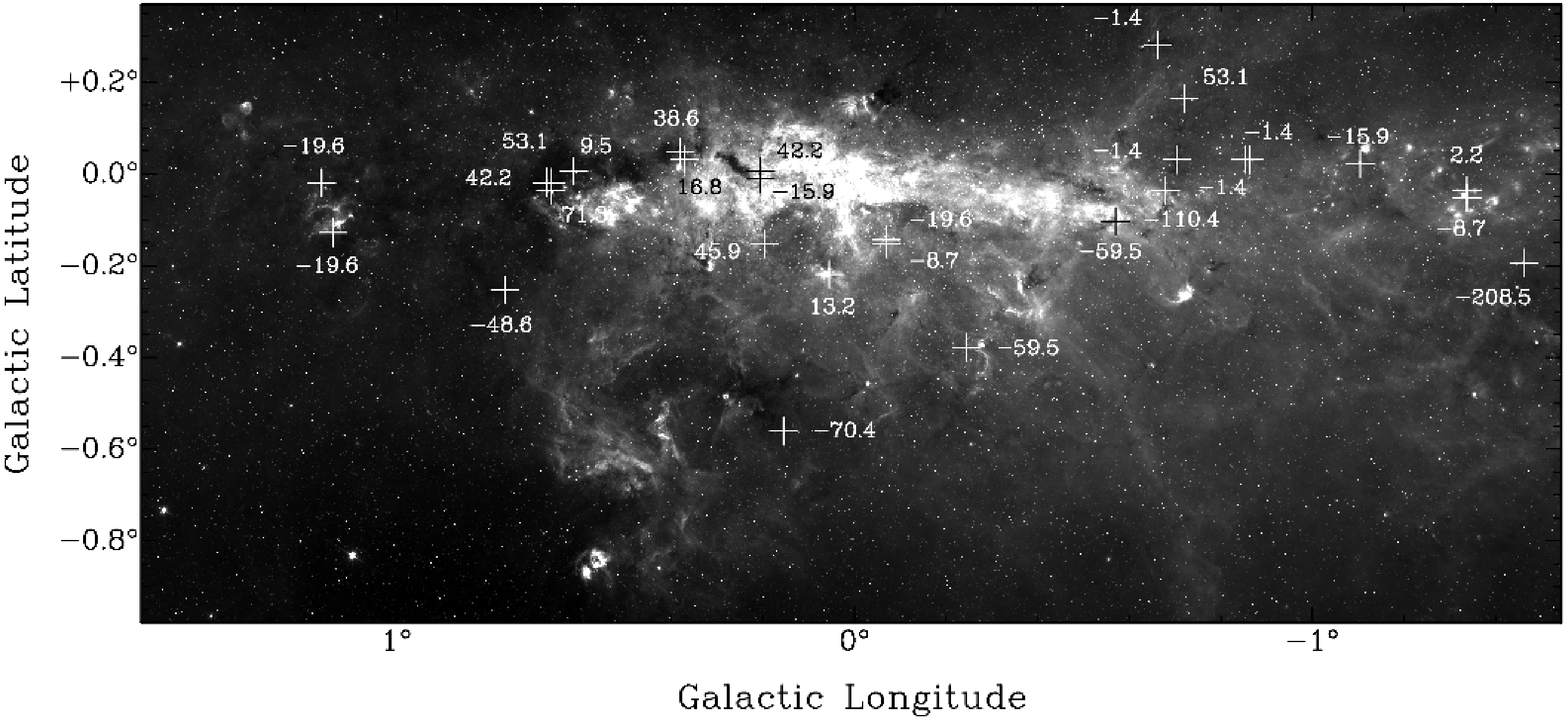}\\
\caption{{\it Top}: {\it Spitzer}/IRAC 8~\um\, image of the Galactic
  center region.  Crosses (+) indicate the positions of the \water\,
  masers identified in the Mopra survey. Each maser is labelled with
  the velocity (in \kms) of the peak maser emission.  {\it Bottom}:
  Same as the {\it top} figure, but with a blank backround instead of
  the 8~\um\, image.
  \label{water_dist_irac}}
\end{figure*}

Using the Clumpfind detection algorithm as described in
Section~\ref{find_water}, we find a total of 37 \water\, masers.
These masers are listed in Table~\ref{water-tot}, which contains the
identification numbers (Column~1), the Galactic coordinates of the
masers (Columns~2 and 3), the velocity of peak emission for each maser
(Column~4), the peak intensities of the masers (Column~5), and the
number of voxels in the source (Column~6).  The peaks range from 52~mK
to 9.54~K, with an average of 1.02~K and a median of 160~mK, and the
velocities range from -208.5 to 129.4~\kms.  Spectra of all 37
detections can be seen in Figure~\ref{first_spec}.  The distribution
of the water masers as a function of Galactic longitude and latitude
can be seen in Figure~\ref{water_dist}.  In
Figure~\ref{water_dist_irac}, the positions of the water masers are
overplotted on an image of the Galactic center region, along with the
velocities of the peak maser emission.  Some masers share the same
positions but have different velocities.  These likely comprise a
cluster of masers, but higher angular resolution observations are
needed to determine their positions more precisely.

\begin{table}
\caption{\label{water-tot}\water\, Masers in the Survey Region}
\begin{tabular}{lcccccc}
\hline\hline
Number  & $\ell$\tablefootmark{a} & $b$\tablefootmark{a} & $v$\tablefootmark{a} & $T_{peak}$ & Number \\
        & (\degreesym) & (\degreesym) & (\kms) & (K) & of Voxels \\
\hline
       1 &   0.663 &  $-$0.036 &   71.3 &      9.54 &         1703 \\
       2 &   0.663 &  $-$0.019 &   53.1 &      6.48 &         1766 \\
       3 &   0.671 &  $-$0.019 &   42.2 &      3.64 &          893 \\
       4 &  $-$0.854 &   0.031 &   $-$1.4 &      5.08 &          372 \\
       5 &  $-$0.862 &   0.031 &   $-$1.4 &      5.01 &          446 \\
       6 &   0.671 &  $-$0.019 &   31.3 &      1.23 &          645 \\
       7 &   0.671 &  $-$0.019 &  100.4 &      0.77 &          673 \\
       8 &   0.380 &   0.047 &   38.6 &      0.78 &          218 \\
       9 &  $-$1.337 &  $-$0.053 &   $-$8.7 &      0.74 &          174 \\
      10 &   0.671 &  $-$0.019 &  111.3 &      0.47 &          243 \\
      11 &   1.138 &  $-$0.128 &  $-$19.6 &      0.46 &          158 \\
      12 &  $-$1.337 &  $-$0.053 &  $-$15.9 &      0.44 &          280 \\
      13 &   0.055 &  $-$0.219 &   13.2 &      0.54 &           93 \\
      14 &  $-$0.570 &  $-$0.103 &  $-$59.5 &      0.21 &          117 \\
      15 &  $-$0.679 &  $-$0.036 & $-$110.4 &      0.16 &           79 \\
      16 &   0.671 &  $-$0.019 &    2.2 &      0.16 &          200 \\
      17 &   0.613 &   0.006 &    9.5 &      0.16 &          104 \\
      18 &  $-$0.704 &   0.031 &   $-$1.4 &      0.11 &          161 \\
      19 &  $-$1.337 &  $-$0.036 &    2.2 &      0.16 &           67 \\
      20 &   0.155 &  $-$0.561 &  $-$70.4 &      0.20 &           76 \\
      21 &  $-$0.862 &   0.031 &  $-$12.3 &      0.16 &           81 \\
      22 &   0.205 &  $-$0.011 &  $-$15.9 &      0.09 &           69 \\
      23 &   1.163 &  $-$0.019 &  $-$19.6 &      0.08 &           35 \\
      24 &   0.763 &  $-$0.253 &  $-$48.6 &      0.09 &           45 \\
      25 &  $-$0.662 &   0.281 &   $-$1.4 &      0.16 &           43 \\
      26 &   0.371 &   0.031 &   16.8 &      0.08 &           59 \\
      27 &   0.196 &  $-$0.153 &   45.9 &      0.06 &           34 \\
      28 &  $-$1.104 &   0.022 &  $-$15.9 &      0.13 &           26 \\
      29 &  $-$0.720 &   0.164 &   53.1 &      0.06 &           26 \\
      30 &   0.671 &  $-$0.019 &  129.4 &      0.09 &           36 \\
      31 &  $-$0.070 &  $-$0.153 &   $-$8.7 &      0.07 &           57 \\
      32 &  $-$0.070 &  $-$0.144 &  $-$19.6 &      0.07 &           21 \\
      33 &  $-$0.245 &  $-$0.378 &  $-$59.5 &      0.10 &           23 \\
      34 &   0.671 &  $-$0.019 &  122.2 &      0.08 &           22 \\
      35 &  $-$1.462 &  $-$0.194 & $-$208.5 &      0.07 &           19 \\
      36 &   0.205 &   0.006 &   42.2 &      0.05 &           24 \\
      37 &   1.705 &  $-$0.486 &  $-$37.7 &      0.06 &           19 \\
\hline
\end{tabular}
\tablefoot{ 
\tablefoottext{a}{Values of $\ell, b,$\, and $v$\, are
    extracted from the Mopra data cube, which has a pixel size of
    0.5\arcmin\, and a channel width of $\sim$~3.6~\kms.}  }
\end{table}

\begin{figure*}
\includegraphics[width=0.32\textwidth,clip=true]{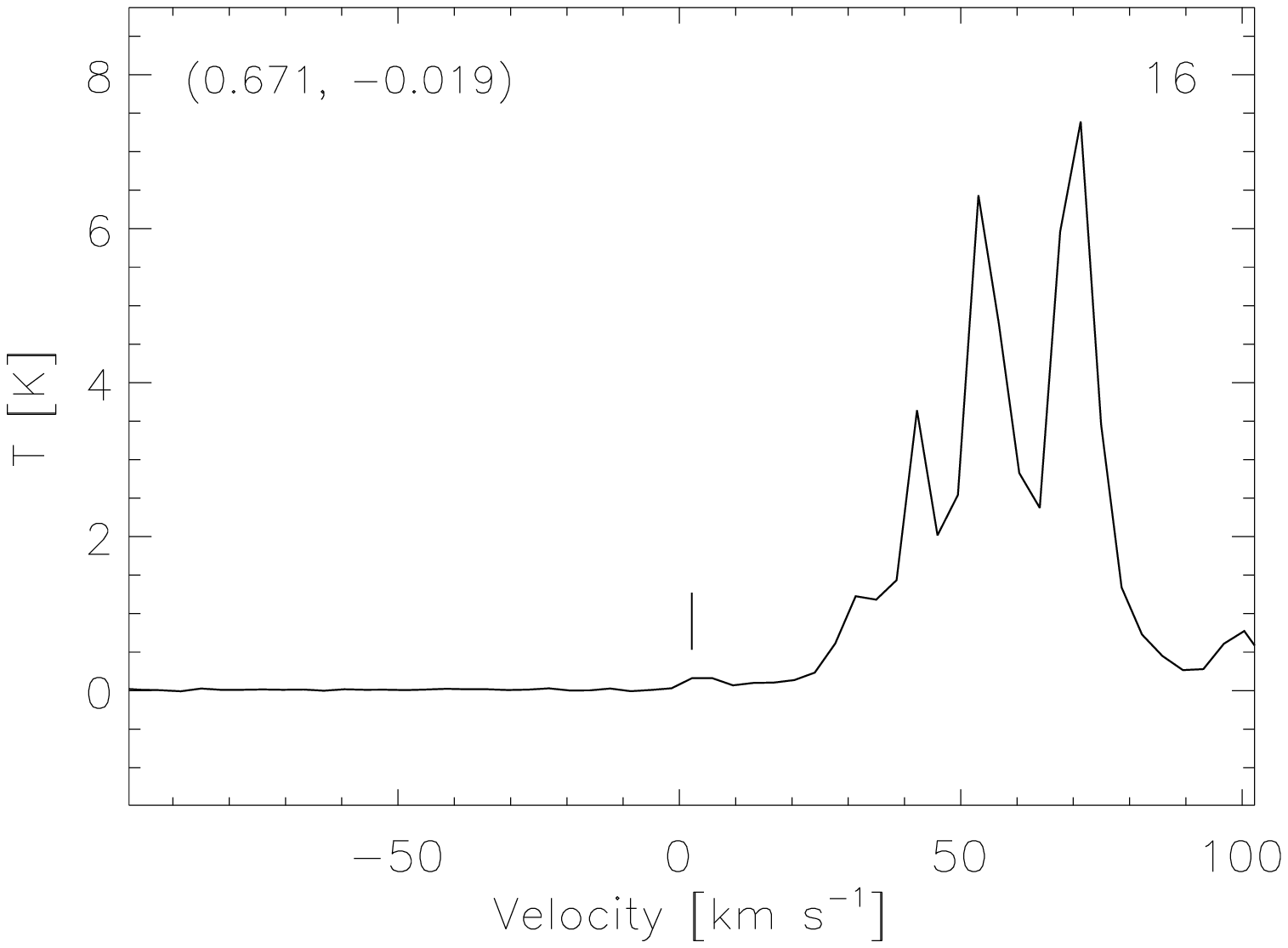}
\includegraphics[width=0.32\textwidth,clip=true]{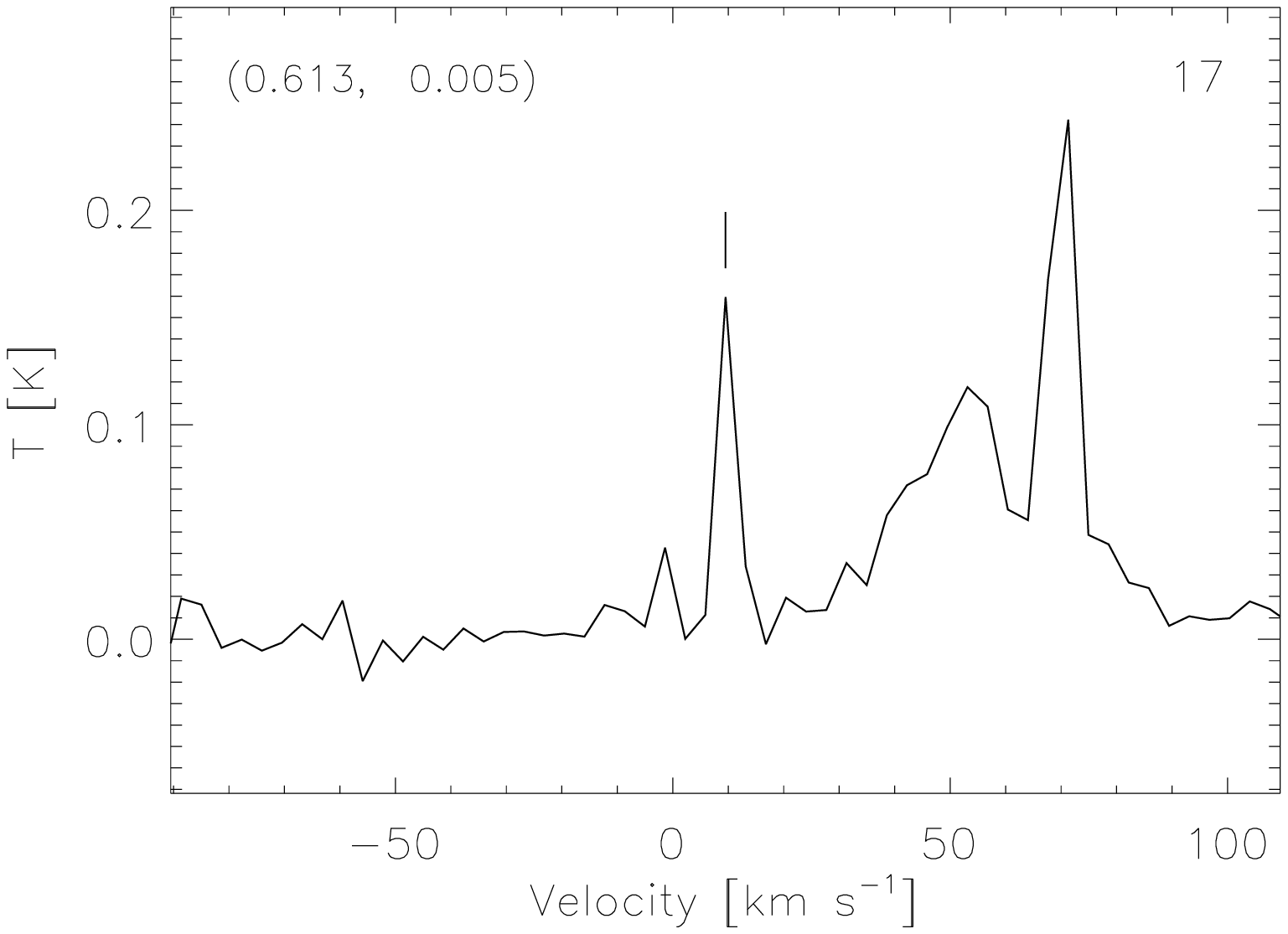}
\includegraphics[width=0.32\textwidth,clip=true]{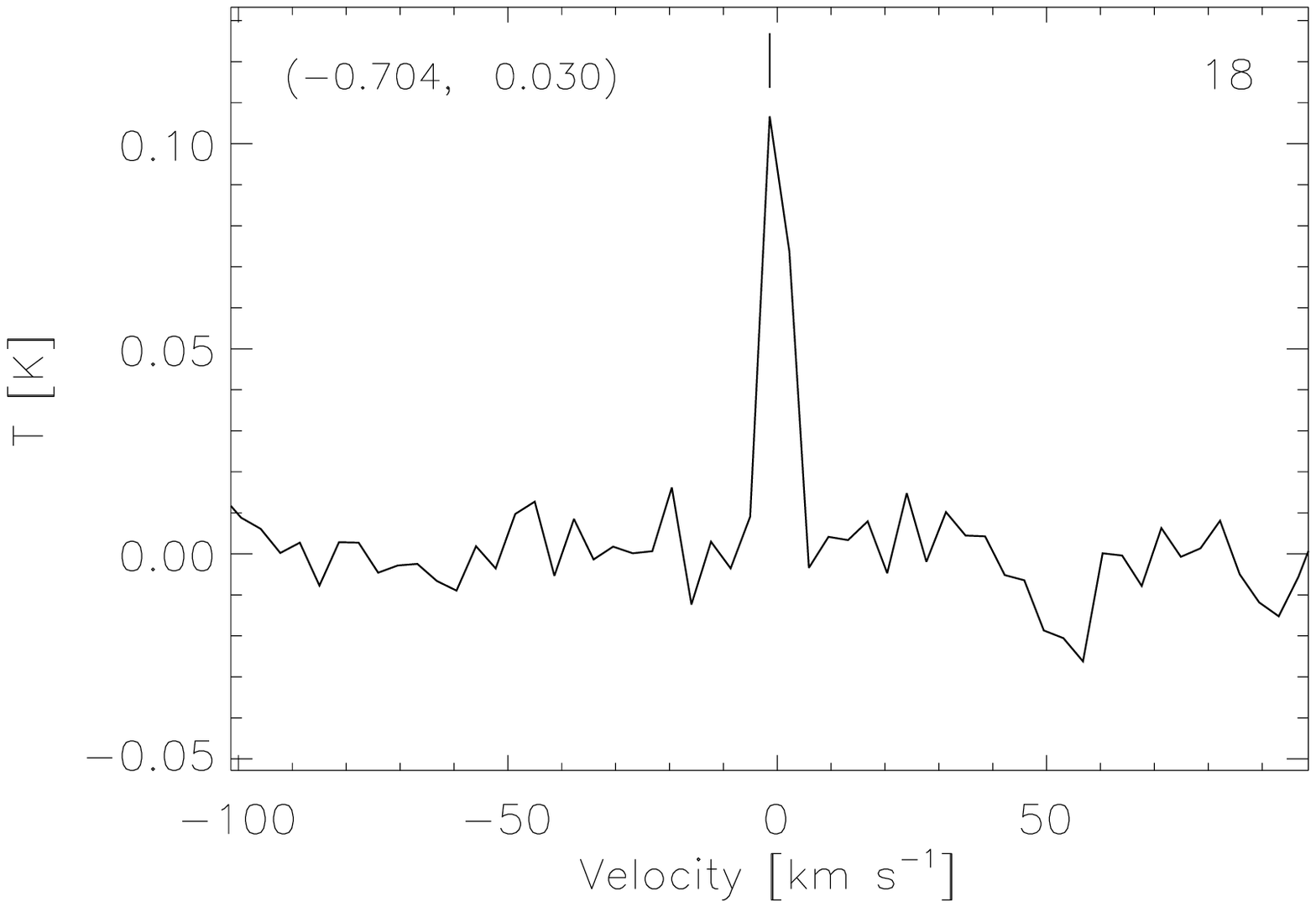}\\
\includegraphics[width=0.32\textwidth,clip=true]{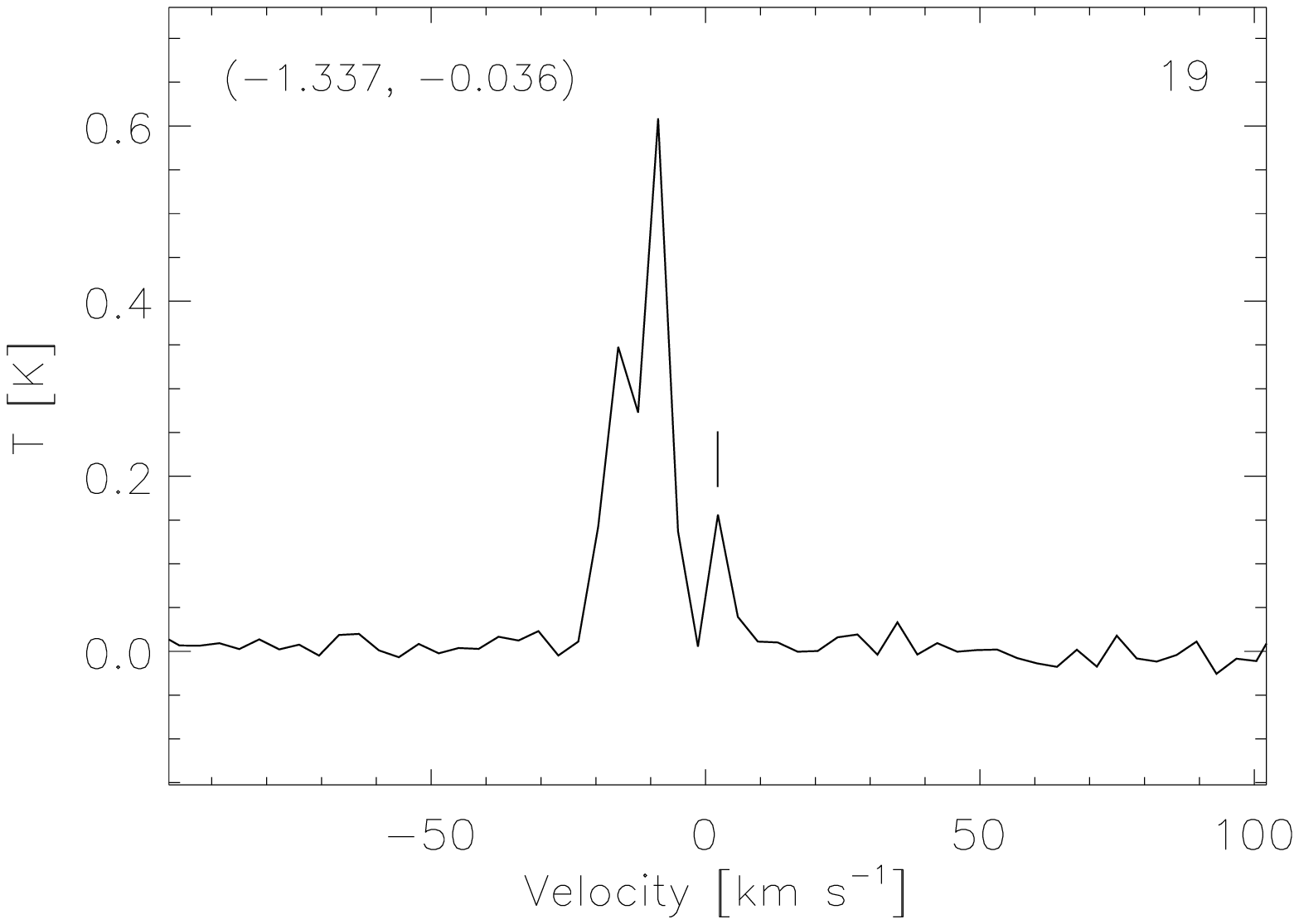}
\includegraphics[width=0.32\textwidth,clip=true]{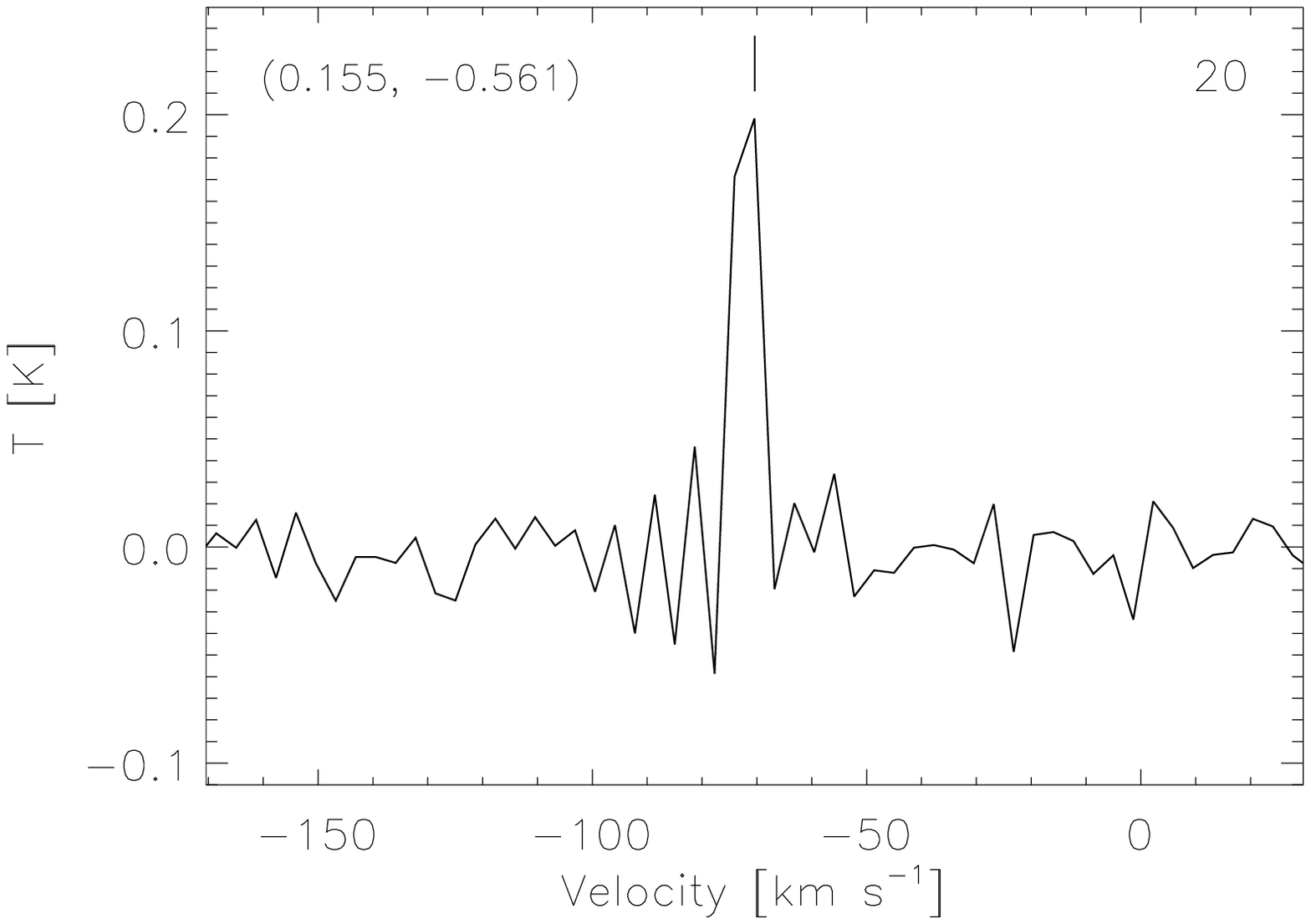}
\includegraphics[width=0.32\textwidth,clip=true]{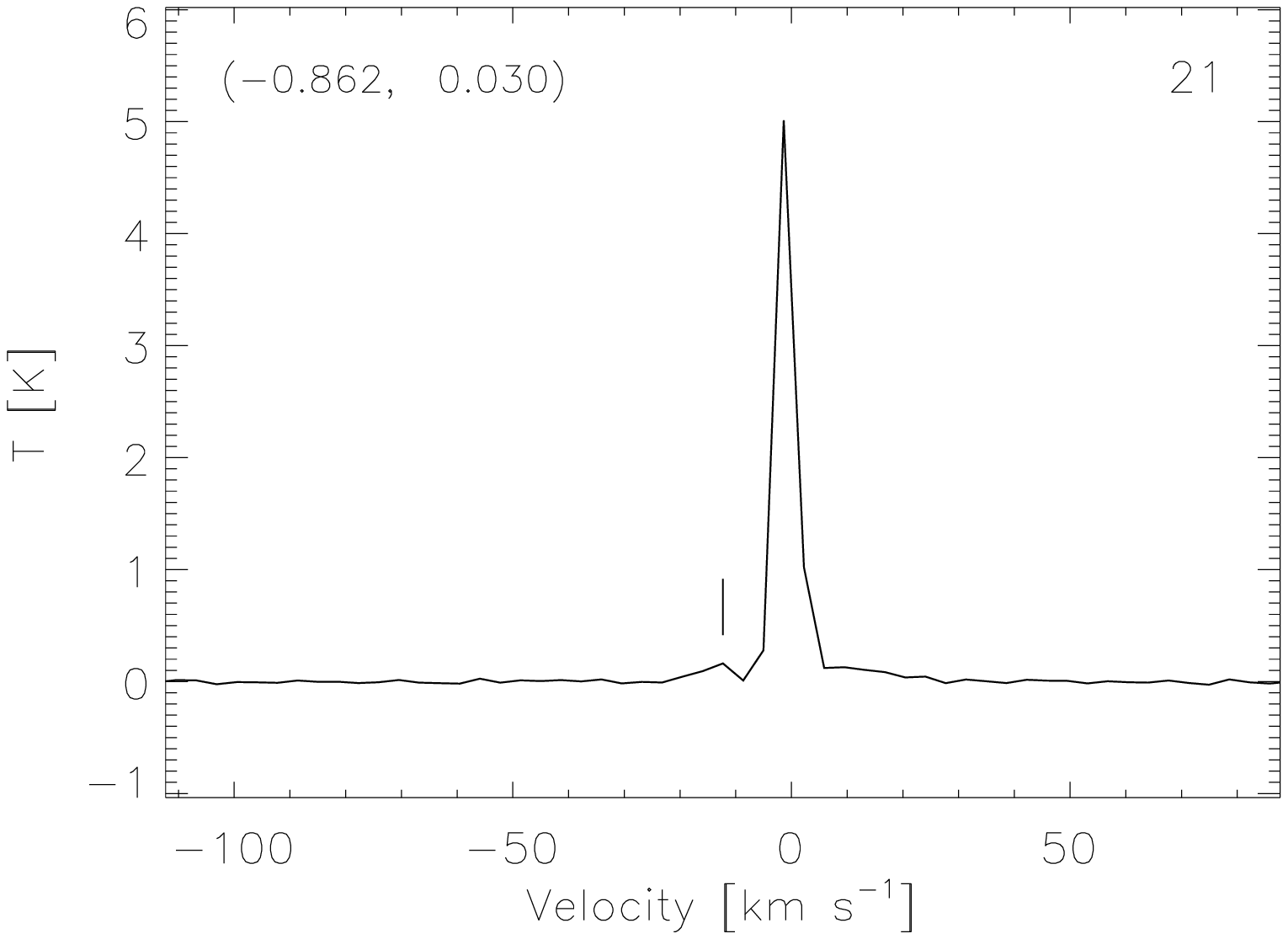}\\
\includegraphics[width=0.32\textwidth,clip=true]{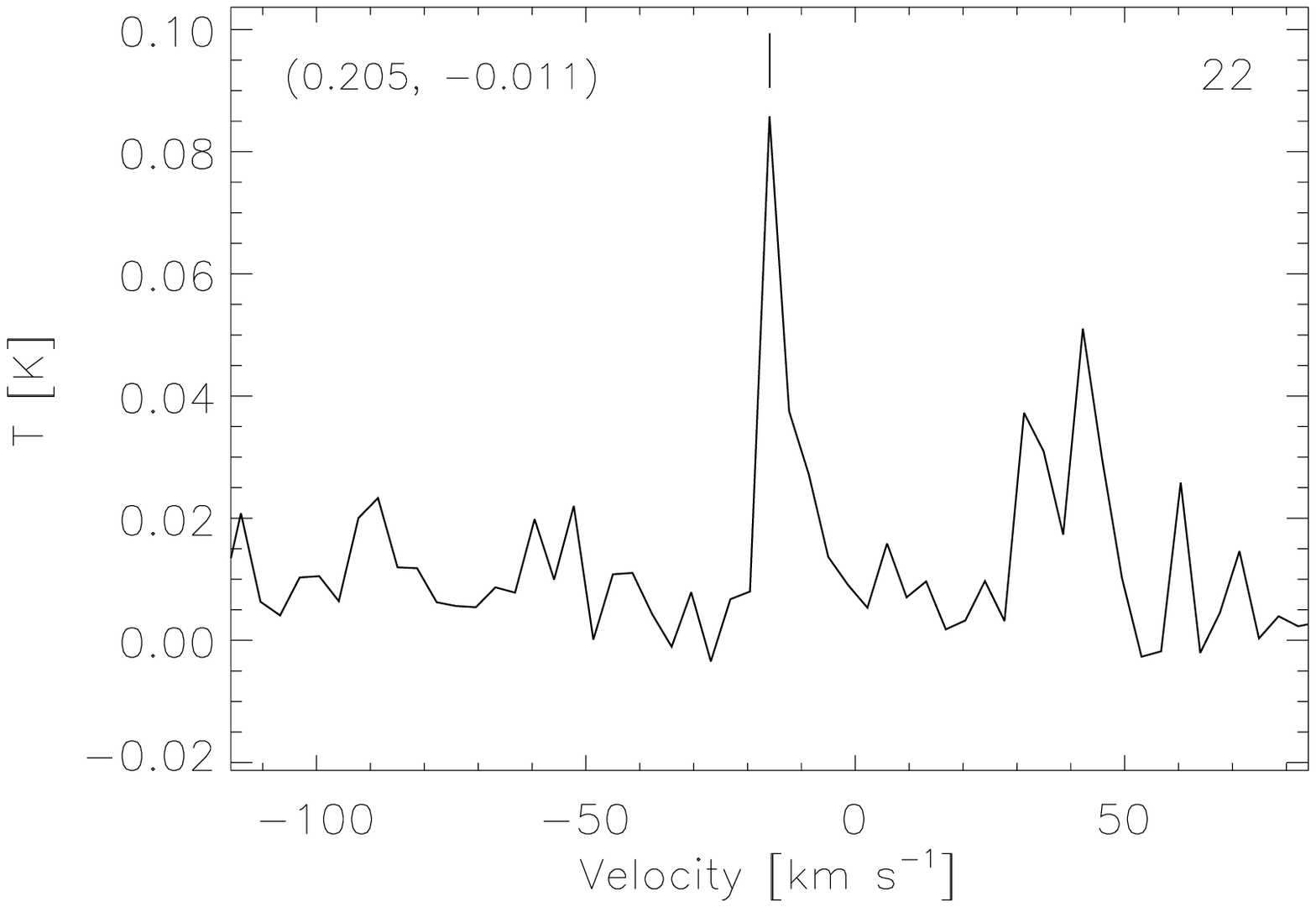}
\includegraphics[width=0.32\textwidth,clip=true]{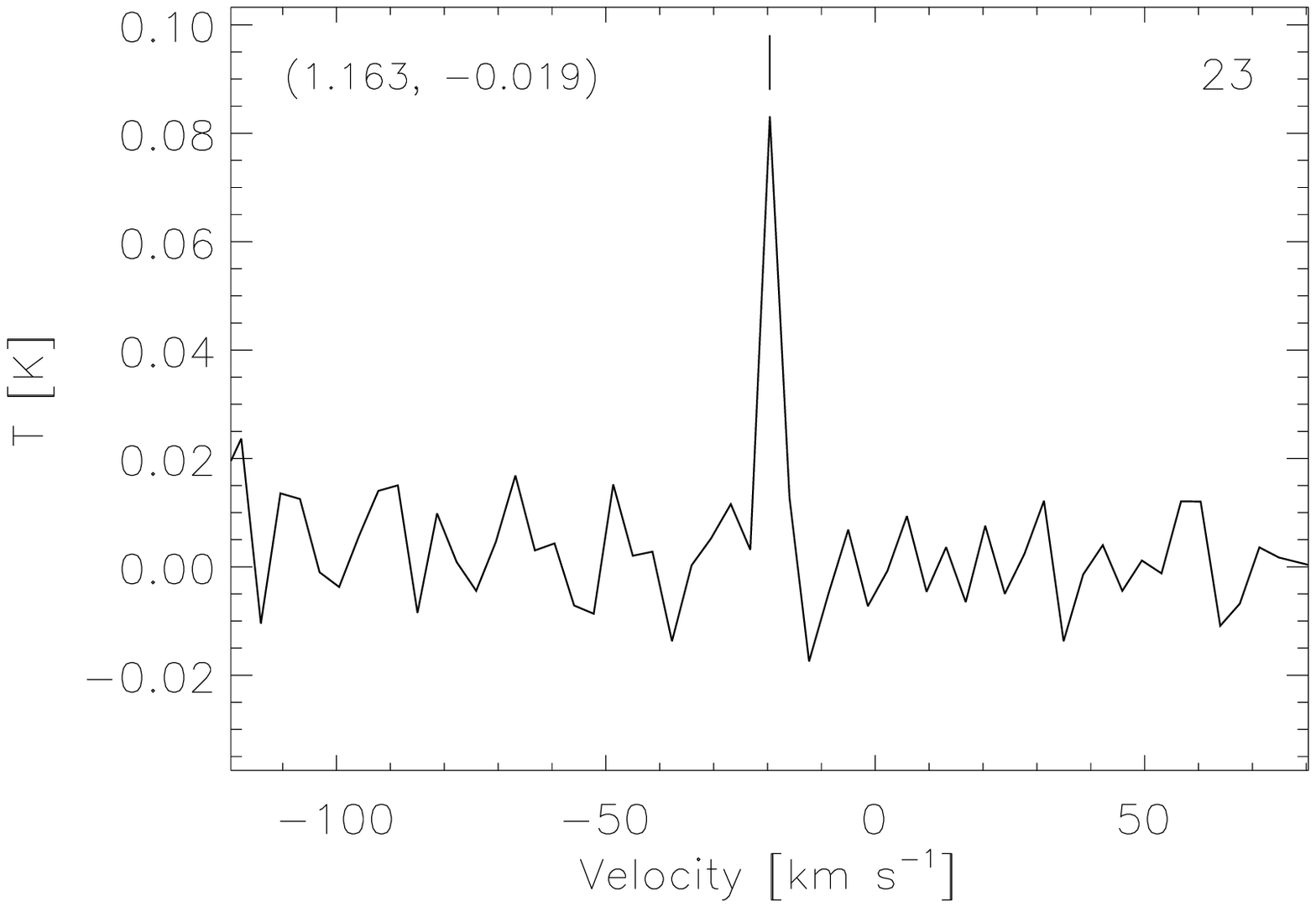}
\includegraphics[width=0.32\textwidth,clip=true]{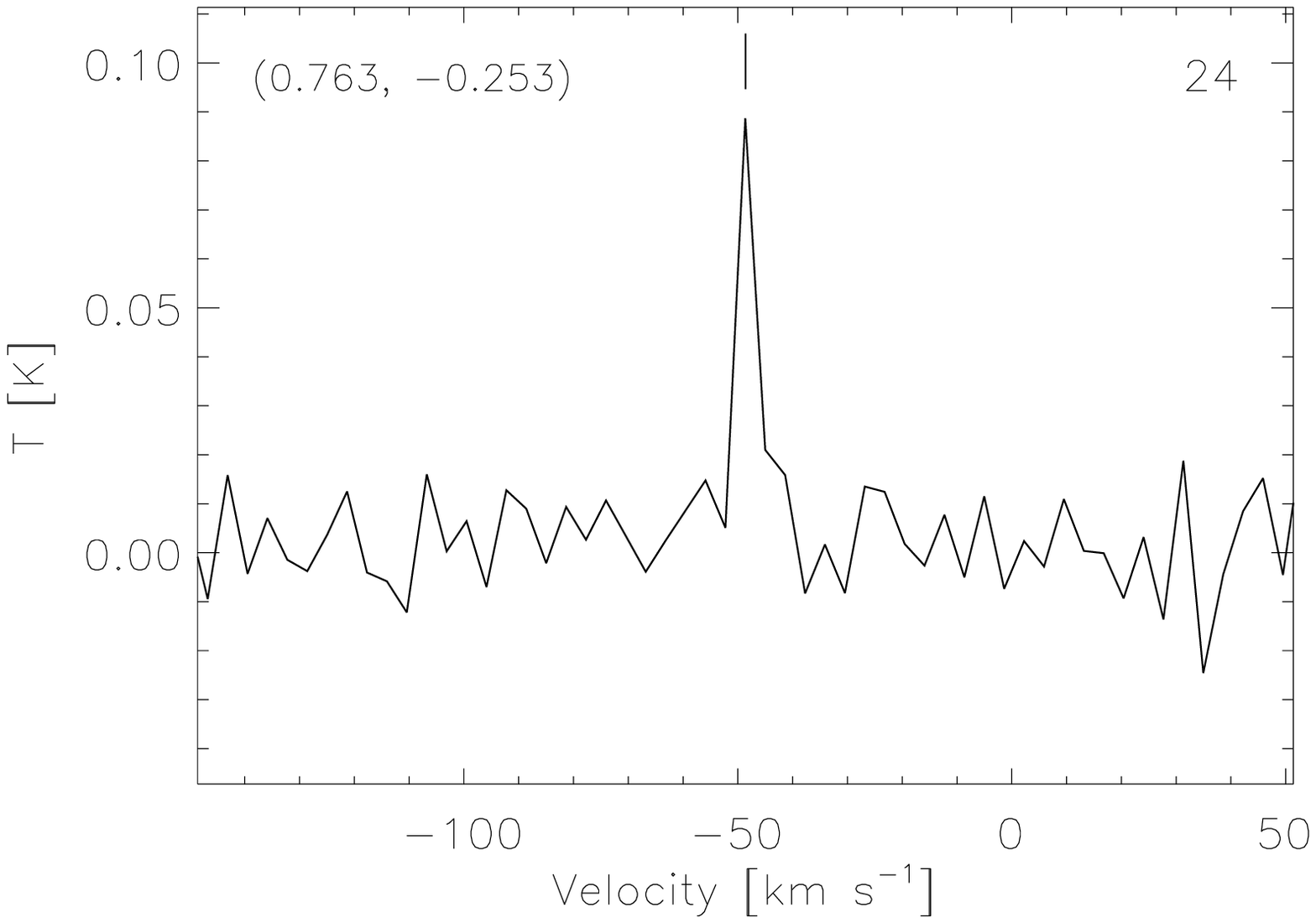}\\
\includegraphics[width=0.32\textwidth,clip=true]{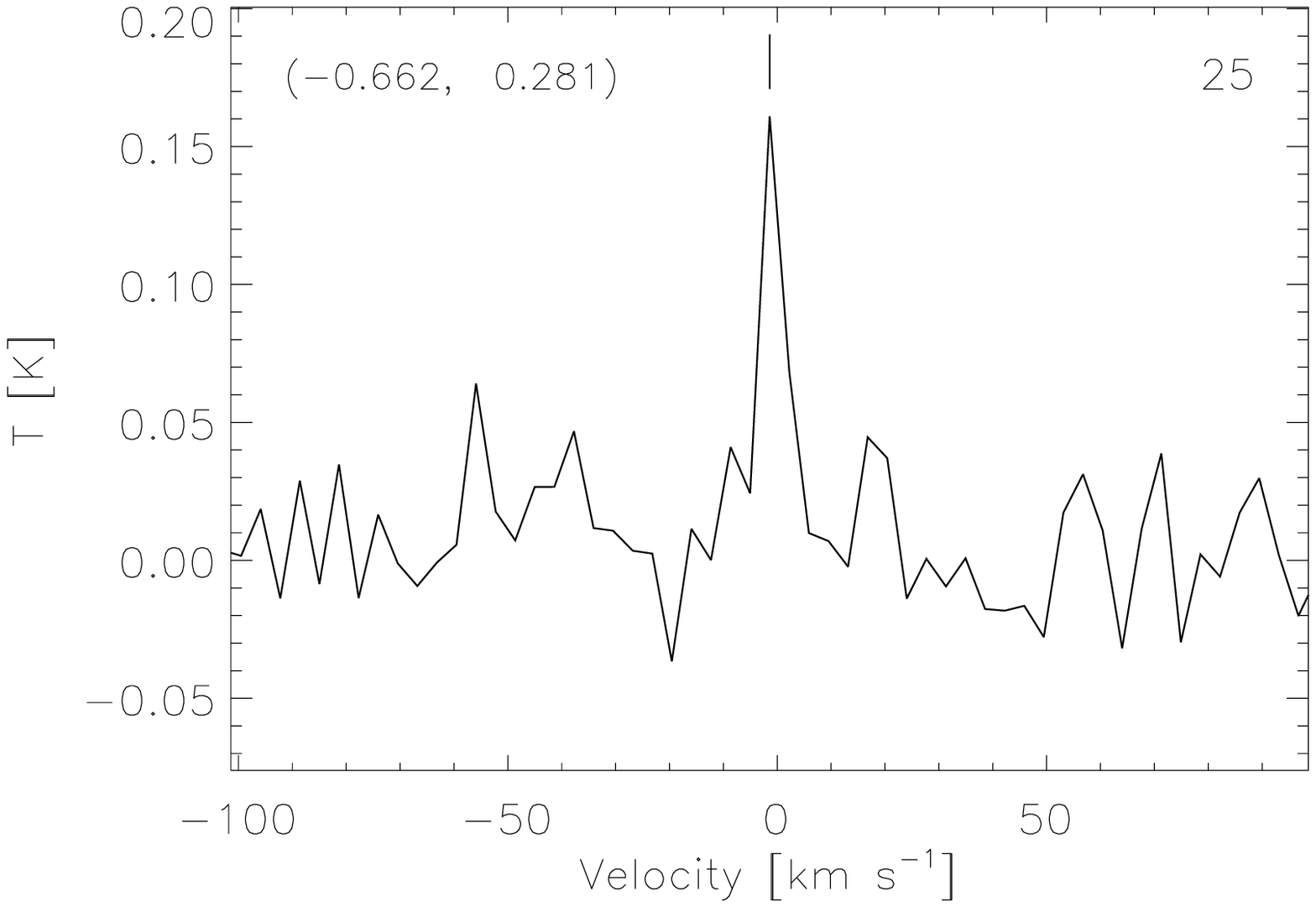}
\includegraphics[width=0.32\textwidth,clip=true]{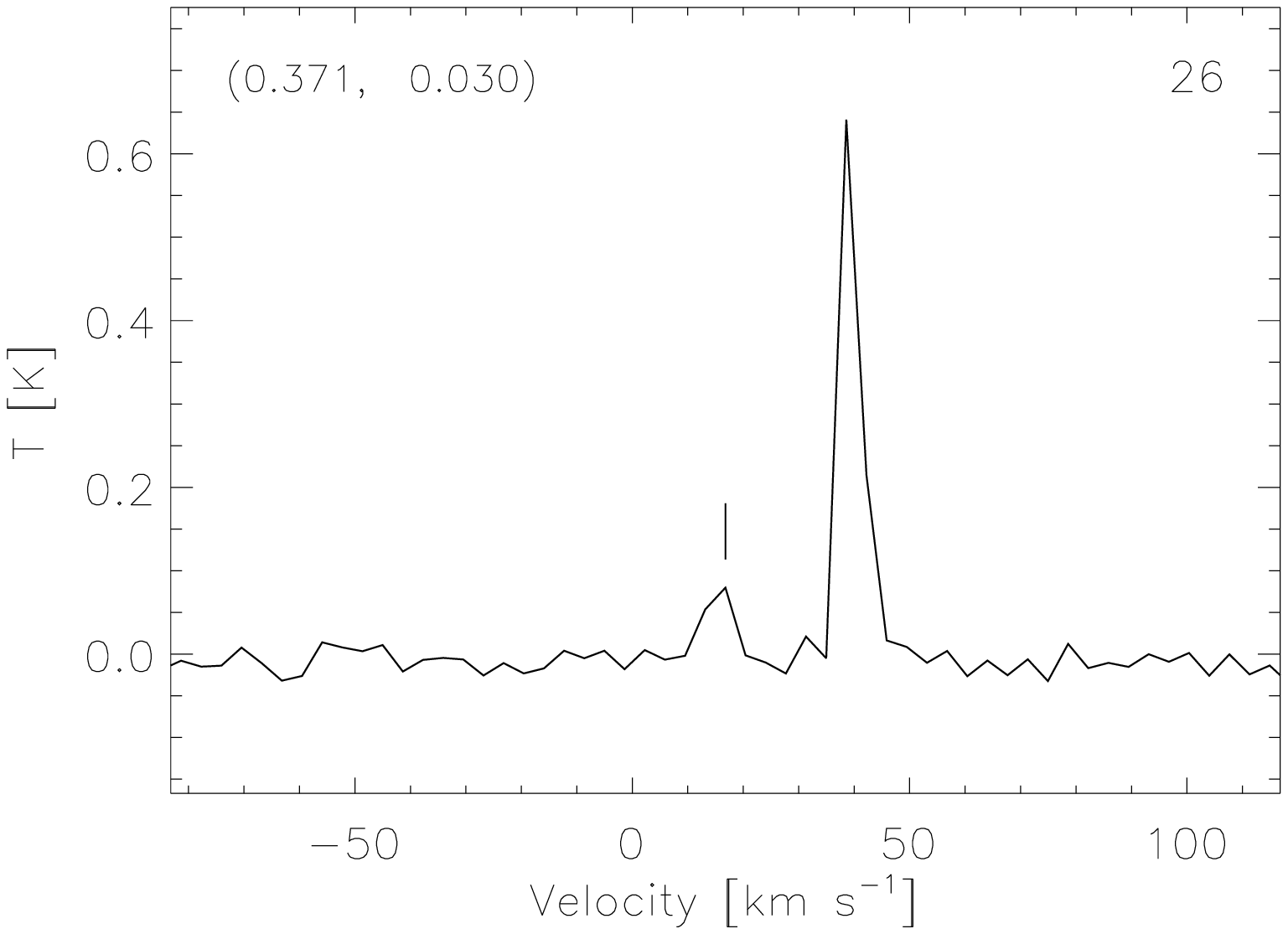}
\includegraphics[width=0.32\textwidth,clip=true]{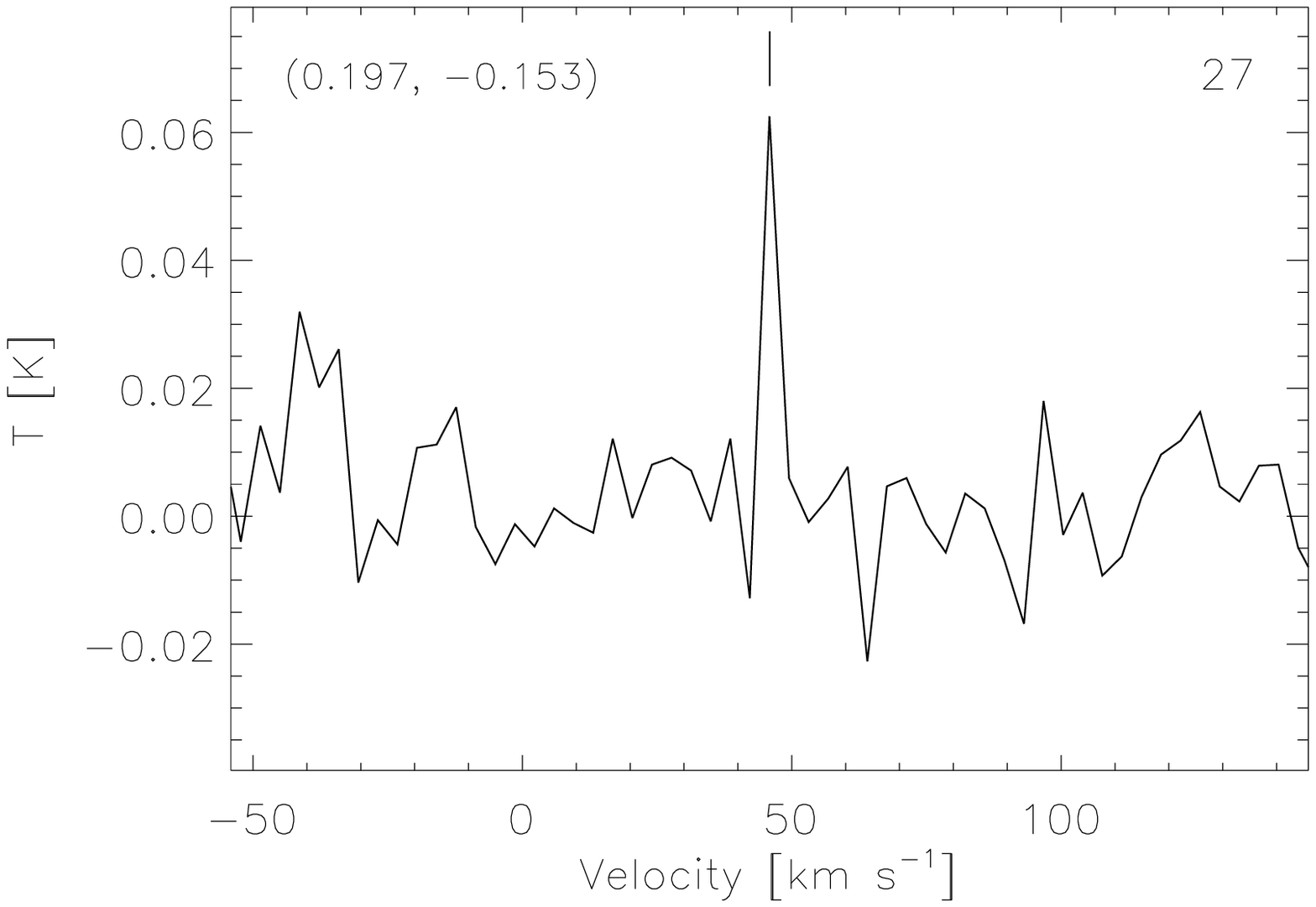}\\
\includegraphics[width=0.32\textwidth,clip=true]{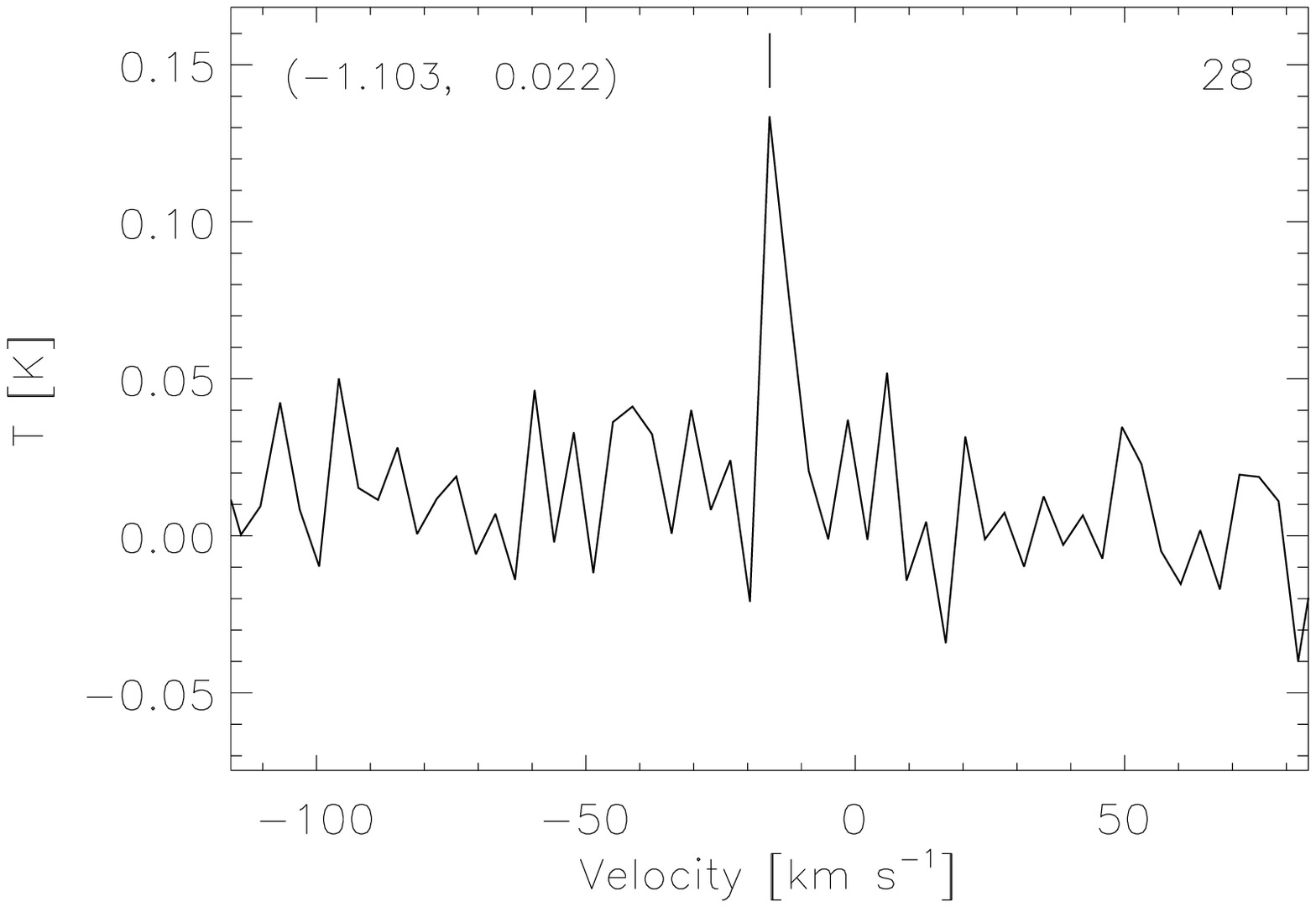}
\includegraphics[width=0.32\textwidth,clip=true]{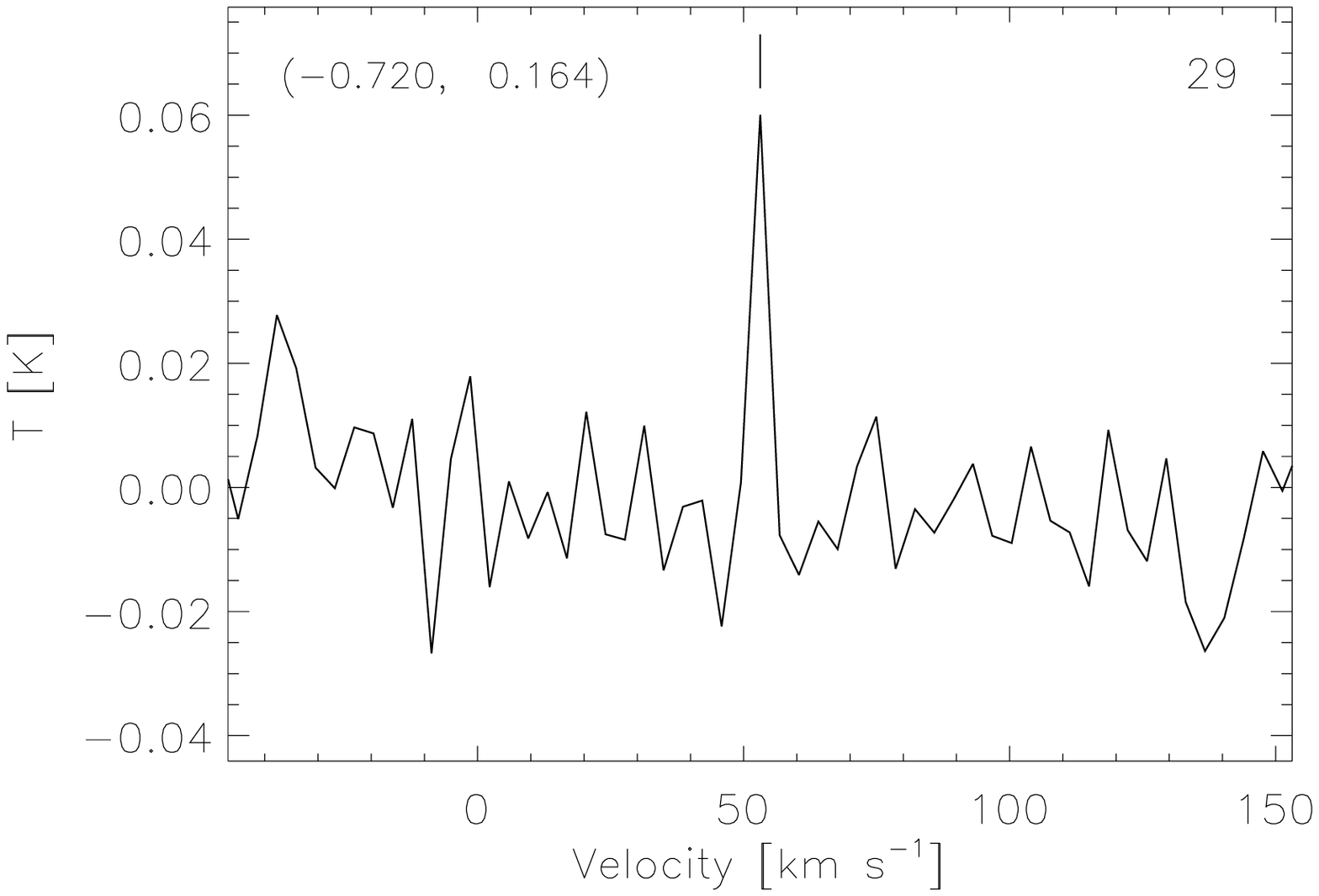}
\includegraphics[width=0.32\textwidth,clip=true]{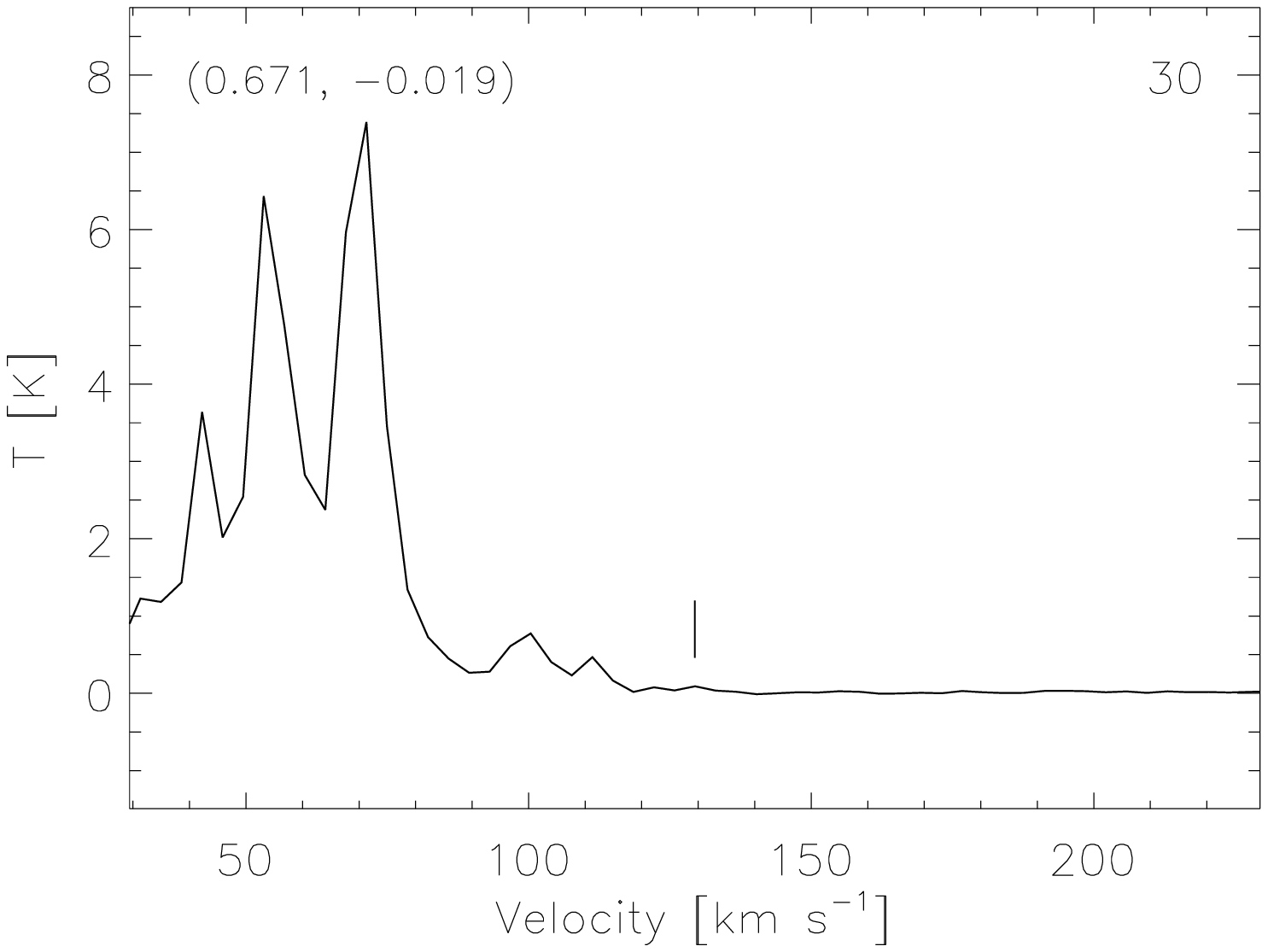}\\
Figure~\ref{first_spec}, cont.
\end{figure*}

\begin{figure*}
\includegraphics[width=0.32\textwidth,clip=true]{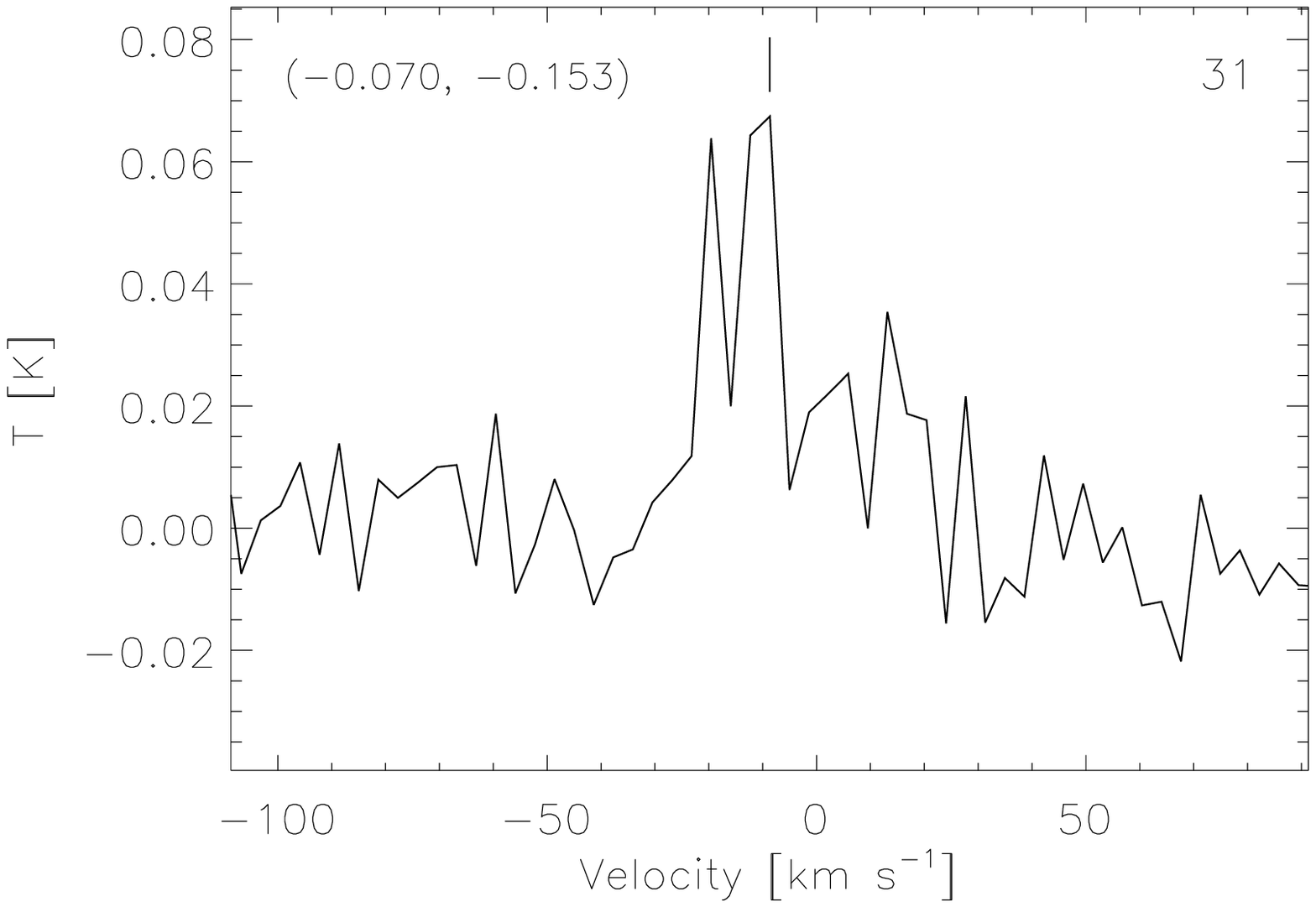}
\includegraphics[width=0.32\textwidth,clip=true]{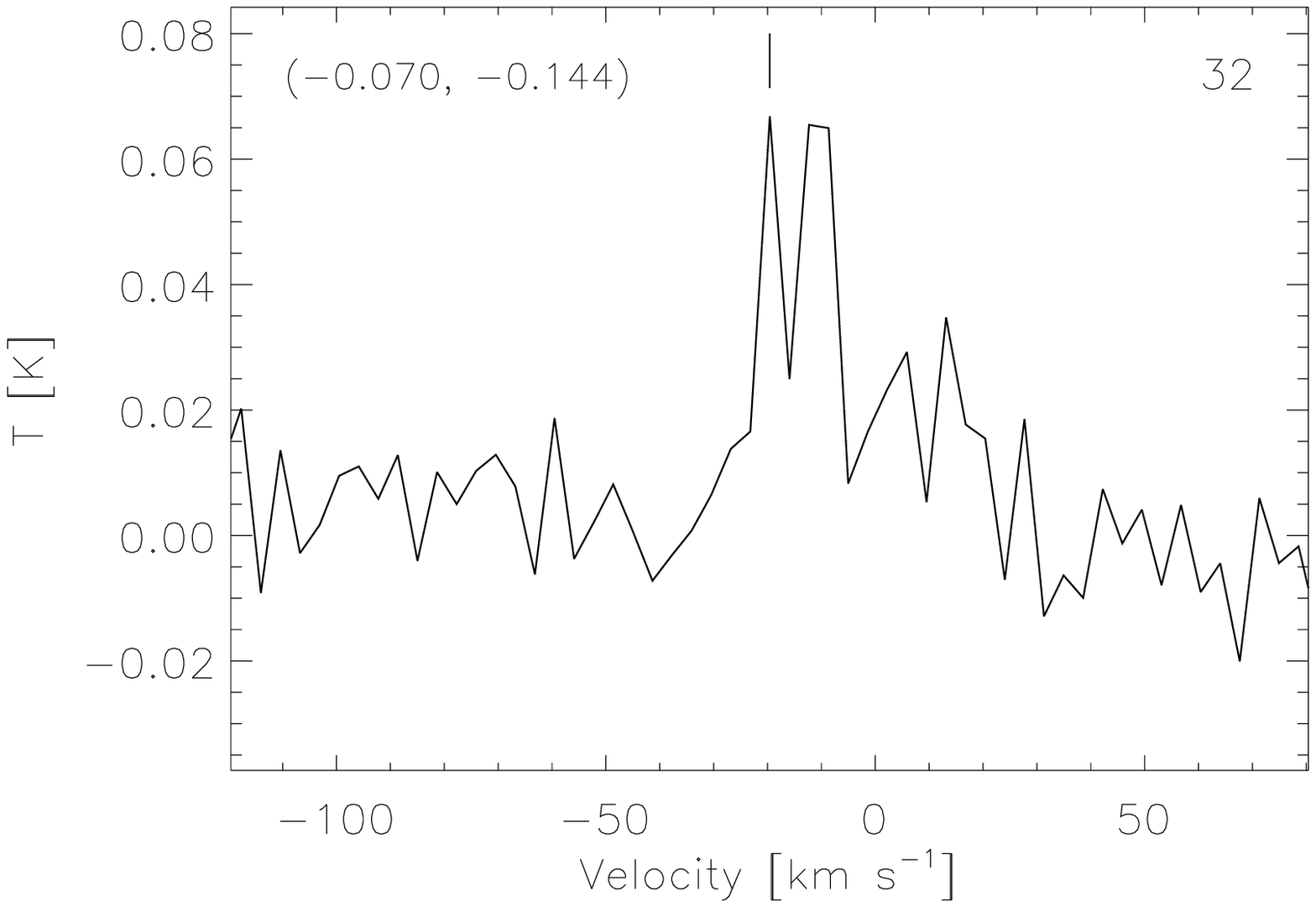}
\includegraphics[width=0.32\textwidth,clip=true]{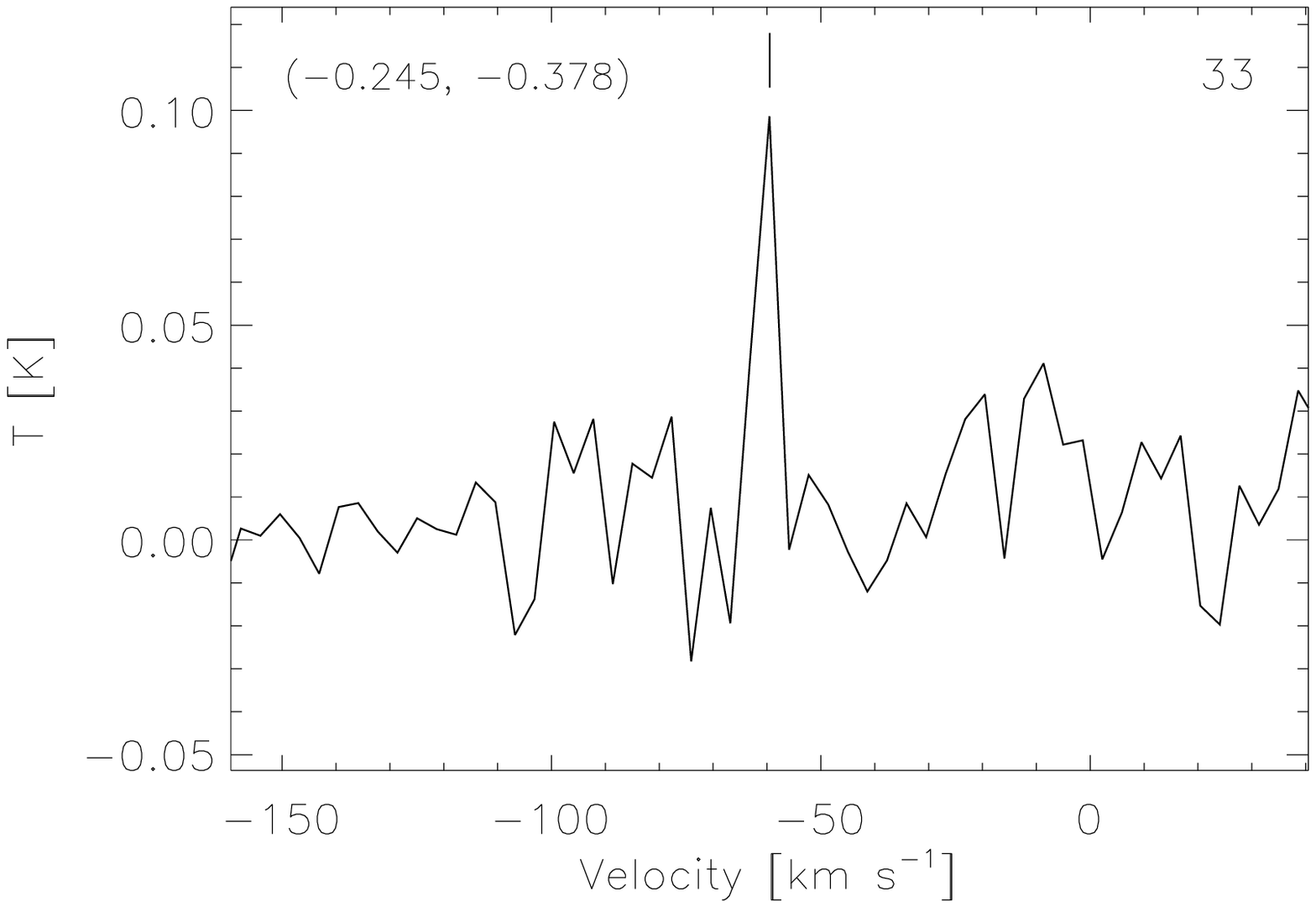}\\
\includegraphics[width=0.32\textwidth,clip=true]{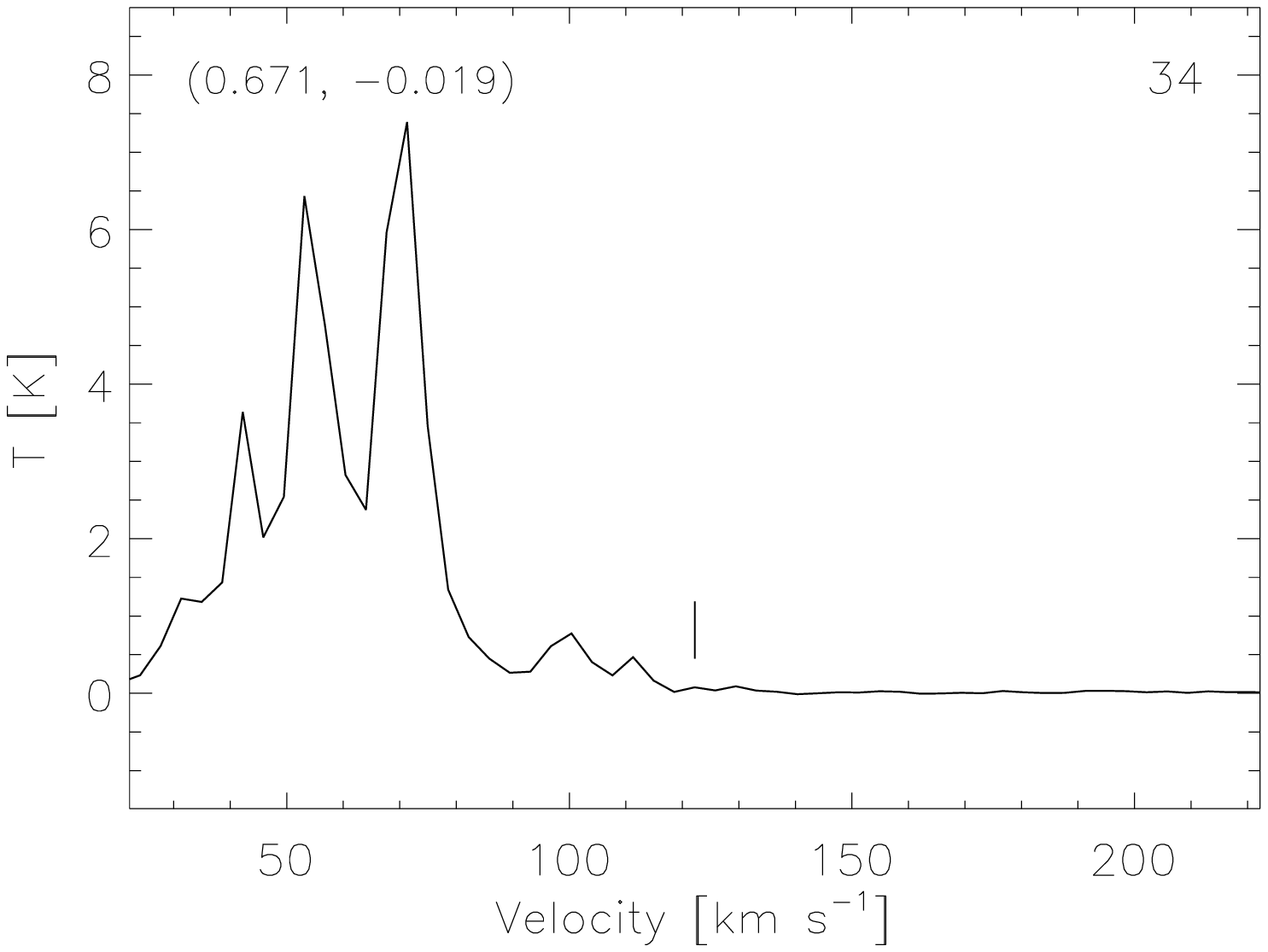}
\includegraphics[width=0.32\textwidth,clip=true]{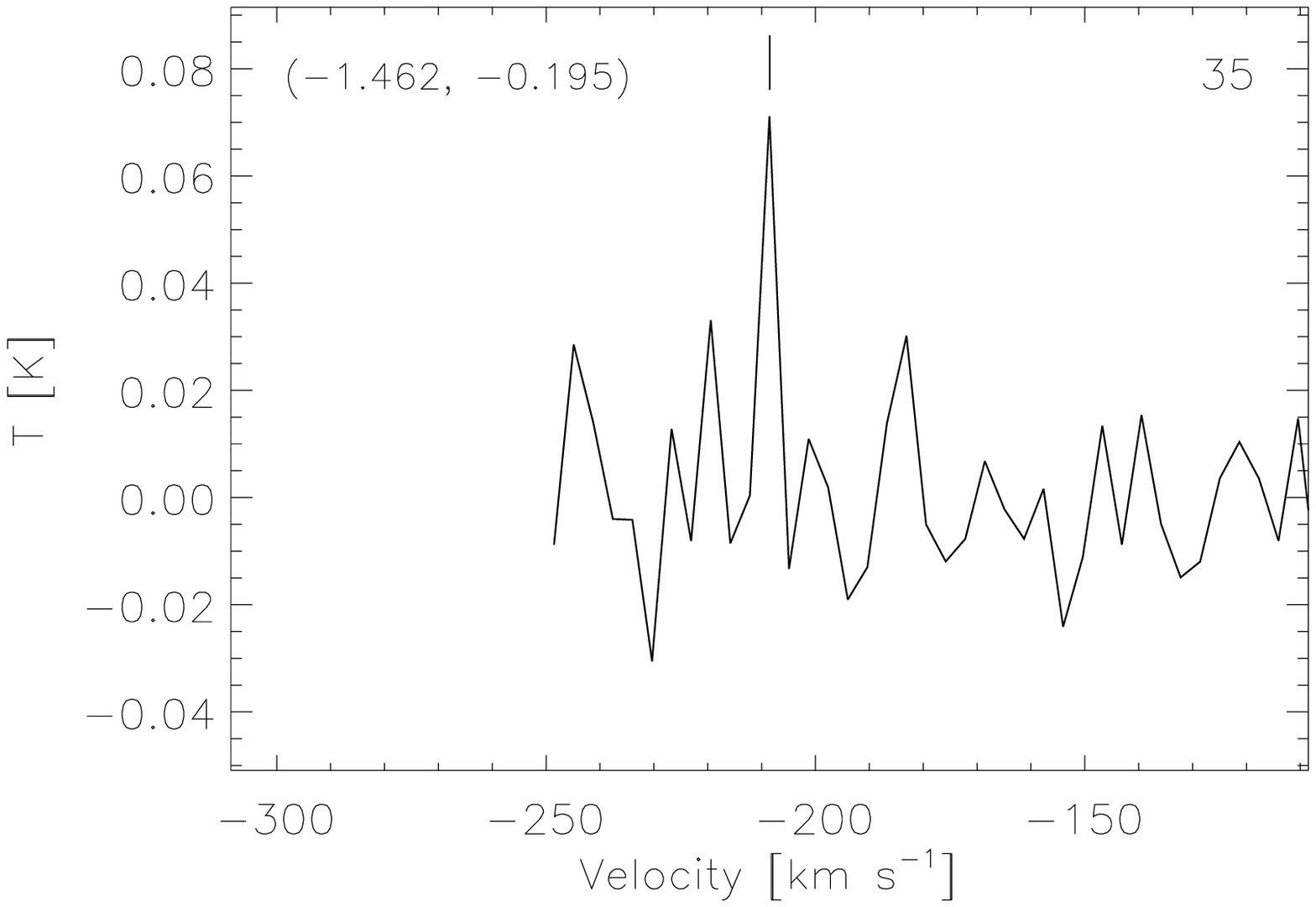}
\includegraphics[width=0.32\textwidth,clip=true]{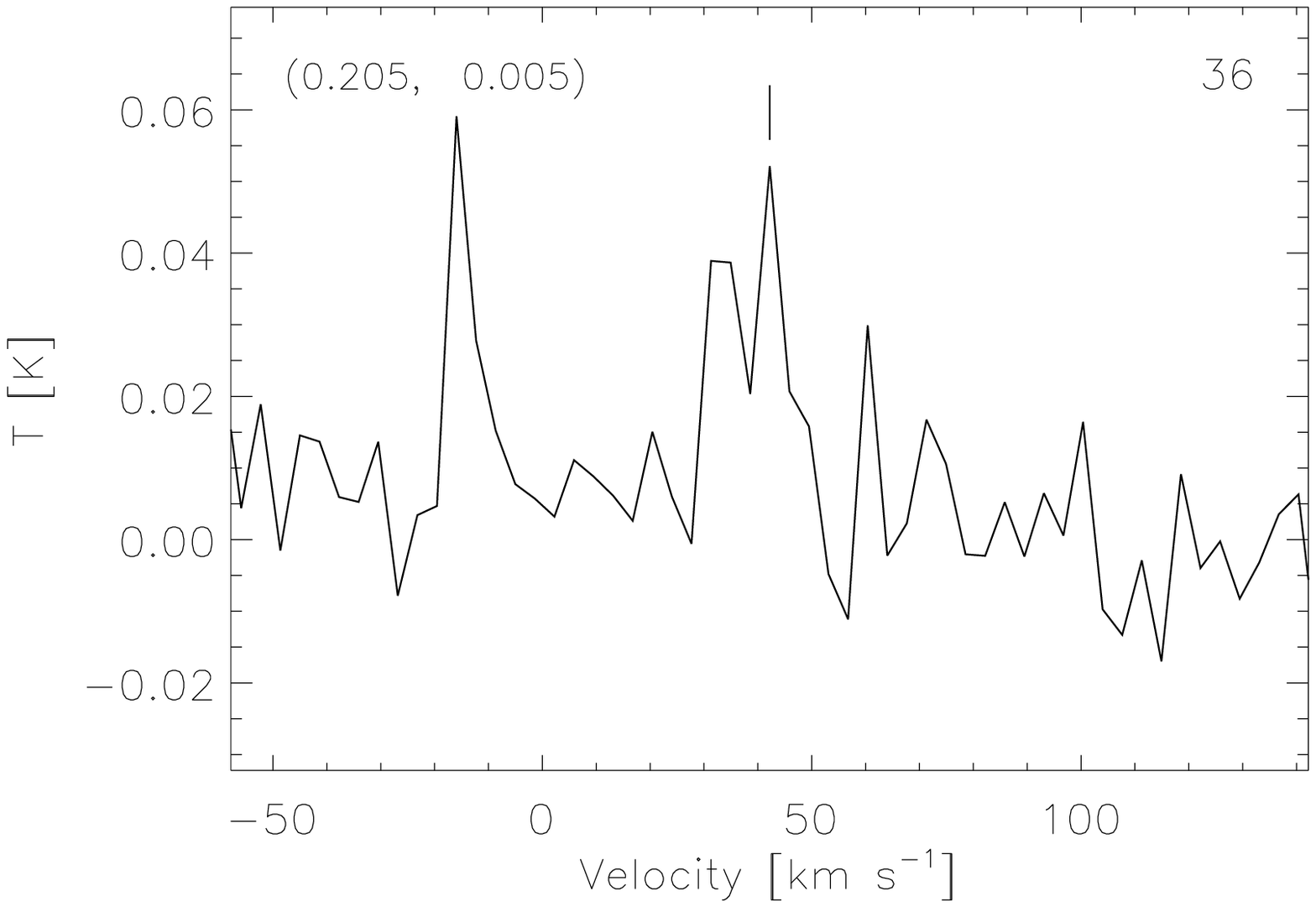}\\
\includegraphics[width=0.32\textwidth,clip=true]{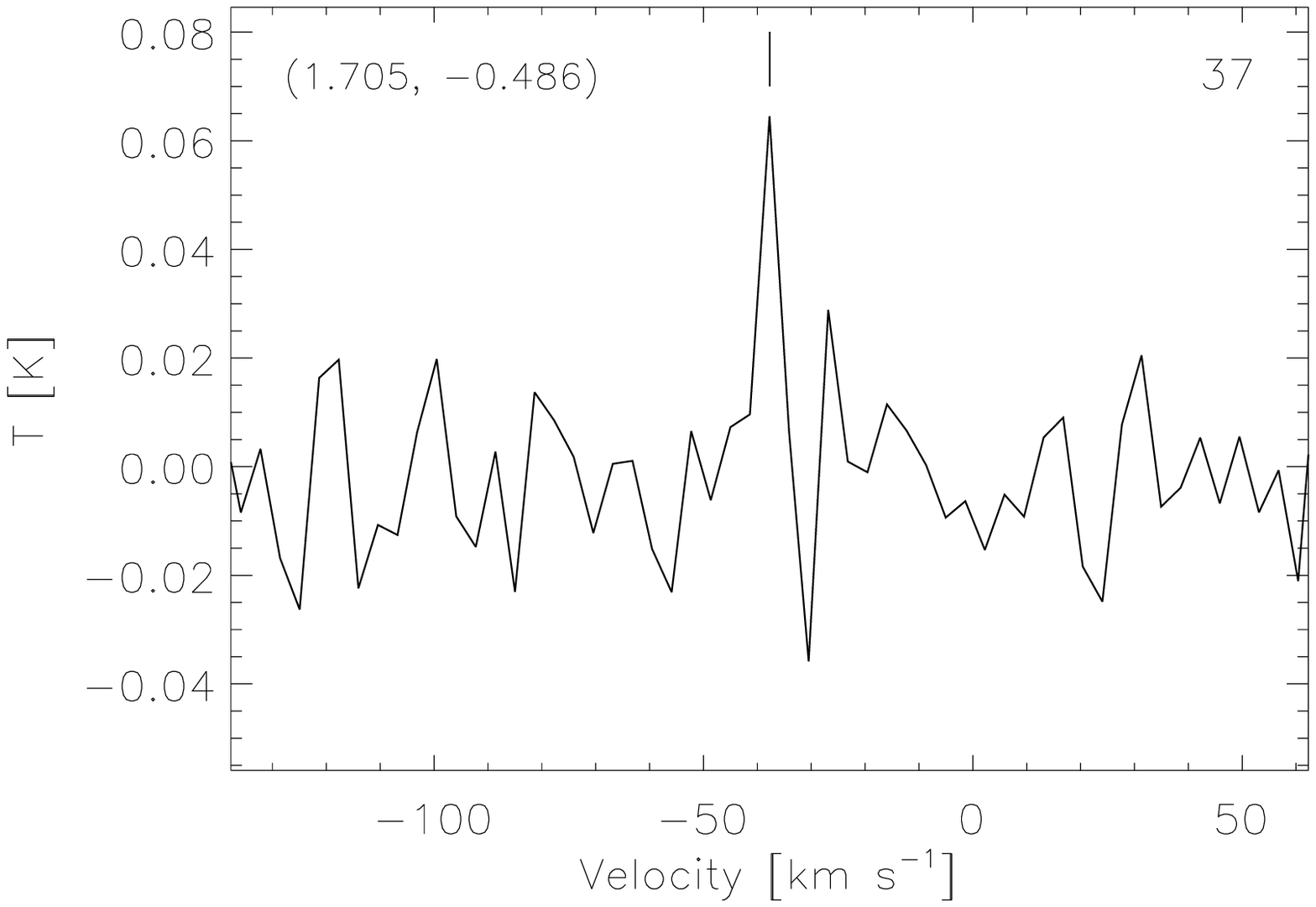}\\
Figure~\ref{first_spec}, cont.
\end{figure*}

\begin{figure}
\includegraphics[width=\hsize,clip=true]{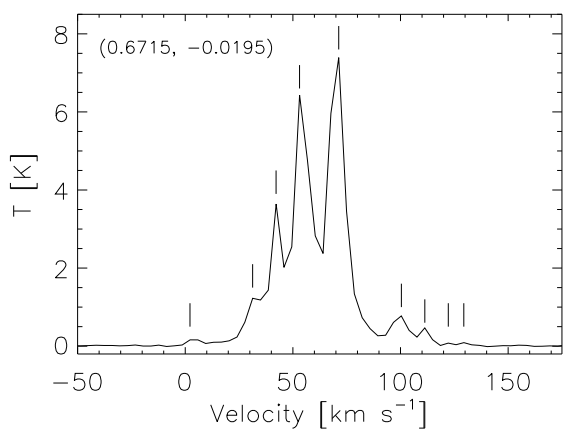}\\
\caption{Mopra \water\, maser spectrum toward Sgr~B.  The position of
  the spectrum (in $\ell, b$) is printed at the top of the plot.  The
  vertical lines indicate the velocities of masers identified by the
  Clumpfind routine.  The velocities of the masers identified are 2.2,
  31.3, 42.2, 53.1, 71.3, 100.4, 111.3, 122.2, and 129.4~\kms. The
  conversion to Jy is 12.3~Jy/K.
 \label{sgrb_spec}}
\end{figure}

The most notable feature in these distributions is the peak of
\water\, masers at $\ell~\sim~0.6$\degreesym, which can be attributed
to the \water\, maser emission from Sgr~B2.  We detect a total of 9
\water\, masers toward Sgr~B, including the three strongest emission
features in the survey.  Figure~\ref{sgrb_spec} shows a spectrum
toward Sgr~B that indicates the velocities of these 9 masers, which are
spread from $\sim~0$ to $\sim$~140~\kms.

\subsubsection{Association of \water\, and \meth\, Masers}

\begin{figure*}
\includegraphics[angle=-90,width=\textwidth,clip=true]{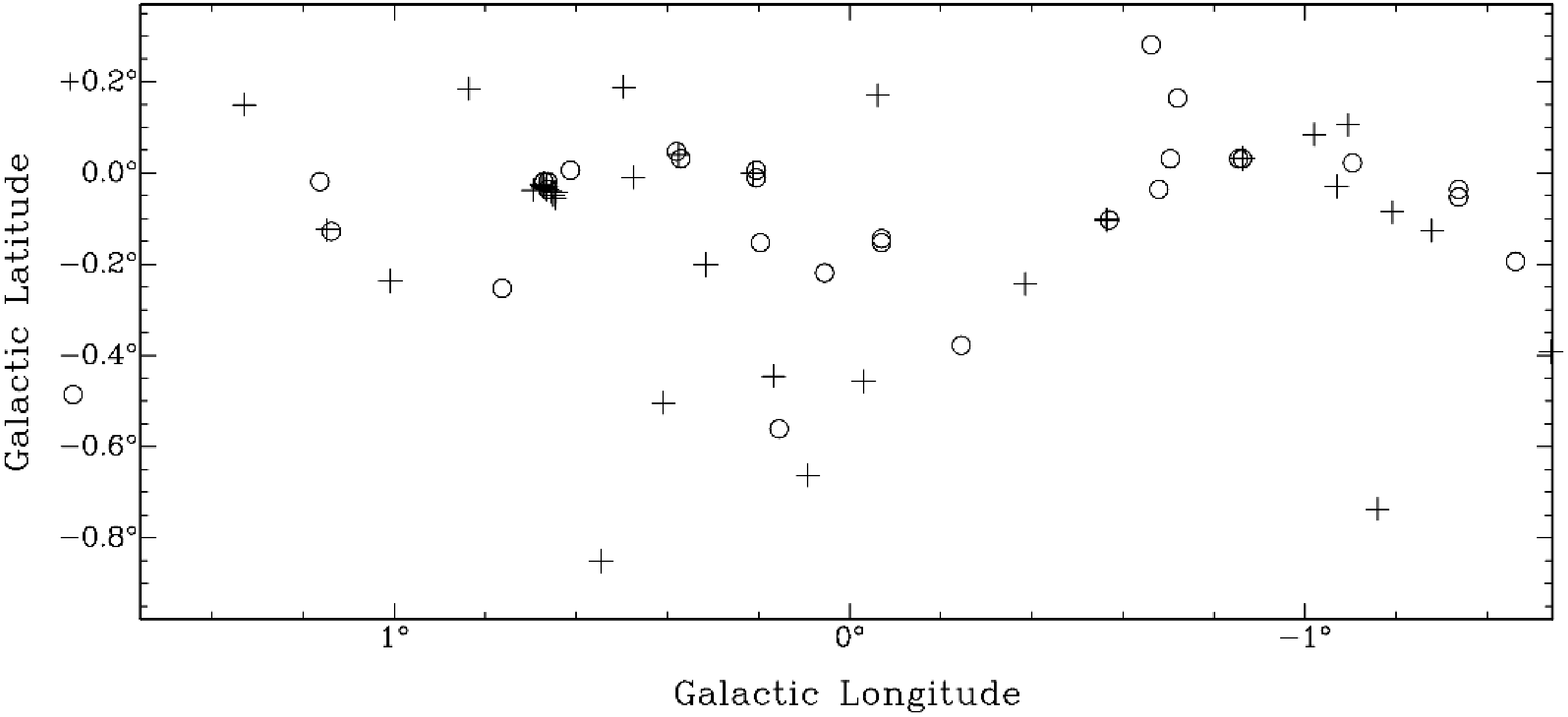}\\
\includegraphics[angle=-90,width=\textwidth,clip=true]{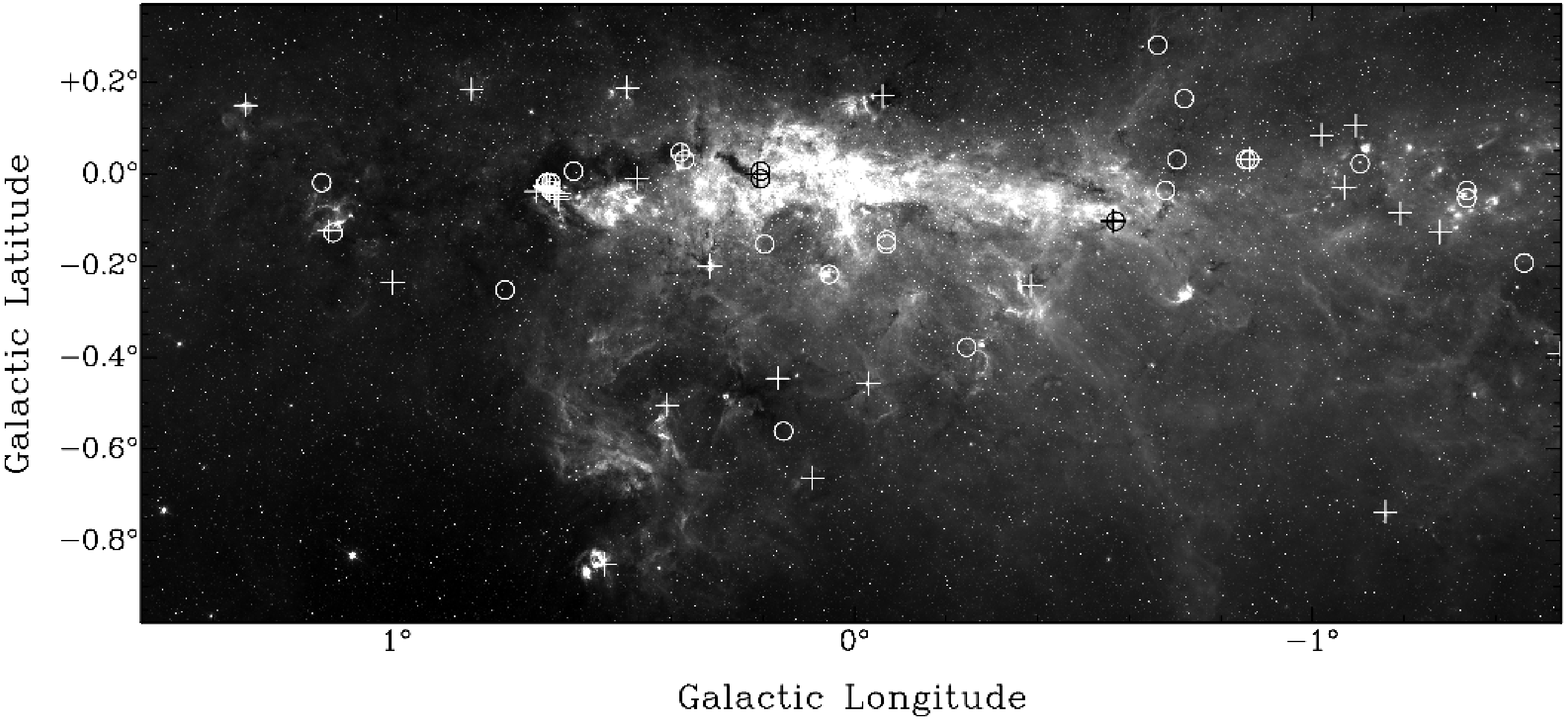}\\
\caption{{\it Top}: {\it Spitzer}/IRAC 8~\um\, image of the Galactic center region showing both \meth\, and
  \water\, masers.  Circles indicate the positions of the \water\,
  masers identified in the Mopra survey. The size of the circle
  represents the FWHM of the Mopra beam at 22~GHz.  Crosses (+) show
  the position of \meth\, masers (the size of the crosses have no
  physical meaning).  We consider a \meth\, maser to be associated
  with a \water\, maser when it falls within the \water\, maser beam.
  For details on these associations, including their velocity
  information, see Table~\ref{meth-water}. {\it Bottom}:
  Same as the {\it top} figure, but with a blank backround instead of
  the 8~\um\, image.
  \label{meth_water_irac}}
\end{figure*}

To associate \water\, and \meth\, masers, we search for \water\,
masers that are within one Mopra beam ($\sim$~144\arcsec) of \meth\,
masers.  Figure~\ref{meth_water_irac} shows the positions of \water\,
and \meth\, masers overlaid on an image of the GC region.  The positions
of the \water\, masers are marked by circles that match the Mopra beam
size.  Crosses mark the positions of \meth\, masers (the size of the
crosses has no physical meaning).  We consider a \water\, and \meth\,
maser to be associated when the center of the cross falls within the
circle marking the Mopra beam at the \water\, maser frequency.

\begin{longtab}
\begin{longtable}{lcc|cccc|cc}
\caption{\label{meth-water} Association of \water\, Masers with \meth\, Masers} \\
\hline\hline
\multicolumn{3}{c}{\meth\, Masers} & \multicolumn{4}{c}{\water\, Masers} & Angular & Green \\
\cline{1-3}
\cline{4-7}
Name  & $v_{low}$\tablefootmark{a} & $v_{high}$\tablefootmark{a} & \# & $\ell$ & $b$ & $v$ & Separation & Source \\
 &  (\kms) & (\kms)  &  & (\degreesym) & (\degreesym) & (\kms) & (\arcsec) & \\
\hline
\endfirsthead
\caption{continued.}\\
\hline\hline
\multicolumn{3}{c}{\meth\, Masers} & \multicolumn{4}{c}{\water\, Masers} & Angular & Green \\
\cline{1-3}
\cline{4-7}
Name  & $v_{low}$\tablefootmark{a} & $v_{high}$\tablefootmark{a} & \# & $\ell$ & $b$ & $v$ & Separation & Source \\
 &  (\kms) & (\kms)  &  & (\degreesym) & (\degreesym) & (\kms) & (\arcsec) & \\
\hline
\endhead
\hline
\endfoot
G359.138$+$0.031 &   $-$7.0 &    1.0 &            4 &  $-$0.854 &   0.030 &   $-$1.4 &           33 & N \\
 & &  &            5 &  $-$0.862 &   0.030 &   $-$1.4 &            4 & N \\
 & &  &           21 &  $-$0.862 &   0.030 &  $-$12.3 &            4 & N \\
G359.436$-$0.104 &  $-$53.0 &  $-$45.0 &           14 &  $-$0.570 &  $-$0.103 &  $-$59.5 &           23 & Y \\
G359.436$-$0.102 &  $-$58.0 &  $-$54.0 &           14 &  $-$0.570 &  $-$0.103 &  $-$59.5 &           24 & Y \\
G0.212$-$0.001 &   41.0 &   50.5 &           22 &   0.205 &  $-$0.011 &  $-$15.9 &           45 & N \\
 & &  &           36 &   0.205 &   0.005 &   42.2 &           34 & N \\
G0.376$+$0.040 &   35.0 &   40.0 &            8 &   0.380 &   0.047 &   38.6 &           30 & Y \\
 & &  &           26 &   0.371 &   0.030 &   16.8 &           38 & Y \\
G0.645$-$0.042 &   46.0 &   53.0 &            1 &   0.663 &  $-$0.036 &   71.3 &           69 & N \\
G0.651$-$0.049 &   46.0 &   49.0 &            1 &   0.663 &  $-$0.036 &   71.3 &           62 & N \\
G0.657$-$0.041 &   48.0 &   56.0 &            1 &   0.663 &  $-$0.036 &   71.3 &           29 & Y \\
G0.665$-$0.036 &   58.0 &   62.0 &            1 &   0.663 &  $-$0.036 &   71.3 &            6 & Y \\
 & &  &            2 &   0.663 &  $-$0.019 &   53.1 &           59 & Y \\
 & &  &            3 &   0.671 &  $-$0.019 &   42.2 &           64 & Y \\
 & &  &            6 &   0.671 &  $-$0.019 &   31.3 &           64 & Y \\
 & &  &            7 &   0.671 &  $-$0.019 &  100.4 &           64 & Y \\
 & &  &           10 &   0.671 &  $-$0.019 &  111.3 &           64 & Y \\
 & &  &           16 &   0.671 &  $-$0.019 &    2.2 &           64 & Y \\
 & &  &           30 &   0.671 &  $-$0.019 &  129.4 &           64 & Y \\
 & &  &           34 &   0.671 &  $-$0.019 &  122.2 &           64 & Y \\
G0.666$-$0.029 &   68.0 &   73.0 &            1 &   0.663 &  $-$0.036 &   71.3 &           29 & N \\
 & &  &            2 &   0.663 &  $-$0.019 &   53.1 &           36 & N \\
 & &  &            3 &   0.671 &  $-$0.019 &   42.2 &           38 & N \\
 & &  &            6 &   0.671 &  $-$0.019 &   31.3 &           38 & N \\
 & &  &            7 &   0.671 &  $-$0.019 &  100.4 &           38 & N \\
 & &  &           10 &   0.671 &  $-$0.019 &  111.3 &           38 & N \\
 & &  &           16 &   0.671 &  $-$0.019 &    2.2 &           38 & N \\
 & &  &           30 &   0.671 &  $-$0.019 &  129.4 &           38 & N \\
 & &  &           34 &   0.671 &  $-$0.019 &  122.2 &           38 & N \\
G0.667$-$0.034 &   49.0 &   56.0 &            1 &   0.663 &  $-$0.036 &   71.3 &           17 & Y \\
 & &  &            2 &   0.663 &  $-$0.019 &   53.1 &           53 & Y \\
 & &  &            3 &   0.671 &  $-$0.019 &   42.2 &           53 & Y \\
 & &  &            6 &   0.671 &  $-$0.019 &   31.3 &           53 & Y \\
 & &  &            7 &   0.671 &  $-$0.019 &  100.4 &           53 & Y \\
 & &  &           10 &   0.671 &  $-$0.019 &  111.3 &           53 & Y \\
 & &  &           16 &   0.671 &  $-$0.019 &    2.2 &           53 & Y \\
 & &  &           30 &   0.671 &  $-$0.019 &  129.4 &           53 & Y \\
 & &  &           34 &   0.671 &  $-$0.019 &  122.2 &           53 & Y \\
G0.672$-$0.031 &   55.0 &   59.0 &            1 &   0.663 &  $-$0.036 &   71.3 &           37 & N \\
 & &  &            2 &   0.663 &  $-$0.019 &   53.1 &           54 & N \\
 & &  &            3 &   0.671 &  $-$0.019 &   42.2 &           43 & N \\
 & &  &            6 &   0.671 &  $-$0.019 &   31.3 &           43 & N \\
 & &  &            7 &   0.671 &  $-$0.019 &  100.4 &           43 & N \\
 & &  &           10 &   0.671 &  $-$0.019 &  111.3 &           43 & N \\
 & &  &           16 &   0.671 &  $-$0.019 &    2.2 &           43 & N \\
 & &  &           30 &   0.671 &  $-$0.019 &  129.4 &           43 & N \\
 & &  &           34 &   0.671 &  $-$0.019 &  122.2 &           43 & N \\
G0.673$-$0.029 &   65.5 &   66.5 &            1 &   0.663 &  $-$0.036 &   71.3 &           46 & N \\
 & &  &            2 &   0.663 &  $-$0.019 &   53.1 &           49 & N \\
 & &  &            3 &   0.671 &  $-$0.019 &   42.2 &           33 & N \\
 & &  &            6 &   0.671 &  $-$0.019 &   31.3 &           33 & N \\
 & &  &            7 &   0.671 &  $-$0.019 &  100.4 &           33 & N \\
 & &  &           10 &   0.671 &  $-$0.019 &  111.3 &           33 & N \\
 & &  &           16 &   0.671 &  $-$0.019 &    2.2 &           33 & N \\
 & &  &           30 &   0.671 &  $-$0.019 &  129.4 &           33 & N \\
 & &  &           34 &   0.671 &  $-$0.019 &  122.2 &           33 & N \\
G0.677$-$0.025 &   70.0 &   77.0 &            1 &   0.663 &  $-$0.036 &   71.3 &           64 & N \\
 & &  &            2 &   0.663 &  $-$0.019 &   53.1 &           54 & N \\
 & &  &            3 &   0.671 &  $-$0.019 &   42.2 &           28 & N \\
 & &  &            6 &   0.671 &  $-$0.019 &   31.3 &           28 & N \\
 & &  &            7 &   0.671 &  $-$0.019 &  100.4 &           28 & N \\
 & &  &           10 &   0.671 &  $-$0.019 &  111.3 &           28 & N \\
 & &  &           16 &   0.671 &  $-$0.019 &    2.2 &           28 & N \\
 & &  &           30 &   0.671 &  $-$0.019 &  129.4 &           28 & N \\
 & &  &           34 &   0.671 &  $-$0.019 &  122.2 &           28 & N \\
G1.147$-$0.124 &  $-$20.5 &  $-$14.0 &           11 &   1.138 &  $-$0.128 &  $-$19.6 &           34 & N \\
\end{longtable}
\tablefoot{
\tablefoottext{a}{For easier comparison with the \water\, masers, here we list the upper and lower bounds
of the \meth\, maser emission, as listed by C10.}
}
\end{longtab}

These results, including the velocity information of the masers, are
also shown in Table~\ref{meth-water}. Columns~1-3 of the table list the
\meth\, maser name (from C10), and the lower and upper velocity limits
of the \meth\, maser emission (from C10).  In Columns~4-7, we list the
properties of the \water\, masers that are assocaited with the \meth\,
masers--the \water\, maser numbers, their Galactic coordinates, and
the velocity of their peak emission.  Column~8 lists the angular
separation between the associated \meth\, and \water\, masers.
Column~9 indicates if the \meth\, maser is associated with a green
source.  If a \meth\, maser is associated with multiple \water\,
masers, then the \meth\, maser information is printed only once, and a
line for each of the \water\, masers is printed.

Because 22~GHz \water\, masers are found in shocks and outflows, their
velocities may not match the velocities of the radiatively excited
6.7~GHz \meth\, masers.  Therefore, we do not use the velocities of
the masers as association criteria.  The velocites of both types of
masers are listed in Table~\ref{meth-water}.

Some of the \water\, masers are associated with more than one
\meth\, maser.  Without higher angular resolution, we are unable to
tell if the masers are exactly coincident or are just nearby to one
another. Nevertheless, we include these associations, along with whether
or not the \meth\, maser is coincident with a green source, to examine
the star formation in the CMZ, as well as to determine candidates for
potential follow-up observations at higher angular resolution.

Of the 175 \meth\, masers in the C10 catalog, 39 fall within the Mopra
\water\, maser survey region.  Of these 39, 15 are associated with
\water\, masers, while 24 are not.  The list of these 15 \meth\,
masers can be found in the first column of Table~\ref{meth-water}.

There are a total of 37 \water\, masers in the Mopra survey region, 18
of which are associated with \meth\, masers.  These \water\, masers,
along with their identification number, can be found in
Table~\ref{meth-water}.  Because \water\, masers can be associated
with more than one \meth\, masers, some of the \water\, masers are
listed more than once in this table.  For completeness, we also
include Table~\ref{meth-nowater}, which lists \meth\, masers that are
not associated with \water\, masers.  The first three columns of
Table~\ref{meth-nowater} contain the name and Galactic coordinates of
the \meth\, masers (from C10), Columns~4 and 5 contain the minimum and
maximum velocities of the maser features (from C10), the peak
intensity of the maser emission is in Column~6, and the last column
contains a flag indicating if the \meth\, maser is associated with a
green source.  Many of the \meth\, masers in this list are outside of
the Mopra \water\, maser survey region.  In addition, we list \water\,
masers that are not associated with \meth\, masers in
Table~\ref{water-nometh}.  The columns in this table are: (1) the
number of the identified \water\, maser, (2) the Galactic longitude of
the maser, (3) the Galactic latitude of the maser, (4) the velocity of
the peak emission from the maser, and (5) the peak emission of the
maser. It is interesting to note that each of the 8 strongest \water\,
masers (numbers 1 through 8 in Table~\ref{water-tot}) are associated
with \meth\, masers, and are thus not included in
Table~\ref{water-nometh}.

\begin{longtab}
\begin{longtable}{lcccccc}
\caption{\label{meth-nowater} \meth\, Masers not Associated with \water\, Masers}\\
\hline\hline
Name  & $\ell$ & $b$ & $v_{low}$\tablefootmark{a} & $v_{high}$\tablefootmark{a} & $S_{peak}$ & Green \\
 & (\degreesym) & (\degreesym) & (\kms) & (\kms) & (Jy) &  Source \\
\hline
\endfirsthead
\caption{continued.}\\
\hline\hline
Name  & $\ell$ & $b$ & $v_{low}$\tablefootmark{a} & $v_{high}$\tablefootmark{a} & $S_{peak}$ & Green \\
 & (\degreesym) & (\degreesym) & (\kms) & (\kms) & (Jy) &  Source \\
\hline
\endhead
\hline
\endfoot
G345.003$-$0.223 & 345.003 &  $-$0.223 &  $-$25.0 &  $-$20.1 & 236 & Y \\
G345.003$-$0.224 & 345.003 &  $-$0.224 &  $-$33.0 &  $-$25.0 & 102 & Y \\
G345.131$-$0.174 & 345.131 &  $-$0.174 &  $-$31.0 &  $-$28.0 & 3.1 & Y \\
G345.198$-$0.030 & 345.198 &  $-$0.030 &   $-$4.0 &    1.0 & 2.53 & N \\
G345.205$+$0.317 & 345.205 &   0.317 &  $-$64.1 &  $-$59.9 & 0.8 & N \\
G345.407$-$0.952 & 345.407 &  $-$0.952 &  $-$15.5 &  $-$14.0 & 2 & N \\
G345.424$-$0.951 & 345.424 &  $-$0.951 &  $-$21.0 &   $-$5.0 & 2.92 & N \\
G345.441$+$0.205 & 345.441 &   0.205 &  $-$13.0 &    2.0 & 2.27 & N \\
G345.487$+$0.314 & 345.487 &   0.314 &  $-$24.0 &  $-$21.5 & 2.5 & N \\
G345.505$+$0.348 & 345.505 &   0.348 &  $-$23.1 &  $-$10.5 & 300 & Y \\
G345.576$-$0.225 & 345.576 &  $-$0.225 & $-$127.2 & $-$122.0 & 0.64 & Y \\
G345.807$-$0.044 & 345.807 &  $-$0.044 &   $-$3.0 &   $-$0.5 & 1 & N \\
G345.824$+$0.044 & 345.824 &   0.044 &  $-$12.0 &   $-$9.0 & 3.17 & Y \\
G345.949$-$0.268 & 345.949 &  $-$0.268 &  $-$22.5 &  $-$21.4 & 1.53 & Y \\
G345.985$-$0.020 & 345.985 &  $-$0.020 &  $-$85.5 &  $-$81.7 & 5.7 & Y \\
G346.036$+$0.048 & 346.036 &   0.048 &  $-$14.5 &   $-$3.9 & 8.99 & N \\
G346.231$+$0.119 & 346.232 &   0.119 &  $-$96.6 &  $-$92.6 & 1.5 & Y \\
G346.480$+$0.221 & 346.480 &   0.221 &  $-$21.0 &  $-$14.0 & 30.15 & Y \\
G346.481$+$0.132 & 346.481 &   0.132 &  $-$11.6 &   $-$4.9 & 2.1 & Y \\
G346.517$+$0.117 & 346.517 &   0.117 &   $-$3.0 &    1.0 & 0.3 & N \\
G346.522$+$0.085 & 346.522 &   0.085 &    4.7 &    6.1 & 1.9 & N \\
G347.230$+$0.016 & 347.230 &   0.016 &  $-$69.9 &  $-$68.0 & 0.86 & N \\
G347.583$+$0.213 & 347.583 &   0.213 & $-$103.8 &  $-$96.0 & 3.18 & N \\
G347.628$+$0.149 & 347.628 &   0.149 &  $-$98.9 &  $-$95.0 & 19.2 & Y \\
G347.631$+$0.211 & 347.631 &   0.211 &  $-$94.0 &  $-$89.0 & 5.81 & N \\
G347.817$+$0.018 & 347.817 &   0.018 &  $-$26.0 &  $-$22.8 & 2.52 & N \\
G347.863$+$0.019 & 347.863 &   0.019 &  $-$37.8 &  $-$28.0 & 6.38 & Y \\
G347.902$+$0.052 & 347.902 &   0.052 &  $-$31.5 &  $-$27.0 & 5.37 & N \\
G348.027$+$0.106 & 348.027 &   0.106 & $-$122.8 & $-$114.3 & 3.07 & Y \\
G348.195$+$0.768 & 348.195 &   0.768 &   $-$2.8 &   $-$0.2 & 4.55 & N \\
G348.550$-$0.979 & 348.550 &  $-$0.979 &  $-$19.0 &   $-$7.0 & 41.1 & Y \\
G348.550$-$0.979n & 348.550 &  $-$0.979 &  $-$23.0 &  $-$14.0 & 22.6 & Y \\
G348.579$-$0.920 & 348.579 &  $-$0.920 &  $-$16.0 &  $-$14.0 & 0.32 & Y \\
G348.654$+$0.244 & 348.654 &   0.244 &   16.5 &   17.5 & 0.82 & Y \\
G348.723$-$0.078 & 348.723 &  $-$0.077 &    9.0 &   12.0 & 2.58 & N \\
G348.703$-$1.043 & 348.704 &  $-$1.044 &  $-$17.5 &   $-$2.5 & 65 & N \\
G348.727$-$1.037 & 348.727 &  $-$1.037 &  $-$12.0 &   $-$6.0 & 80.78 & N \\
G348.884$+$0.096 & 348.885 &   0.097 &  $-$79.0 &  $-$73.0 & 12.18 & Y \\
G348.892$-$0.180 & 348.892 &  $-$0.180 &    1.0 &    2.0 & 2.7 & N \\
G349.067$-$0.017 & 349.067 &  $-$0.017 &    6.0 &   16.0 & 2.3 & Y \\
G349.092$+$0.105 & 349.092 &   0.105 &  $-$78.0 &  $-$74.0 & 33.3 & Y \\
G349.092$+$0.106 & 349.092 &   0.106 &  $-$83.0 &  $-$78.0 & 9.9 & Y \\
G349.151$+$0.021 & 349.151 &   0.021 &   14.1 &   25.0 & 3.36 & N \\
G349.579$-$0.679 & 349.579 &  $-$0.679 &  $-$26.0 &  $-$24.0 & 1.9 & N \\
G349.799$+$0.108 & 349.798 &   0.108 &  $-$65.5 &  $-$57.4 & 3 & Y \\
G349.884$+$0.231 & 349.884 &   0.231 &   13.5 &   17.5 & 6.96 & Y \\
G350.015$+$0.433 & 350.015 &   0.433 &  $-$37.0 &  $-$29.0 & 7.2 & Y \\
G350.104$+$0.084 & 350.104 &   0.084 &  $-$69.0 &  $-$67.5 & 9.9 & N \\
G350.105$+$0.083 & 350.105 &   0.083 &  $-$76.0 &  $-$61.0 & 13.6 & N \\
G350.116$+$0.084 & 350.116 &   0.084 &  $-$69.0 &  $-$67.0 & 10.3 & N \\
G350.116$+$0.220 & 350.116 &   0.220 &    3.0 &    5.0 & 2.78 & N \\
G350.189$+$0.003 & 350.189 &   0.003 &  $-$65.0 &  $-$62.0 & 1.07 & N \\
G350.299$+$0.122 & 350.299 &   0.122 &  $-$70.0 &  $-$61.0 & 31.34 & Y \\
G350.340$+$0.141 & 350.340 &   0.141 &  $-$60.0 &  $-$57.5 & 2.5 & N \\
G350.344$+$0.116 & 350.344 &   0.116 &  $-$66.0 &  $-$55.0 & 19.9 & N \\
G350.356$-$0.068 & 350.356 &  $-$0.068 &  $-$68.5 &  $-$66.0 & 1.44 & Y \\
G350.470$+$0.029 & 350.470 &   0.029 &  $-$11.0 &   $-$5.5 & 1.44 & N \\
G350.520$-$0.350 & 350.520 &  $-$0.349 &  $-$25.0 &  $-$22.0 & 1.67 & Y \\
G350.686$-$0.491 & 350.686 &  $-$0.491 &  $-$15.0 &  $-$13.0 & 17.85 & Y \\
G350.776$+$0.138 & 350.776 &   0.138 &   34.5 &   39.0 & 0.65 & N \\
G351.161$+$0.697 & 351.161 &   0.697 &   $-$7.0 &   $-$2.0 & 17.02 & Y \\
G351.242$+$0.670 & 351.242 &   0.670 &    2.0 &    3.0 & 0.74 & N \\
G351.251$+$0.652 & 351.251 &   0.652 &   $-$7.5 &   $-$6.0 & 0.99 & Y \\
G351.382$-$0.181 & 351.382 &  $-$0.181 &  $-$69.0 &  $-$58.0 & 19.66 & Y \\
G351.417$+$0.645 & 351.417 &   0.645 &  $-$12.0 &   $-$6.0 & 3423 & Y \\
G351.417$+$0.646 & 351.417 &   0.646 &  $-$12.0 &   $-$7.0 & 1840 & Y \\
G351.445$+$0.660 & 351.445 &   0.660 &  $-$14.0 &    1.0 & 129 & Y \\
G351.581$-$0.353 & 351.581 &  $-$0.353 & $-$100.0 &  $-$88.0 & 47.5 & Y \\
G351.611$+$0.172 & 351.611 &   0.172 &  $-$46.0 &  $-$31.5 & 4.2 & N \\
G351.688$+$0.171 & 351.688 &   0.171 &  $-$47.5 &  $-$35.0 & 41.54 & Y \\
G351.775$-$0.536 & 351.775 &  $-$0.536 &   $-$9.0 &    3.0 & 231 & Y \\
G352.083$+$0.167 & 352.083 &   0.167 &  $-$68.2 &  $-$63.6 & 6.77 & Y \\
G352.111$+$0.176 & 352.111 &   0.176 &  $-$61.0 &  $-$50.0 & 7.46 & Y \\
G352.133$-$0.944 & 352.133 &  $-$0.944 &  $-$18.8 &   $-$5.6 & 16.32 & Y \\
G352.517$-$0.155 & 352.517 &  $-$0.155 &  $-$52.0 &  $-$49.0 & 9.69 & N \\
G352.525$-$0.158 & 352.525 &  $-$0.158 &  $-$62.0 &  $-$52.0 & 0.7 & N \\
G352.584$-$0.185 & 352.584 &  $-$0.185 &  $-$92.6 &  $-$79.7 & 6.38 & N \\
G352.604$-$0.225 & 352.604 &  $-$0.225 &  $-$85.0 &  $-$81.0 & 3.3 & Y \\
G352.855$-$0.201 & 352.855 &  $-$0.200 &  $-$54.1 &  $-$50.1 & 1.29 & N \\
G353.216$-$0.249 & 353.216 &  $-$0.249 &  $-$25.0 &  $-$15.0 & 5.14 & N \\
G353.273$+$0.641 & 353.273 &   0.641 &   $-$7.0 &   $-$3.0 & 8.3 & Y \\
G353.363$-$0.166 & 353.363 &  $-$0.166 &  $-$80.1 &  $-$78.3 & 2.79 & N \\
G353.370$-$0.091 & 353.370 &  $-$0.091 &  $-$56.0 &  $-$43.4 & 1.35 & N \\
G353.378$+$0.438 & 353.378 &   0.438 &  $-$16.5 &  $-$14.0 & 0.97 & N \\
G353.410$-$0.360 & 353.410 &  $-$0.360 &  $-$23.0 &  $-$19.0 & 116 & Y \\
G353.429$-$0.090 & 353.429 &  $-$0.090 &  $-$63.9 &  $-$45.0 & 13.39 & N \\
G353.464$+$0.562 & 353.464 &   0.562 &  $-$52.7 &  $-$48.7 & 11.88 & Y \\
G353.537$-$0.091 & 353.537 &  $-$0.091 &  $-$59.0 &  $-$54.0 & 2.51 & Y \\
G354.206$-$0.038 & 354.205 &  $-$0.038 &  $-$37.5 &  $-$35.0 & 1.11 & N \\
G354.308$-$0.110 & 354.308 &  $-$0.110 &   11.0 &   19.5 & 3.44 & N \\
G354.496$+$0.083 & 354.495 &   0.083 &   17.5 &   27.5 & 8.41 & N \\
G354.615$+$0.472 & 354.615 &   0.472 &  $-$27.0 &  $-$12.5 & 166 & Y \\
G354.701$+$0.299 & 354.701 &   0.300 &   98.0 &  104.0 & 1.29 & N \\
G354.724$+$0.300 & 354.724 &   0.300 &   91.0 &   95.0 & 12.58 & N \\
G355.184$-$0.419 & 355.184 &  $-$0.419 &   $-$2.0 &   $-$0.5 & 1.35\tablefootmark{b} & Y \\
G355.343$+$0.148 & 355.343 &   0.148 &    4.0 &    7.0 & 1.24 & N \\
G355.344$+$0.147 & 355.344 &   0.147 &   19.0 &   21.0 & 10.17 & N \\
G355.346$+$0.149 & 355.346 &   0.149 &    9.0 &   12.5 & 7.39 & N \\
G355.538$-$0.105 & 355.538 &  $-$0.105 &   $-$3.5 &    5.0 & 1.25 & Y \\
G355.545$-$0.103 & 355.545 &  $-$0.103 &  $-$31.0 &  $-$27.5 & 1.22 & Y \\
G355.642$+$0.398 & 355.642 &   0.398 &   $-$9.0 &   $-$6.9 & 1.44 & N \\
G355.666$+$0.374 & 355.666 &   0.374 &   $-$4.5 &    0.6 & 2.47 & N \\
G356.054$-$0.095 & 356.054 &  $-$0.095 &   15.6 &   17.7 & 0.52 & N \\
G356.662$-$0.263 & 356.662 &  $-$0.263 &  $-$57.0 &  $-$44.0 & 8.38 & Y \\
G357.558$-$0.321 & 357.558 &  $-$0.321 &   $-$5.5 &    0.0 & 2.16 & N \\
G357.559$-$0.321 & 357.558 &  $-$0.321 &   15.0 &   18.0 & 2.01 & N \\
G357.922$-$0.337 & 357.922 &  $-$0.337 &   $-$5.5 &   $-$4.0 & 0.97 & Y \\
G357.924$-$0.337 & 357.924 &  $-$0.337 &   $-$4.5 &    3.0 & 2.34 & Y \\
G357.965$-$0.164 & 357.965 &  $-$0.163 &   $-$9.0 &    3.0 & 2.74 & Y \\
G357.967$-$0.163 & 357.967 &  $-$0.163 &   $-$6.0 &    0.0 & 47.5 & Y \\
G358.371$-$0.468 & 358.371 &  $-$0.468 &   $-$1.0 &   13.0 & 44.01 & Y \\
G358.386$-$0.483 & 358.386 &  $-$0.483 &   $-$7.0 &   $-$5.0 & 6.95 & Y \\
G358.460$-$0.391 & 358.460 &  $-$0.391 &   $-$0.5 &    4.0 & 47.73 & Y \\
G358.460$-$0.393 & 358.460 &  $-$0.393 &   $-$8.5 &    6.0 & 11.19 & Y \\
G358.721$-$0.126 & 358.721 &  $-$0.126 &    8.8 &   13.9 & 2.99 & N \\
G358.809$-$0.085 & 358.809 &  $-$0.085 &  $-$60.3 &  $-$50.5 & 6.86 & Y \\
G358.841$-$0.737 & 358.841 &  $-$0.737 &  $-$30.0 &  $-$17.0 & 10.94 & Y \\
G358.906$+$0.106 & 358.906 &   0.106 &  $-$20.5 &  $-$16.5 & 1.7 & N \\
G358.931$-$0.030 & 358.931 &  $-$0.029 &  $-$22.0 &  $-$14.5 & 5.9 & N \\
G358.980$+$0.084 & 358.980 &   0.084 &    5.0 &    7.0 & <0.2 & Y \\
G359.615$-$0.243 & 359.615 &  $-$0.243 &   14.0 &   27.0 & 38.62 & Y \\
G359.938$+$0.170 & 359.938 &   0.170 &   $-$1.5 &    0.2 & 2.34 & Y \\
G359.970$-$0.457 & 359.969 &  $-$0.457 &   20.0 &   24.1 & 2.39 & Y \\
G0.092$+$0.663 &   0.092 &  $-$0.663 &   10.0 &   25.0 & 18.86 & Y \\
G0.167$-$0.446 &   0.167 &  $-$0.446 &    9.5 &   17.0 & 1.33 & Y \\
G0.315$-$0.201 &   0.315 &  $-$0.201 &   14.0 &   27.0 & 62.6 & Y \\
G0.316$-$0.201 &   0.316 &  $-$0.201 &   20.0 &   22.0 & 0.58 & Y \\
G0.409$-$0.504 &   0.409 &  $-$0.504 &   24.5 &   27.0 & 2.61 & Y \\
G0.475$-$0.010 &   0.475 &  $-$0.009 &   23.0 &   31.0 & 3.14 & N \\
G0.496$+$0.188 &   0.496 &   0.188 &  $-$12.0 &    2.0 & 24.51 & N \\
G0.546$-$0.852 &   0.546 &  $-$0.851 &    8.0 &   20.0 & 61.92 & Y \\
G0.647$-$0.055 &   0.647 &  $-$0.055 &   49.0 &   52.0 & 2.0\tablefootmark{b} & N \\
G0.695$-$0.038 &   0.695 &  $-$0.038 &   64.0 &   75.0 & 32.33 & N \\
G0.836$+$0.184 &   0.836 &   0.185 &    2.0 &    5.0 & 6.64 & N \\
G1.008$-$0.237 &   1.008 &  $-$0.237 &    1.0 &    7.0 & 13.57 & N \\
G1.329$+$0.150 &   1.329 &   0.150 &  $-$13.5 &  $-$11.0 & 2.08 & N \\
G1.719$-$0.088 &   1.719 &  $-$0.088 &   $-$9.0 &   $-$4.5 & 7.82 & N \\
G2.143$+$0.009 &   2.143 &   0.009 &   54.0 &   65.0 & 7.08 & Y \\
G2.521$-$0.220 &   2.521 &  $-$0.220 &   $-$7.5 &    5.0 & 1.02 & Y \\
G2.536$+$0.198 &   2.536 &   0.198 &    2.0 &   20.5 & 29.4 & Y \\
G2.591$-$0.029 &   2.591 &  $-$0.029 &   $-$9.5 &   $-$4.0 & 1.76 & N \\
G2.615$+$0.134 &   2.615 &   0.134 &   93.5 &  104.0 & 1.22 & N \\
G2.703$+$0.040 &   2.703 &   0.040 &   91.5 &   98.0 & 8.97 & N \\
G3.253$+$0.018 &   3.253 &   0.019 &   $-$1.5 &    3.5 & 3.54 & N \\
G3.312$-$0.399 &   3.312 &  $-$0.399 &    0.0 &   10.0 & 1.17 & Y \\
G3.442$-$0.348 &   3.442 &  $-$0.348 &  $-$35.5 &  $-$34.5 & 1.06 & N \\
G3.502$-$0.200 &   3.502 &  $-$0.200 &   43.0 &   45.5 & 1.57 & N \\
G3.910$+$0.001 &   3.910 &   0.001 &   15.0 &   24.5 & 5.04 & Y \\
G4.393$+$0.079 &   4.393 &   0.079 &    0.0 &    9.0 & 6.74 & N \\
G4.434$+$0.129 &   4.434 &   0.129 &   $-$1.5 &    8.0 & 3.29 & N \\
G4.569$-$0.079 &   4.569 &  $-$0.079 &    9.0 &   10.0 & 0.44 & N \\
G4.586$+$0.028 &   4.586 &   0.028 &   15.0 &   27.0 & 1.16 & N \\
G4.676$+$0.276 &   4.676 &   0.276 &   $-$5.5 &    6.0 & 2.06 & N \\
G4.866$-$0.171 &   4.866 &  $-$0.171 &    5.0 &    6.0 & 0.56 & N \\
G5.618$-$0.082 &   5.618 &  $-$0.082 &  $-$28.0 &  $-$18.5 & 3.37 & Y \\
G5.630$-$0.294 &   5.630 &  $-$0.294 &    9.0 &   22.0 & 1.28 & Y \\
G5.657$+$0.416 &   5.657 &   0.416 &   13.0 &   22.0 & 1.75 & Y \\
G5.677$-$0.027 &   5.677 &  $-$0.027 &  $-$14.5 &  $-$11.0 & 0.79 & N \\
G5.885$-$0.393 &   5.885 &  $-$0.393 &    6.0 &    7.5 & 0.48 & Y \\
G5.900$-$0.430 &   5.900 &  $-$0.429 &    0.0 &   10.6 & 6.2 & N \\
\end{longtable}
\tablefoot{
\tablefoottext{a}{For easier comparison with the \water\, masers, here we list the upper and lower bounds
of the \meth\, maser emission, as listed by C10.}
\tablefoottext{b}{For \meth\, masers not detected in the C10 follow-up data, we use their survey cube data
or the value they list from \citet{houg95}.}
}
\end{longtab}

\begin{table}
\caption{\label{water-nometh}\water\, Masers not Associated with \meth\, Masers}
\begin{tabular}{lcccccc}
Number  & $\ell$\tablefootmark{a} & $b$\tablefootmark{a} & $v$\tablefootmark{a} & $T_{peak}$ \\
 & (\degreesym) & (\degreesym) & (\kms) & (K) \\
\hline\hline
           9 &  $-$1.337 &  $-$0.053 &   $-$8.7 &   0.74 \\ 
          12 &  $-$1.337 &  $-$0.053 &  $-$15.9 &   0.45 \\ 
          13 &   0.055 &  $-$0.220 &   13.2 &   0.54 \\ 
          15 &  $-$0.678 &  $-$0.036 & $-$110.4 &   0.16 \\ 
          17 &   0.613 &   0.005 &    9.5 &   0.16 \\ 
          18 &  $-$0.704 &   0.030 &   $-$1.4 &   0.11 \\ 
          19 &  $-$1.337 &  $-$0.036 &    2.2 &   0.16 \\ 
          20 &   0.155 &  $-$0.561 &  $-$70.4 &   0.20 \\ 
          23 &   1.163 &  $-$0.019 &  $-$19.6 &   0.08 \\ 
          24 &   0.763 &  $-$0.253 &  $-$48.6 &   0.09 \\ 
          25 &  $-$0.662 &   0.281 &   $-$1.4 &   0.16 \\ 
          27 &   0.197 &  $-$0.153 &   45.9 &   0.06 \\ 
          28 &  $-$1.103 &   0.022 &  $-$15.9 &   0.13 \\ 
          29 &  $-$0.720 &   0.164 &   53.1 &   0.06 \\ 
          31 &  $-$0.070 &  $-$0.153 &   $-$8.7 &   0.07 \\ 
          32 &  $-$0.070 &  $-$0.144 &  $-$19.6 &   0.07 \\ 
          33 &  $-$0.245 &  $-$0.378 &  $-$59.5 &   0.10 \\ 
          35 &  $-$1.462 &  $-$0.195 & $-$208.5 &   0.07 \\ 
          37 &   1.705 &  $-$0.486 &  $-$37.7 &   0.07 \\ 
\hline
\end{tabular}
\tablefoot{
\tablefoottext{a}{Values of $\ell, b,$\, and $v$\, are
    extracted from the Mopra data cube, which has a pixel size of
    0.5\arcmin\, and a channel width of $\sim$~3.6~\kms.}  
}
\end{table}

\section{Discussion \label{discussion}}

\subsection{Association Rate as a Function of Galactic Longitude}
To search for trends in the association rate of 6.7~GHz masers and
4.5~\um\, emission, we binned the sources (and association rates) by
Galactic Longitude.  The results are shown in Figure~\ref{hit_pct}.
The association rate varies from bin to bin (from 20~$\pm$~14\% to
85~$\pm$~25\%), but we find no strong trend with Galactic longitude.
There appears to a decrement of associations at
$\ell~\sim~-5$\degreesym, and enhancements of associations at
$\ell~\sim~-2$\degreesym and $-8$\degreesym.  Because of the large
error bars and the large bin-to-bin variations, we consider the likely
cause of these to be the results of small number statistics and
binning.

\subsection{Association Rate as a Function of Galactic Latitude}
We find that \meth\, masers near the Galactic midplane ($|b|~<~$10\arcmin) are
less likely (34.8~$\pm$~6.4\%) to be associated with green sources than
those above or below it (62.9~$\pm$~8.4\%).  While this could be a real
effect, and masers outside the midplane could simply have higher
association rates with 4.5~\um\, sources than those in the midplane,
we find it much more likely to be an optical depth effect of having
two star formation tracers at very different wavelengths/frequencies.

The Galactic plane is optically thin at 6.7~GHz, making it possible to
detect 6.7~GHz \meth\, masers throughout the Galaxy, including the far
side of the Galactic plane.  This is not the case, however, for
4.5~\um\, emission.  At this much shorter wavelength, extinction plays
a much larger role \citep[see, e.g.,][]{pand87}, preventing the detection of green sources on the
far side of the Galaxy (behind, e.g., the molecular clouds on
the near side of the Galaxy).  Because the extinction is higher in the
Galactic plane, we are likely to have a cleaner line of sight to
6.7~GHz masers that reside above or below the Galactic plane.  Thus,
we find a lower association rate of 4.5~\um\, emission with 6.7~GHz
\meth\, masers for sources that are in the midplane of the Galaxy
($l~<$~10\arcmin) than those above or below it.

\subsection{Comparing the Galactic Center with the Galactic Disk\label{gal_cen}}
To see how star formation in the Galactic center region may differ
from that of the disk, we compare the 6.7~GHz \meth\, association rate
with enhanced 4.5~\um\, emission.  As described in
Section~\ref{dist-results}, we make three different distance esimates based
on the positions of the masers, the velocities of the masers, and the
combination of these two criteria.

Using Galactic coordinates alone ($|\ell|~<$~1.3\degreesym,
$|b|~<$~10\arcmin) to estimate the distances to the \meth\, masers,
and assuming that all masers that fit the distance criteria are indeed
in the GC (and not the foreground), we find that 34.8~$\pm$~12.3\% are
associated with green sources in the GC, and that 51.3~$\pm$~5.8\% of
disk sources are associated with green sources.  The association rate
for the disk includes all Galactic latitudes.  Because the GC is in
the Galactic midplane, however, we should compare the GC association
rate to the association rate in the rest of the midplane
(6~\degreesym$<~\ell~<~345$\degreesym), excluding the GC longitudes
($|\ell|~<$~1.3\degreesym).  When we do this, we find that the disk
midplane association rate is 34.9~$\pm$~7.4\%, very close to the
association rate for the GC.  Thus, we find no differences between the
two regions using position alone as a distance indicator.

When we use only the source velocity to estimate the locations of the
masers, we find nearly identical association rates between the \meth\,
masers and the enhanced 4.5~\um\, emission sources in the GC
(48.1~$\pm$~13.3\%) and the disk (49.3~$\pm$~5.8\%).  Because the
positions of the masers are not considered in this distance estimate,
no adjustment for the midplane is necessary.  The similar association
rates again point to no significant difference between the GC and disk
regions.

\begin{figure}
\centering
\includegraphics[angle=0,width=\hsize,clip=true]{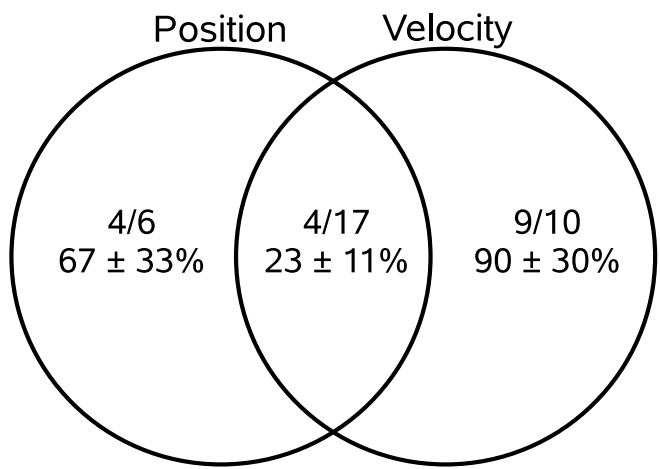}\\
\caption{Diagram showing green source detection rates based on two different
  distance estimates.  Sources placed in the Galactic center based on position
  all have $|\ell|~<$~1.3\degreesym\, and $|b|~<$~10\arcmin.  Distance
  estimates based on velocity are taken from C10.  For comparison, the green
  source detection rates for masers that are not at the distance of the
  Galactic center (based on position or velocity) is 48.6~$\pm$~5.9\%\,
  (69/142). \label{venn}}
\end{figure}

Using both positional and velocity criteria to place sources at the
distance of the Galactic center, we find a detection rate
(23.5~$\pm$~11.8\%) that is lower than that of all other midplane sources
(34.9~$\pm$~7.4\%).  While the association rate in the GC is lower
using both distance criteria, the large error bars keep this from
being a significant result.  When we look at the sources placed in the
GC region by only {\it one} of the two indicators (position {\it or}
velocity), we find that they have high association
rates, 66.7~$\pm$~33.3\%\, for those sources whose postions match the
GC, and 90.0~$\pm$~30.0\%\, for those whose velocities match the GC.
Figure~\ref{venn}\, showns a diagram with these association
rates.  These elevated rates could simply be a result of small number
statistics, since when taking into account their error bars, they are
consistent with the disk association rate, especially at larger
Galactic latitudes.  Nevertheless, this difference remains intriguing.

As described in Section~\ref{water_data}, the Mopra \water\, survey
covered the region $-1.5$\degreesym~$<~\ell~<~2$\degreesym,
$-0.6$\degreesym~$<~b~<~0.5$\degreesym\, with a velocity coverage of
$-$250 to 300~\kms.  This covers the entire CMZ, with only limited
coverage outside of the CMZ.  Nevertheless, we can compare
the \water\, masers in the CMZ to those outside the CMZ, as determined
by their positions.  We find that of the 27 \water\, masers in the
CMZ, 18 are associated with \meth\, masers (67~$\pm$~16\%).  Outside
of the CMZ, we find that none of the 10 \water\, masers are associated
with \meth\, masers.  Thus, there appears to be a significant
difference between the \water\, masers in these regions.  As discussed
above, the \water\, maser emission from Sgr~B2 is the dominant feature
of the survey.  Even if we remove the 9 Sgr~B2 \water\, masers (all of
which are associated with \meth\, masers), the difference persists
(50~$\pm$~17\%).

It is important to note that \water\, masers are also found toward AGB
stars.  These \water\, masers tend to be less luminous than those
found toward high-mass star forming regions \citep{pala93,casw11}.
Indeed, in a 0.5~deg$^2$ \water\, maser survey of the Galactic center,
\citet{casw11}\, found that only 3 of their 27 detected masers could
be definitely attributed to AGB stars.  Using this detection rate, of
the 37 \water\, masers detected in our Mopra survey, $\sim$~4 may be
associated with evolved AGB stars rather than sites of star formation.
The \water\, masers associated with \meth\, masers or green sources
are very unlikely to also be associated with AGB stars.  Thus, the
effect of the small number of AGB \water\, masers in our sample is
negligible.

The fact that we do not find any \meth\, masers toward the \water\,
masers outside the CMZ is intriguing, and may be related to a
different timescale of star formation inside the CMZ when compared to
the Galactic disk.  While still somewhat speculative, some recent work
\citep{elli07,bree10} has proposed an evolutionary sequence of \meth\,
masers in which Class~I \meth\, masers, generated by protostellar
outflows, are formed before radiatively excited Class~II \meth\,
masers (e.g., the 6.7~GHz \meth\, maser).  In addition, some
observations have suggested that the 22~GHz \water\, maser also arises
in an early phase of star formation and overlaps with the Class~I
\meth\, masers, in particular toward green sources
\citep[e.g.,][]{cham09}.  Thus, if we assume that the 22~GHz \water\,
masers arise at an earlier evolutionary state than the 6.7~GHz \meth\,
masers, then perhaps the much higher correlation rate of the two
masers in the GC (67\%\, vs 0\%) corresponds to a longer overlap
period of the two stages, and possibly a compressed timescale of star
formation in the GC relative to that of the disk.  This speculative
result can be tested using higher-resolution \water\, maser data to
verify the GC correlations, a larger number of \water\, and \meth\,
masers in the GC (from more sensitive surveys), as well as large
co-spatial surveys of \water\, and \meth\, masers in the Galactic
plane to better establish the correlation rate in the disk.

\subsection{Asymmetry in the \water\, Maser Distribution around the Galactic Center \label{asymmetry}}

\citet{yuse09} found an asymmetry in the distribution of young stellar
object (YSO) candidates as probed by 24~\um\, sources, with a larger
number of YSO candidates with $\ell~<~0$ relative to $\ell~>~0$.  The
distribution of \water\, masers as a function of Galactic longitude
seen in Figure~\ref{water_dist} shows no obvious asymmetry with
respect to $\ell~=~0$, with 19 masers having $\ell~>~0$ and 18 masers
having $\ell~<~0$.  The most striking feature in this distribution of
\water\, masers is the peak at $\sim$~$\ell~=~0.7$, which is almost
entirely due to masers in Sgr~B2 (10 of the 11 masers in the bin have
coordinates that place them within Sgr~B2).  If we remove these 10 masers
from the distribution, we find that there are 11 \water\, masers with
$\ell~>~0$, and 16 with $\ell~<~0$, consistent with the asymmetry
found by \citet{yuse09}.  One explanation for the asymmetry offered by
\citet{yuse09} was that higher extinction due to dense clouds may be
responsible for the asymmetry in the distribution of 24~\um\, sources.
Extinction is not a problem at the frequency of the \water\, masers,
suggesting that the asymmetry may be a real feature.  The larger numbers
of both 24~\um\, sources and \water\, masers at $\ell~<~0$ indicate
an enhancement of on-going star formation in this region.

\section{Conclusions \label{conclusions}}
We present the results of our study of star formation in the CMZ and
how it compares to the Galactic disk.  We make our comparison through
the correlation of 6.7~GHz \meth\, masers, 22~GHz \water\, masers, and
green sources.  As part of this study, we developed an automated
algorithm to identify regions of extended, enhanced 4.5~\um\, emission
toward 6.7~GHz \meth\, masers.  This new technique can be a powerful
tool to trace regions of star formation provided the column density
along the line of sight is not too high.  Moreover, we used the Mopra
telescope to survey the 22~GHz \water\, maser emission in the CMZ,
resulting in the detection of 37 inidividual water masers.

We find that the association rate of \meth\, masers with green sources
as a function of Galactic longitude is relatively flat.  In Galactic
latitude, on the other hand, we find that there is a significant
decrease in the association rate along the Galactic midplane, which is
likely an infrared extinction effect.  Using two different distance
estimates, we do not detect any significant difference in the
association rates of green sources with \meth\, masers between the
Galactic center and the disk.  This suggests that once the star
formation process has begun, its observational signatures are similar
in the GC and the Galaictic disk, despite the different physical
conditions of the gas in the two regions.

We find that the \water\, masers in our survey are much more likely to
be associated with a \meth\, maser if they are located in the Galactic
center rather than the Galactic disk.  Indeed, we find zero \water\,
masers (out of 10) associated with \meth\, masers outside the Galactic
center region.  This may indicate a different timescale of star
formation in the GC when compared to the Galactic disk.  Future high
angular resolution observations of the newly detected \water\, masers
are needed to better establish the positions of the \water\, masers
and test this result.

\begin{acknowledgements}
  Acknowledgements to be included.
\end{acknowledgements}



\begin{thebibliography}{}

\bibitem[Ao et al.(2013)]{ao13} Ao, Y., Henkel, C., Menten, K.~M., et
  al.\ 2013, \aap, 550, A135

\bibitem[Arendt et al.(2008)]{aren08} Arendt, R.~G., et al.\ 
2008, \apj, 682, 384 

\bibitem[Benjamin et al.(2003)]{benj03} Benjamin, R.~A., et al. 2003, \pasp,
  115, 953

\bibitem[Breen et al.(2010)]{bree10} Breen, S.~L., Ellingsen, S.~P., Caswell,
  J.~L., \& Lewis, B.~E.\ 2010, \mnras, 401, 2219

\bibitem[Caswell(1997)]{casw97} Caswell, J.~L.\ 1997, \mnras, 289, 203

\bibitem[Caswell et al.(2010)]{casw10} Caswell, J.~L., et al.
  2010, \mnras, 404, 1029 

\bibitem[Caswell et al.(2011)]{casw11} Caswell, J.~L., Breen, S.~L.,
  \& Ellingsen, S.~P.\ 2011, \mnras, 410, 1283

\bibitem[Chambers et al.(2009)]{cham09} Chambers, E.~T., Jackson, J.~M.,
  Rathborne, J.~M., \& Simon, R.\ 2009, \apjs, 181, 360

\bibitem[Chambers et al.(2011)]{cham11} Chambers, E.~T., 
  Yusef-Zadeh, F., \& Roberts, D.\ 2011, \apj, 733, 42 

\bibitem[Chen et al.(2009)]{chen09} Chen, X., Ellingsen, S.~P., \& Shen,
  Z.-Q.\ 2009, \mnras, 396, 1603

\bibitem[Claussen et al.(1998)]{clau98} Claussen, M.~J., Marvel,
  K.~B., Wootten, A., \& Wilking, B.~A.\ 1998, \apjl, 507, L79

\bibitem[Cox \& Laureijs(1989)]{cox89} Cox, P., \& Laureijs, R. 1989, The
  Center of the Galaxy, 136, 121

\bibitem[Cragg et al.(1992)]{crag92} Cragg, D.~M., Johns, 
  K.~P., Godfrey, P.~D., \& Brown, R.~D.\ 1992, \mnras, 259, 203 

\bibitem[Cyganowski et al.(2008)]{cyga08} Cyganowski, C.~J., et al.\ 2008,
  \aj, 136, 2391

\bibitem[Cyganowski et al.(2009)]{cyga09} Cyganowski, C.~J., Brogan, C.~L.,
  Hunter, T.~R., \& Churchwell, E.\ 2009, \apj, 702, 1615

\bibitem[De Buizer \& Vacca(2010)]{debu10} De Buizer, J.~M., \& Vacca,
  W.~D.\ 2010, \aj, 140, 196

\bibitem[Elitzur et al.(1989)]{elit89} Elitzur, M., Hollenbach, D.~J.,
  \& McKee, C.~F.\ 1989, \apj, 346, 983

\bibitem[Ellingsen(2005)]{elli05} Ellingsen, S.~P.\ 2005, \mnras, 359, 1498

\bibitem[Ellingsen et al.(2007)]{elli07} Ellingsen, S.~P., Voronkov, M.~A.,
  Cragg, D.~M., Sobolev, A.~M., Breen, S.~L., \& Godfrey, P.~D. 2007, IAU
  Symposium, 242, 213

\bibitem[Fazio et al.(2004)]{2004ApJS..154...10F} Fazio, G.~G., Hora,
  J.~L., Allen, L.~E., et al.\ 2004, \apjs, 154, 10

\bibitem[Fazio et al.(2004)]{fazi04} Fazio, G.~G., Hora, J.~L., Allen,
  L.~E., et al.\ 2004, \apjs, 154, 10

\bibitem[Ferri{\`e}re et al.(2007)]{ferr07} Ferri{\`e}re, K., Gillard,
  W., \& Jean, P.\ 2007, \aap, 467, 611

\bibitem[Foster et al.(2011)]{fost12} Foster, J.~B., Jackson, J.~M., Chambers,
  E.~T., \& Stojimirovic, I.\ 2010, \apj, submitted.

\bibitem[Furuya et al.(2001)]{furu01} Furuya, R.~S., Kitamura, Y.,
  Wootten, H.~A., Claussen, M.~J., \& Kawabe, R.\ 2001, \apjl, 559,
  L143

\bibitem[Genzel et al.(1978)]{genz78} Genzel, R., Downes, D., Moran,
  J.~M., et al.\ 1978, \aap, 66, 13

\bibitem[Gwinn(1994)]{gwin94} Gwinn, C.~R.\ 1994, \apj, 429, 241

\bibitem[Hollenbach et al.(2013)]{holl13} Hollenbach, D., Elitzur, M.,
  \& McKee, C.~F.\ 2013, \apj, 773, 70

\bibitem[Houghton \& Whiteoak(1995)]{houg95} Houghton, S., \&
  Whiteoak, J.~B.\ 1995, \mnras, 273, 1033

\bibitem[H\"uttemeister et al.(1998)]{hutt98}
  Huettemeister, S., Dahmen, G., Mauersberger, R., et al.\ 1998, \aap,
  334, 646

\bibitem[Immer et al.(2012)]{imme12} Immer, K., Menten, K.~M.,
  Schuller, F., \& Lis, D.~C.\ 2012, \aap, 548, A120

\bibitem[Jones et al.(2011)]{jone11} Jones, P.~A., Burton, M.~G.,
  Tothill, N.~F.~H., \& Cunningham, M.~R.\ 2011, \mnras, 411, 2293

\bibitem[Jones et al.(2012)]{jone12} Jones, P.~A., Burton, M.~G.,
  Cunningham, M.~R., et al.\ 2012, \mnras, 419, 2961

\bibitem[Jones et al.(2013)]{jone13} Jones, P.~A., Burton, M.~G.,
  Cunningham, M.~R., Tothill, N.~F.~H., \& Walsh, A.~J.\ 2013, \mnras,
  433, 221

\bibitem[Kennicutt(1998)]{kenn98} Kennicutt, R.~C., Jr.\ 1998, \araa, 36, 189 

\bibitem[Longmore et al.(2012)]{long12} Longmore, S.~N., Rathborne,
  J., Bastian, N., et al.\ 2012, \apj, 746, 117

\bibitem[Marston et al.(2004)]{mars04} Marston, A.~P., Reach, 
  W.~T., Noriega-Crespo, A., et al.\ 2004, \apjs, 154, 333 

\bibitem[Mart{\'{\i}}n-Pintado et al.(2000)]{mart00}
  Mart{\'{\i}}n-Pintado, J., de Vicente, P., Rodr{\'{\i}}guez-Fern{\'a}ndez,
  N.~J., Fuente, A., \& Planesas, P.\ 2000, \aap, 356, L5

\bibitem[Menten(1991)]{ment91} Menten, K.~M.\ 1991, \apjl, 380, L75

\bibitem[Mills \& Morris(2013)]{mill13} Mills, E.~A.~C., \& Morris,
  M.~R.\ 2013, \apj, 772, 105

\bibitem[Pandey \& Mahra(1987)]{pand87} Pandey, A.~K., \& Mahra,
  H.~S.\ 1987, \mnras, 226, 635

\bibitem[Minier et al.(2003)]{mini03} Minier, V., Ellingsen, S.~P., Norris,
  R.~P., \& Booth, R.~S.\ 2003, \aap, 403, 1095

\bibitem[Molinari et al.(2011)]{moli11} Molinari, S., Bally, 
J., Noriega-Crespo, A., et al.\ 2011, \apjl, 735, L33 

\bibitem[Morris \& Serabyn(1996)]{morr96} Morris, M., \& Serabyn, E.\ 1996,
  \araa, 34, 645

\bibitem[Noriega-Crespo et al.(2004)]{nori04} Noriega-Crespo, 
  A., Morris, P., Marleau, F.~R., et al.\ 2004, \apjs, 154, 352 

\bibitem[Norman \& Silk(1979)]{norm79} Norman, C., \& Silk, J.\ 1979,
  \apj, 228, 197

\bibitem[Odenwald \& Fazio(1984)]{oden84} Odenwald, S.~F., \&
  Fazio, G.~G.\ 1984, \apj, 283, 601

\bibitem[Oka et al.(2005)]{oka05} Oka, T., Geballe, T.~R., Goto, M.,
  Usuda, T., \& McCall, B.~J.\ 2005, \apj, 632, 882

\bibitem[Palagi et al.(1993)]{pala93} Palagi, F., Cesaroni, R.,
  Comoretto, G., Felli, M., \& Natale, V.\ 1993, \aaps, 101, 153

\bibitem[Pierce-Price et al.(2000)]{pier00} Pierce-Price, D., et al. 2000,
  \apjl, 545, L121

\bibitem[Ram{\'{\i}}rez et al.(2008)]{rami08} Ram{\'{\i}}rez, S.~V., Arendt,
  R.~G., Sellgren, K., Stolovy, S.~R., Cotera, A., Smith, H.~A., \&
  Yusef-Zadeh, F.\ 2008, \apjs, 175, 147

\bibitem[Rathborne et al.(2005)]{rath05} Rathborne, J.~M., Jackson, J.~M.,
  Chambers, E.~T., Simon, R., Shipman, R., \& Frieswijk, W. 2005, \apjl, 630,
  L181

\bibitem[Rathborne et al.(2006)]{rath06} Rathborne, J.~M., Jackson, J.~M., \&
  Simon, R. 2006, \apj, 641, 389

\bibitem[Riquelme et al.(2010)]{riqu10} Riquelme, D., Bronfman,
  L., Mauersberger, R., May, J., \& Wilson, T.~L. 2010, \aap, 523, A45

\bibitem[Stolovy et al.(2006)]{stol06} Stolovy, S., et al.\ 2006, Journal of
  Physics Conference Series, 54, 176

\bibitem[Urquhart et al.(2010)]{urqu10} Urquhart, J.~S., Hoare, M.~G.,
  Purcell, C.~R., et al.\ 2010, \pasa, 27, 321

\bibitem[Walsh et al.(2001)]{wals01} Walsh, A.~J., Bertoldi, F., Burton,
  M.~G., \& Nikola, T.\ 2001, \mnras, 326, 36

\bibitem[Walsh et al.(2011)]{wals11} Walsh, A.~J., Breen, S.~L.,
  Britton, T., et al.\ 2011, \mnras, 416, 1764

\bibitem[Williams et al.(1994)]{will94} Williams, J.~P., de Geus,
  E.~J., \& Blitz, L.\ 1994, \apj, 428, 693

\bibitem[Yusef-Zadeh et al.(2007)]{yuse07} Yusef-Zadeh, F., et al.\ 2007, IAU
  Symposium, 242, 366

\bibitem[Yusef-Zadeh et al.(2009)]{yuse09} Yusef-Zadeh, F., Hewitt,
  J.~W., Arendt, R.~G., et al.\ 2009, \apj, 702, 178

\bibitem[Yusef-Zadeh et al.(2013a)]{yuse13a} Yusef-Zadeh, F., Cotton,
  W., Viti, S., Wardle, M., \& Royster, M.\ 2013a, \apjl, 764, L19

\bibitem[Yusef-Zadeh et al.(2013b)]{yuse13b} Yusef-Zadeh, F., et
  al. 2013b, JPhCh, in press


\end{thebibliography}
\end{document}